# Statistical Models of Ionospheric Variability and Irregularities in the Topside Ionosphere Based on the Swarm Satellite Data

Daria Kotova[(1)], Alan Wood[(2)], Eelco Doornbos[(3)], Jaroslav Urbář[(4)], Luca Spogli[(5)], Yaqi Jin[(1)], Lucilla Alfonsi[(5)], Gareth Dorrian[(6)], Mainul Hoque[(7)], Kasper van Dam[(3)], Elisabetta Iorfida[(8)], and Wojciech J. Miloch[(1)]
(1) Department of Physics, University of Oslo, Norway
(2) Met Office, Exeter, UK
(3) The Royal Netherlands Meteorological Institute (KNMI), The Netherlands
(4) Institute of Atmospheric Physics CAS, Czech Republic
(5) Istituto Nazionale di Geofisica e Vulcanologia, Italy
(6) Space Environment and Radio Engineering (SERENE) group, University of Birmingham, UK
(7) German Aerospace Center (DLR), Germany
(8) European Space Agency (ESA), Noordwijk, The Netherlands

## Abstract

The ionosphere is a highly complex plasma containing electron density structures with a wide range of spatial scales. Coupling of the ionosphere with the Earth's magnetosphere and the solar wind, as well as to the neutral atmosphere, makes the ionosphere highly dynamic and highly dependent on the driving processes. Thus, modelling the ionosphere and capturing its full dynamic range considering all spatiotemporal scales is challenging.

Swarm is the European Space Agency's (ESA) first constellation mission for Earth Observation, comprising multiple satellites in low Earth orbit. During the Swarm-VIP-Dynamic project, a suite of statistical models has been developed using observations from Swarm and proxies for heliogeophysical processes. The statistical modelling technique of Generalised Linear Modelling was used to create models for both the electron density and the variability of the plasma structures at horizontal spatial scales between 7.5 km and 100 km. Separate models were created for low, middle, auroral and polar latitudes. The models make predictions based on explanatory variables, which act as proxies for the underlying physical processes. The performance of the models of the electron density approached the theoretical best values for some of the goodness-of-fit statistics. This suggests that the modelling method is appropriate for the task undertaken. The models of ionospheric variability at larger spatial scales (~100 km) also perform well, however the model performance decreases at smaller spatial scales. This suggests that there are physical processes missing from the models. Possible candidates are instability processes or driving forces of the ionosphere by wave activity from below, neither of which are captured by the models.

## 1. Introduction

Solar extreme ultraviolet radiation ionizes the upper layers of the atmosphere, called the Earth's ionosphere. An additional source of ionization is the precipitation of charged particles, which occurs predominantly in polar regions and also causes the aurora borealis and australis (Kelley, 2009). The drivers of the ionosphere are highly variable, leading to variability and structuring of the ionosphere (e.g., Moen et al., 2013; Heelis & Maute, 2020). Coupling of the ionosphere with the Earth's magnetosphere and solar wind, as well as coupling to the neutral atmosphere, make the phenomena in the ionosphere highly dynamic. Ionospheric variability can be observed at different spatio-temporal scales, ranging from planetary down to decameter spatial scales, and from solar cycle down to daily, hourly, and even minutely temporal variations (Tsunoda, 1988). Thus, modelling of the whole ionosphere and capturing its full dynamic range is a challenging task.

The ionosphere forms part of the broader geospace system and responds sensitively to variations in solar and geomagnetic activity. Enhanced geomagnetic activity increases particle precipitation and field-aligned currents, drives plasma convection, triggers a diverse spectrum of instabilities and turbulence, and injects energy into the thermosphere (Hargreaves, 1992). The ionosphere also receives energy from the lower regions of the atmosphere, in the form of waves and tides (Jackson et al., 2019). Thus, it represents the ideal environment to study the signatures of both space weather and terrestrial processes, together with the interplay between them.

Ionospheric irregularities represent fluctuations in plasma density compared to background values and span a wide range of scales, from centimetres to hundreds of kilometres (Fejer & Kelley, 1980). Their formation results from a complex interplay of ionospheric dynamical processes. These include electrodynamical phenomena, transport processes, instabilities and turbulence, which are driven and influenced by space weather events and interactions with the neutral atmosphere. Ionospheric irregularities cause scintillation and phase disturbances in radio signals and can severely degrade the accuracy, integrity and availability of Global Navigation Satellite Systems (GNSS) (Kintner et al., 2007). Additionally, high-frequency (HF), very high-frequency (VHF), and ultra-high-frequency (UHF) radio communications, reliant on ionospheric reflection and refraction, are susceptible to these irregularities. During intense geomagnetic storms, communication blackouts, navigation errors, and increased radiation hazards may occur, underscoring the critical need for models capable of characterising the ionosphere in both quiet and disturbed conditions (Basu et al., 2001). From a societal perspective, reliable ionospheric specification and forecasting are essential, because ionospheric variations can disrupt a wide range of technological systems.

Different models have been developed over the years to characterise ionospheric plasma. The background ionosphere can be often modelled by, for example, empirical models, such as NeQuick-2 (Nava et al., 2011), International Reference Ionosphere (Bilitza et al, 2022) or Neustrelitz Electron Density Model (NEDM, Hoque et al. 2020). Another category of models based on physical interactions, such as the Thermosphere-Ionosphere-Electrodynamics General Circulation Model (TIE-GCM) or Global Ionosphere Thermosphere Model (GITM), are often used to understand the steady state ionosphere (Dickinson et al., 1981; Richmond et al., 1992; Ridley et al., 2006; Qian et al., 2014). TIE-GCM simulates the global ionosphere-thermosphere system by solving the coupled momentum, energy and continuity equations. While physically comprehensive, such models do not typically predict plasma structuring which is related to fast phenomena and instability and turbulent processes. Data assimilation systems, such as the Advanced Ensemble electron density (Ne) Assimilation System (AENeAS; Elvidge & Angling, 2019), combine physics-based modelling with observational constraints from sources such as GNSS total electron content (TEC) data, ionosondes and radio occultation. Their performance depends strongly on data availability and does not fully capture the smallest plasma-irregularity scales. Finally, climatological approaches use observations gathered over years to predict the behaviour of a system under similar conditions (Spogli et al., 2009; Alfonsi et al., 2011; Prikryl et al., 2011, De Franceschi et al., 2019), but this class of models often struggles to capture rarely encountered conditions. More recently, Jin et al. (2023) developed an empirical model of ROTI in the Arctic region using the Empirical Orthogonal Function (EOF) method. The model is built based on 12 years of ROTI maps from ground-based GNSS scintillation receivers and is able to well capture the ionospheric variability driven by various sources such as solar radiation, solar wind and IMF orientations.

Satellite missions have played an increasingly important role in advancing ionospheric modelling since a wide variety of spatio-temporal scales can be addressed with global coverage. ESA's Swarm mission provides high-quality, multi-instrument in-situ observations of electron density, magnetic field and thermospheric density at high temporal resolution and across a range of local times and seasons, allowing for both space weather and climatological studies (see, e.g., Jin et al., 2019). These measurements enable investigations of ionospheric processes including equatorial plasma bubbles, polar cap patches, auroral boundaries, turbulence, and demonstrate the strength of in-situ measurements in understanding ionospheric processes in detail (as reviewed by Wood et al., 2022). Thus, Swarm provides an exceptional foundation for improved empirical and semi-empirical modelling of ionospheric structure at multiple spatial scales.

Building on this capability, the Swarm-VIP project developed a first generation statistical models of electron density and multi-scale ionospheric variability at the horizontal spatial scales of 20 km, 50 km and 100 km using Generalised Linear Modelling (GLM) (Wood et al., 2024). These models covered four latitude regions (equatorial, midlatitude, auroral and polar) and captured certain climatological features of the topside ionosphere while remaining independent of the Swarm data used to construct them. The models had clear successes. The goodness-of-fit statistics showed that they did not exhibit significant bias, the precision showed that they captured some, but not all, of the variability of the ionospheric plasma and it was shown that they can replicate some, but not all, of the climatological features of the topside ionosphere (Spogli et al., 2024). One of the primary reasons for the limitations in model performance was due to the duration of the available Swarm dataset, which offered only limited longitudinal coverage of simultaneous sampling during all seasons, local times and solar activity levels. It was postulated that this was, in part, due to a missing process in the models being the neutral atmosphere. This led Wood et al. (2024) to conclude that a thermospheric data product was needed in the models, but at a higher temporal (and hence spatial) resolution than was currently available.

The Swarm-VIP-Dynamic project addresses these limitations by expanding the Swarm data coverage into the ascending and maximum phases of solar cycle 25, incorporating newly available accelerometer-based observations with a high (10 second) temporal, and hence a high spatial, resolution from which the thermospheric density can be inferred (DNSxACC) (Siemes et al., 2016), and developing a more flexible statistical-modelling framework. This work includes (i) a new database spanning almost 10 years from which the models can be developed, (ii) evaluation of statistical distributions of the variables which are to be predicted and their transformations, (iii) region- and hemisphere-specific model formulation, (iv) incorporation of new heliogeophysical proxies, (v) more realistic representation of the Equatorial Ionisation Anomaly (EIA), and (vi) the construction of three model families: full-physics models (v3.1), models based on version 3.1, but excluding Swarm-dependent explanatory variables(v3.2), and operational near real-time models (v3.3).

A central advancement of the Swarm-VIP-Dynamic project is the expanded capability to model ionospheric structuring at much smaller spatial scales than was previously possible. While the original Swarm-VIP models captured variability at scale sizes of ~20-100 km using parameters available in the IPIR product (Jin et al., 2022), the Swarm-VIP-Dynamic project incorporates additional high-resolution quantities that enable the characterisation of small-scale plasma irregularities down to sub-kilometer scales. This includes the 16 Hz faceplate (FP) electron-density measurements, used to construct a metric of the fine scale variability in the ionosphere. This variability occurs on scales down to ~500 m.\Such scales are relevant to GNSS Fresnel filtering and therefore essential for assessing scintillation impacts. Furthermore, the project introduces the one-dimensional spectral slope *p*, derived from the power spectral density (PSD) of the 16 Hz plasma data. This parameter has long been recognised as a key-ingredient in LEO-based modelling of scintillation indices (Wernik et al., 2007). These additions allow the Swarm-VIP-Dynamic models to explore how the dominant geophysical drivers of ionospheric variability differ across scale sizes. The models pave the way to possible implementation of scintillation modelling entirely based on Swarm data for GNSS applications.

This paper describes the development of a suite of statistical models (Swarm-VIP-Dynamic models), evaluates its performance across latitudes and spatial scales, and assesses the feasibility of running these models in near real-time operational settings.

By leveraging the unique strengths of the Swarm mission and by systematically addressing the limitations identified in earlier work, the Swarm-VIP-Dynamic models provide a substantially improved characterisation of the topside ionosphere and its dynamic nature and variability, with clear relevance for space-weather monitoring and GNSS-based technological systems.

## 2. Dataset

The databases use six official Level-2 Swarm data products which are available at ftp://swarm-diss.eo.esa.int, namely:

- Ionospheric Plasma IRregularities (IPIR) characterised by the Swarm satellites (IPDxIRR_2F, where 2F means Level-2 Fast-Track)
- Swarm 2 Hz Langmuir Probe extended dataset
- Swarm field-aligned currents (FAC_TMS_2F)
- Swarm accelerometer-based observations (ACCxCAL), which are accelerometer-derived non-gravitational acceleration as the L2 product
- Swarm thermospheric density, derived from GPS (DNSxPOD) and from accelerometer (DNSxACC)

In addition, the Swarm-VIP-Dynamic team has developed a version of SW_EXTD_EFIAMUS_FP based on 16 Hz measurements of the plasma density from the faceplate (hereafter called MUSIC, from the MUlti-Scale Irregularities produCt, https://swarm-diss.eo.esa.int/#swarm/Advanced/Plasma_Data/TDS_EFI_MUS_FP). These datasets collectively capture plasma density, small-scale irregularities, spacecraft potential, field-aligned currents, and thermospheric drag conditions, offering a multi-instrument perspective on ionospheric variability. Databases were rebuilt using seven Swarm C products (namely IPDxIRR, 2 Hz Langmuir Probe extended Dataset, DNSxPOD, FAC_TMS, ACCxCAL, DNSxACCand MUSIC) available on the ESA servers as of 1st December 2024, with ACCxCAL and DNSxACC extended to 31st May 2024 (to include the May 2024 storm event, which was essential for capturing extreme conditions relevant to the model's capability of describing physical processes). Here we adopted a less restrictive approach by allowing days to be included even if one or more of Swarm products (2 Hz Langmuir Probe extended Dataset, DNSxPOD, FAC_TMS, ACCxCAL, DNSxACC) were missing, provided that the IPIR product was present. The final temporal span is 19th July 2014 – 31th May 2024, with 12 missing days due to absent IPDxIRR data product; additional FAC_TMS unavailability (325 days) and a mid-timestamp bug (where 29 days were not used, as the database build code cannot deal with the situation where the IPIR timestamp is exactly half way between two FAC_TMS timestamps) reduced dataset to 29 days, but a conscious decision was taken not to fix bug. The MUSIC data product (Jin et al., 2026) is not available continuously, but all available data have been used.

To represent the external forcing of the ionosphere, the database also incorporated a suite of heliogeophysical proxies. These included solar radiation fluxes, solar wind parameters, and geomagnetic indices, many of which had been utilised in earlier Swarm-VIP modelling (Wood et al., 2024). The full list of explanatory variables trialled is given in Annex A where is introduced an additional solar-activity proxy for physically meaningful representations of the thermospheric heating state, termed F10.7_eff. This proxy was created by evaluating the NRLMSIS 2.0 model along the Swarm trajectories, computing the modelled daily mean density along the orbit, and iteratively adjusting the daily F10.7 and 81-day average F10.7 inputs to the MSIS model, setting both to the same value, until the output obtained a good fit with the observed daily mean density from the DNSxPOD product. The resulting "effective" F10.7, or F10.7_eff for short, is a time series of daily values, just like the true F10.7, but only available during the duration of the Swarm mission, where the DNSxPOD product is available. In principle also thermosphere density data derived from other missions could be used to extend and update the time series. The F10.7_eff proxy can be swapped with the true F10.7 measurements when generating and evaluating models. The premise is that the adjusted value that leads to a best fit with the observed average thermosphere state could also lead to a better representation of the ionosphere state.

At equatorial latitudes, one further explanatory variable was trialled, namely a function to represent the crests in the EIA. This function was $\left|9.5 - |\mathrm{MLAT}|\right|$ for the electron density and $\left|12 - |\mathrm{MLAT}|\right|$ for other dependent variables. In essence, the position of the maximum value of the electron density was taken to be at 9.5 MLAT and the position of the maximum gradient in the electron density was taken to be at 12 MLAT. These positions were based on a climatology of the Swarm data.

Because GLM frameworks require independence among samples for valid estimation of parameter uncertainties, the full 2-16 Hz Swarm data streams were reduced using the downsampling procedure developed during Swarm-VIP (detailed in Wood et al., 2024). This method selects statistically independent points by sampling at intervals proportional to ionospheric spatial correlation scales and the satellite's orbital velocity. For example, in polar cap regions the database retains one sample every 142 seconds, with the start phase of the window chosen randomly to avoid systematic sampling of any particular local time or geophysical configuration. This ensured that the training, optimisation and evaluation datasets would not inherit spatiotemporal autocorrelations that could bias goodness-of-fit assessments (see for more details Wood et al, 2024).

All data products were subjected to rigorous cleaning based on the flags or quality indicators provided by ESA (more detailed information can be found in the Supplementary material section Cleaning of Databases, Table S1). In the IPIR product, electron density, rate of change of density (ROD), and ROD index (RODI) values flagged as anomalous (40000) or as filler (99999) were removed, and certain mid-latitude and low-latitude intervals affected by Langmuir probe (LP) artefacts around 09 and 15 MLT were identified using a rate of change of TEC index (ROTI)-based filter (Kotova et al., 2022) and excluded. Similarly, coordinates and solar zenith angle (SZA) in the 2 Hz LP dataset were discarded whenever all coordinate fields simultaneously equalled zero (indicating filler data). We also filtered out negative densities which according to the product handbook are likely due to errors inherent in the radiation pressure model and acceleration determination using GPS data (ref to handbook: https://swarmhandbook.earth.esa.int/catalogue/SW_DNSxPOD_2_). This choice means that intervals of the lowest thermospheric densities will be slightly under-represented in the dataset compared to observations. It also means that intervals where the true value of the thermospheric density is low are more likely to be represented by an observation that is higher than the true value, than one which is lower than the true value. This may introduce a slight bias into the models, and any user of the models should be aware of this limitation. First-round cleaning removed entire rows with any anomaly; second-round cleaning retained rows and set only the problematic fields to NaN, maximising usable timestamps for model training/optimisation/evaluation.

Latitudinal partitioning follows IPDxIRR region flags, with explicit MLAT (magnetic latitudes, i.e. quasi-dipole latitude) bounds. The methodology used to determine the ionospheric region was described by Jin et al., (2022). The additional code was created to calculate the geomagnetic co-ordinates and the SZA for the Swarm satellites at a given timestamp (*CoSZA*) meant that geomagnetic co-ordinates could be used to split up the database. The most equatorward observations which could realistically be expected were assumed to be those observed during the most extreme event, which was the May 2024 superstorm. Data were taken, at 1 second resolution, from 9th – 13th May 2024 inclusive. The lowest latitudes assigned to the polar, auroral and mid latitude regions were 51˚, 29˚ and 27˚ MLAT respectively, with each of these values rounded down to the nearest integer. Therefore, the following limits were applied:

- Polar latitudes: 51˚ - 90˚ MLAT
- Auroral latitudes: 29˚ - 90˚ MLAT
- Mid latitudes: 27˚ - 75˚ MLAT
- Equatorial latitudes: 0˚ - 40˚ MLAT

Only a very small fraction of the database (~2% at mid-latitudes, 0% for polar; less than 0.001% for auroral, 0.2% for equatorial) had to be removed on the basis of this additional classification, confirming that the bounds set by the IPIR region flag were physically appropriate and internally consistent.

The databases needed to be broken into three subsets:

- Training: To train the model.
- Optimisation: To re-fit the model.
- Evaluation: To evaluate the performance of the model by testing against Swarm data and to evaluate the performance of the model against external datasets.

Each subset needs to be representative of the whole database. For example, if all the times of higher solar activity appeared in just one subset, then this would be a poor choice. It is not sensible to randomly assign days to different subsets as, if a major storm lasts three days, and day 1 and 3 are in the training dataset, then a reasonably good estimate could be made of the conditions on day 2 without any need for a model. If day 2 was part of the evaluation dataset, then model performance could be overestimated (Smirnov et al., 2023). To avoid storm-contamination and LT/longitude sampling biases, the mission was segmented into ~130-day LT-cycling sections, with the breaks between sections set by when successive northbound orbits of Swarm C crossed the 12 LT boundary at the equator. Each segment contained all longitudes and all local times and each of the other sections were randomly assigned to the training, optimisation and evaluation subsets, in the ratio 3:1:1, based on their representation of quiet and disturbed geomagnetic conditions, seasonal coverage, availability of external datasets, and specific scientific priorities. Evaluation sections were chosen for cross-dataset comparisons (e.g., April 2023 storm events; August 2019 as quiet interval; see paper by Urbar et al., 2026), with additional subsets chosen to ensure coverage of all seasons and local times, as well as intervals of both higher and lower solar activity. Optimisation sections selected semi-randomly to ensure seasonal and activity coverage, a major geomagnetic storm, and times of lower and higher solar activity. This careful segmentation ensured that the models were trained on a representative range of physical conditions while maintaining strict independence between training, optimisation and evaluation phases (see Table 1).

**Table 1:** Dates and times for the start of the sections into which the Swarm data were divided, together with the mean value of the F10.7 cm solar radio flux for this interval, the occurrence of major geomagnetic storms and details of the subset to which the data were assigned.

| Month | Day | Year | DOY | Hour | Minute | Section | Subset | Geomagnetic storms | F107 median |
|---|---|---|---|---|---|---|---|---|---|
| 7 | 16 | 2014 | 197 | 0 | 0 | 1 | Training | | 127.35 |
| 9 | 18 | 2014 | 261 | 5 | 29 | 2 | Optimisation | | 147.2 |
| 1 | 26 | 2015 | 26 | 16 | 10 | 3 | Training | March 2015 events | 125 |
| 6 | 11 | 2015 | 162 | 16 | 25 | 4 | Training | June 2015 storm | 105.4 |
| 10 | 25 | 2015 | 299 | 2 | 22 | 5 | Training | | 107 |
| 3 | 2 | 2016 | 62 | 8 | 40 | 6 | Training | | 88.35 |
| 7 | 14 | 2016 | 196 | 21 | 54 | 7 | Training | | 82.6 |
| 11 | 28 | 2016 | 333 | 14 | 55 | 8 | Training | | 74.8 |
| 4 | 8 | 2017 | 98 | 10 | 37 | 9 | Optimisation | | 74.1 |
| 8 | 19 | 2017 | 231 | 11 | 58 | 10 | Optimisation | September 2017 events | 74.2 |
| 12 | 30 | 2017 | 364 | 23 | 23 | 11 | Evaluation | | 69.4 |
| 5 | 14 | 2018 | 134 | 9 | 22 | 12 | Training | | 69.6 |
| 9 | 25 | 2018 | 268 | 12 | 32 | 13 | Evaluation | | 69.6 |
| 2 | 1 | 2019 | 32 | 18 | 28 | 14 | Evaluation | | 70.25 |
| 6 | 17 | 2019 | 168 | 6 | 46 | 15 | Evaluation | | 67.3 |
| 10 | 31 | 2019 | 304 | 17 | 54 | 16 | Optimisation | | 70.7 |
| 3 | 7 | 2020 | 67 | 22 | 57 | 17 | Training | | 69.3 |
| 7 | 19 | 2020 | 201 | 23 | 28 | 18 | Optimisation | | 72.85 |
| 12 | 3 | 2020 | 338 | 4 | 18 | 19 | Training | | 75.35 |
| 4 | 13 | 2021 | 103 | 8 | 31 | 20 | Training | | 75.95 |
| 8 | 23 | 2021 | 236 | 2 | 37 | 21 | Training | November 2021 storm | 87.7 |
| 1 | 3 | 2022 | 3 | 17 | 56 | 22 | Training | | 114.3 |
| 5 | 18 | 2022 | 138 | 5 | 36 | 23 | Training | | 122.7 |
| 9 | 30 | 2022 | 273 | 4 | 26 | 24 | Training | | 141.05 |
| 2 | 7 | 2023 | 37 | 12 | 28 | 25 | Evaluation | April 2023 events | 158.15 |
| 6 | 22 | 2023 | 173 | 12 | 34 | 26 | Training | | 157.1 |
| 11 | 6 | 2023 | 310 | 21 | 38 | 27 | Training | | 156.45 |
| 3 | 15 | 2024 | 75 | 4 | 34 | 28 | Training | May 2024 events & March 2024 storm | 173.75 |

## 3. Modelling Method: Overview

The statistical modelling approach used in the present work is an extension of GLM framework developed in the Swarm-VIP project. A full description of the original method is provided in Wood et al. (2024), where GLMs were used to relate ionospheric plasma parameters to a set of heliogeophysical proxies. It is postulated that the explanatory variable influences the dependent variable, and so the dependent variable can be predicted from the explanatory variable. Briefly, a GLM expresses a transformed expectation of the dependent variable $y$ as a linear combination of explanatory variables, allowing for dependent variable has non-Gaussian distributions and flexible link functions:

$$g(E(y)) = \beta_0 + \beta_1 \cdot x_1 + \cdots + \beta_n \cdot x_n,$$

here $g(E(y))$ is a function of the expected value of the dependent variable $y$, which is to be predicted, $x_1 \cdots x_n$ is the explanatory variable and $\beta_0 \cdots \beta_n$ are empirically determined constants known as the parameter estimates. The careful selection of the link function and the underlying distribution of the dependent variable was therefore a foundational requirement before any physical interpretation could be attempted

The Swarm-VIP-Dynamic project retains this GLM foundation but introduces several methodological refinements motivated by the limitations identified in earlier work.

### 3.1. Implementation: Choice of Dependent Variables

Five dependent variables have been selected for modelling in Swarm-VIP-Dynamic. These were chosen to investigate the variability of the ionospheric plasma at a range of scale sizes. They were:

- **Electron density:** Taken from the IPDxIRR data product, although these values are copied directly from the LP files. The measurement is of the ion density, and a quasi-neutral plasma is assumed.
- **Grad_Ne@100km:** The electron density gradient in a running window calculated via linear regression over 27 data points for the 2 Hz electron density data (taken from the IPDxIRR data product).
- **RODI10s:** Rate Of change of Density Index (RODI) is the standard deviation of ROD over 10 seconds (taken from the IPDxIRR data product).
- **RODI1s_FP:** Rate RODI over 1 seconds (16 data points for the 16 Hz electron data). This is taken from the MUSIC dataset.
- **p:** One-dimensional spectral index $p$ calculated using log–log least squares fit over frequency range 0.1-8 Hz in the PSD of the linearly detrended plasma density over a 10 s window centred around timestamp (160 data points for the 16 Hz electron data). This is taken from the MUSIC dataset.

Each of these dependent variables was modelled separately, with separate models created for each parameter and each region. Separate models were also created for each hemisphere, except for the equatorial region where one model was fitted for each dependent variable to avoid a discontinuity at the equator. Later in the text, we primarily use the word "model" in a general sense, as the fitting and optimization steps are identical, even though these steps produce a set of related but distinct models for each region and parameter.

### 3.2. Choice of distribution and transformation for the dependent variables

In a GLM, the dependent variable is not required to follow a normal distribution, but the distribution must be known. In Swarm-VIP, extensive effort was dedicated to determining suitable combinations of distributions and transformations. More than 5,000 quantile-quantile (q-q) plots were evaluated and an appropriate distribution for each dependent variable in each latitudinal region was determined. It is not necessary to determine the *best* distribution for each dependent variable, simply to determine a *good* distribution. In Swarm-VIP it was judged to be more important to get a distribution that worked well across all latitudinal sectors for a given dependent variable (to allow comparisons between models) than to have a different, and possibly slightly better, distribution in different regions. Swarm-VIP-Dynamic retained the same principle: the goal is not to identify an optimal distribution for each individual region, but rather to adopt a distribution transformation pair that is sufficiently good and internally consistent across regions. The Gamma distribution was ultimately selected for the majority of variability metrics, while a normal distribution was used for electron density (Wood et al., 2024).

In Swarm-VIP, various transformations (logarithms, exponentials, $n^{th}$ powers, and $n^{th}$ roots) were tested to stabilise the dependent variable and achieve better distributional behaviour. The $n^{th}$ root transformation emerged as the most robust and transferable across latitudes (Wood et al., 2024). Swarm-VIP-Dynamic adopted the same approach but restricted the trialled transformations to $n^{th}$ roots up to the $9^{th}$ root, leveraging the earlier findings.

The expanded database, which includes additional years of observations and improved temporal coverage across geomagnetic conditions, required a complete re-evaluation of the distributions and transformations. Not all distributions could be trialled for all dependent variables, for example the lognormal distribution requires all of the dependent values to be greater than zero, a condition not fulfilled for the spectral slope *p*. The distributions trialled for each dependent variable are shown in Table 2. A total of 2,541 new q-q plots were inspected manually. The resulting distribution–transformation combinations are summarised in Table 3. The goal was not to identify a perfect distribution for each case, but to *secure one* that is sufficiently representative and consistent across regions as presented in Figure 1.

**Table 2:** The distributions trialled for the dependent variables. Values of zero were excluded from RODI10s and RODI1s_FP. Had these not been excluded, then the distributions marked as * could not have been trailled.

| Distribution | Ne | \|Grad_Ne@100km\| | RODI10s | RODI1s_FP | Slope |
|---|---|---|---|---|---|
| Birnbaum Saunders | - | Yes | Yes* | Yes* | - |
| Burr | - | Yes | Yes* | Yes* | - |
| Exponential | Yes | Yes | Yes | Yes | - |
| ExtremeValue | Yes | Yes | Yes | Yes | Yes |
| Gamma | Yes | Yes | Yes | Yes | - |
| HalfNormal | Yes | Yes | Yes | Yes | - |
| Inverse Gaussian | Yes | Yes | Yes* | Yes* | - |
| Logistic | Yes | Yes | Yes | Yes | Yes |
| Loglogistic | Yes | Yes | Yes* | Yes* | - |
| Lognormal | Yes | Yes | Yes* | Yes* | - |
| Nakagami | Yes | Yes | Yes* | Yes* | - |
| Normal | Yes | Yes | Yes* | Yes* | Yes |
| Rician | - | Yes | Yes* | Yes* | - |
| tLocationScale | - | Yes | Yes | Yes | Yes |
| Weibull | Yes | Yes | Yes* | Yes* | - |

**Table 3:** The transformations applied to the dependent variables used to represent the ionospheric plasma and the variability in this plasma, together with the distributions chosen in modelling. Those distributions marked as * were the best available, but there were limitations in terms of how well these represented the dependent variable. ** A gamma distribution was used for the equatorial region, a normal distribution was used for the other latitudinal regions.

| Dependent variable | Distribution | Transformation applied to dependent variable | | | |
|---|---|---|---|---|---|
| | | Polar | Auroral | Mid-latitude | Equatorial |

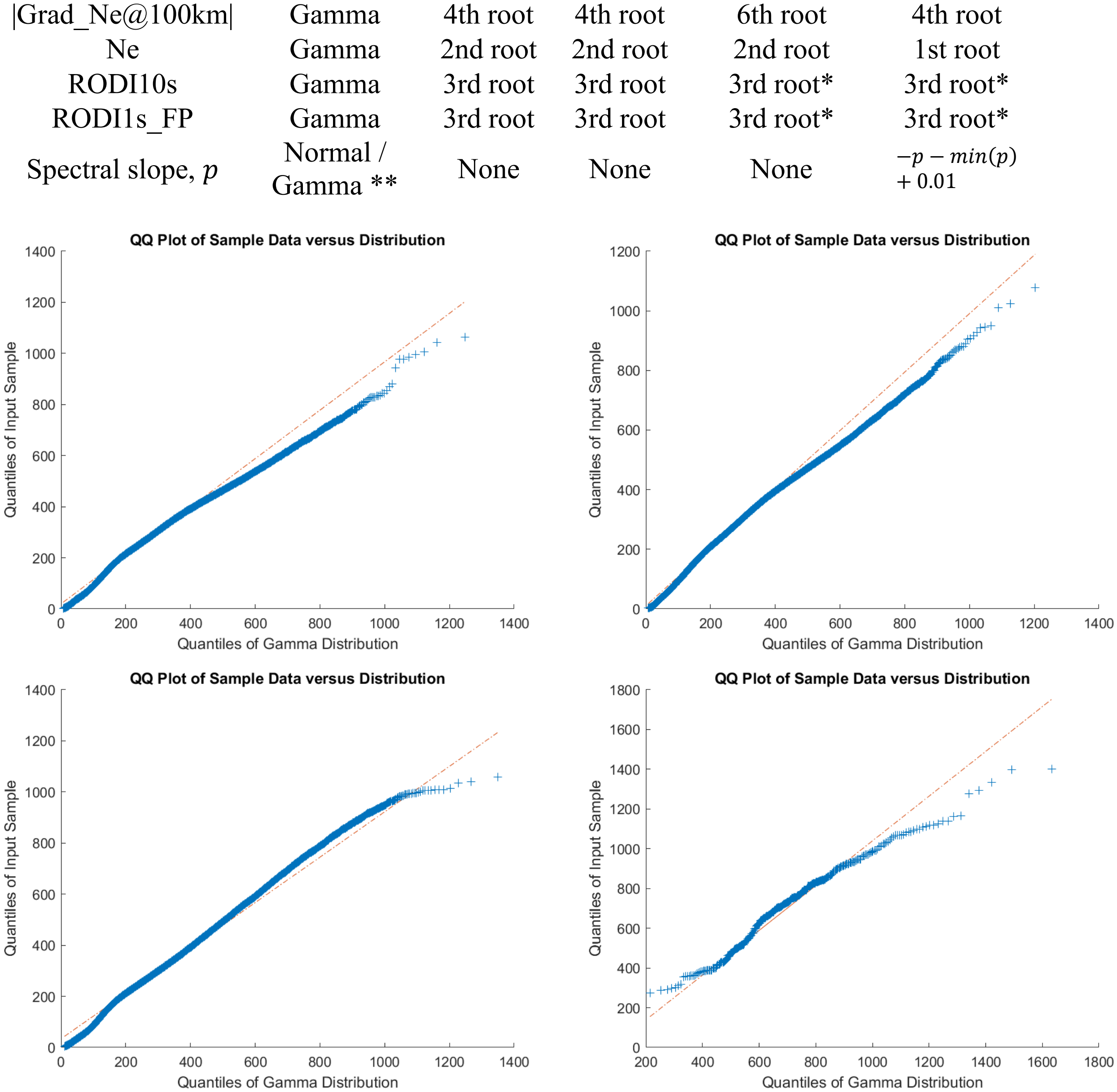

| | | | | | |
|---|---|---|---|---|---|
| \|Grad_Ne@100km\| | Gamma | 4th root | 4th root | 6th root | 4th root |
| Ne | Gamma | 2nd root | 2nd root | 2nd root | 1st root |
| RODI10s | Gamma | 3rd root | 3rd root | 3rd root* | 3rd root* |
| RODI1s_FP | Gamma | 3rd root | 3rd root | 3rd root* | 3rd root* |
| Spectral slope, $p$ | Normal / Gamma ** | None | None | None | $-p - min(p) + 0.01$ |

**Figure 1:** Quantile-quantile plots for the transformations applied to the electron density, together with the distributions chosen for the four latitudinal regions. The top left panel shows the polar region, the top right panel shows the auroral region, the bottom left panel shows the midlatitude region and the lower right panel shows the equatorial region.

The spectral slope $p$, which was not modelled in Swarm-VIP, required special treatment. In the polar and auroral regions, the distribution was best approximated by a normal distribution without transformation (see Fig. 2), although q-q plots revealed that the data likely reflect a mixture of underlying physical processes (see Fig. 3). While it is possible to model the dependent variable using two different distributions (one for higher values of $p$ and one for lower values of $p$). However, it is not advantageous to do so, as the resulting model would require the user to know in what region the prediction would fall before running the model. It is also possible to create a bespoke distribution for this dependent variable, but that is beyond the scope of the present project.

The range of values on the x- and y-scale for the q-q plot suggest that, if the normal distribution is chosen for this dependent variable, then the model will predict an appropriate range of values. However, the kink in the quantiles (blue crosses) between region 1 and region 2 means that the model may exhibit a small bias towards underprediction of at higher values of the spectral slope and to overprediction at lower values of this dependent variable (see Fig. 3). As shown by Jin et al. (2026), the spectral slope steeper than -1 is more meaningful, because flat PSD usually comes from quiet and smoother ionosphere without clear irregularities.

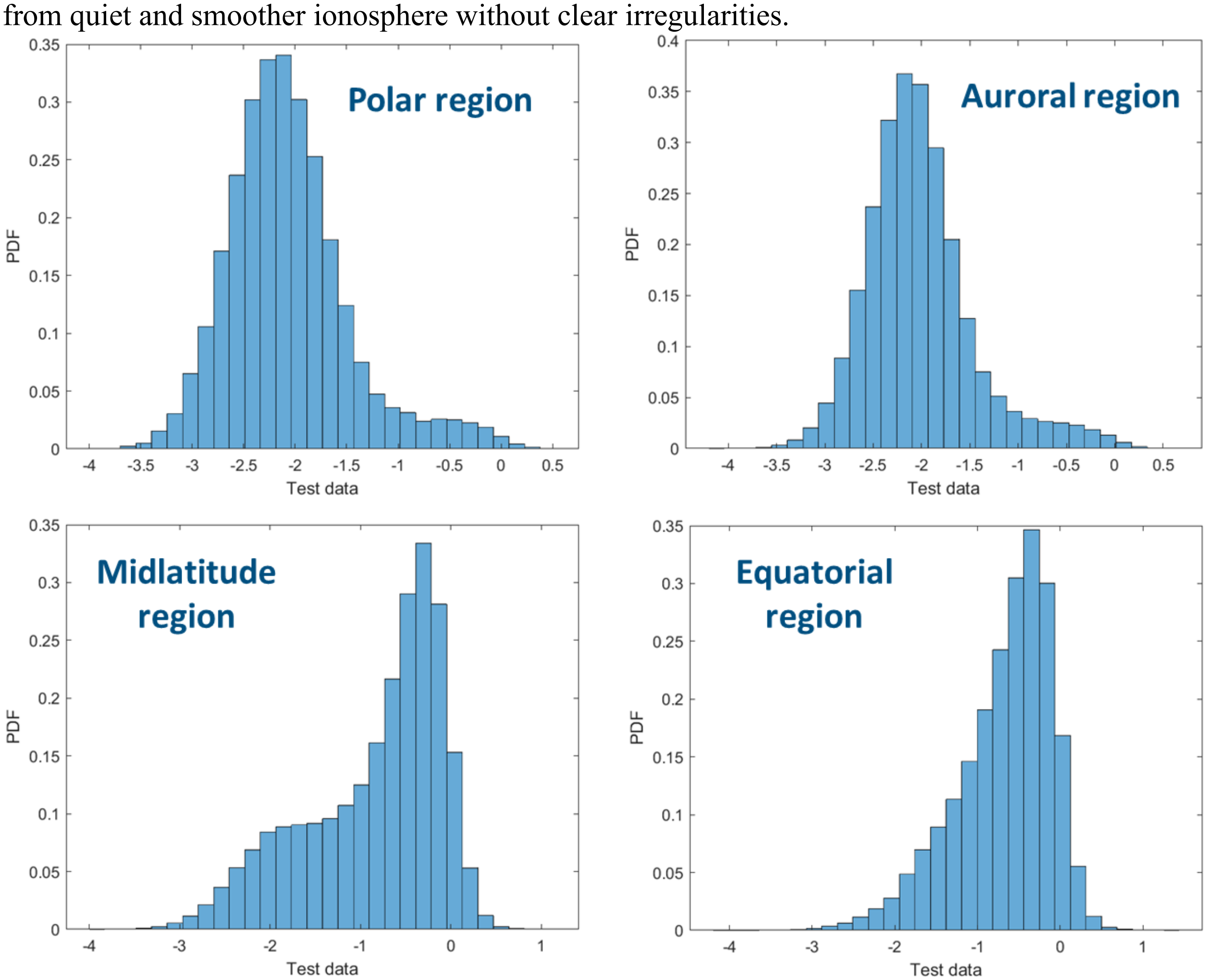

**Figure 2:** Probability Density Functions (PDFs) for the spectral slope ($p$) in the polar, auroral, mid latitude and equatorial regions.

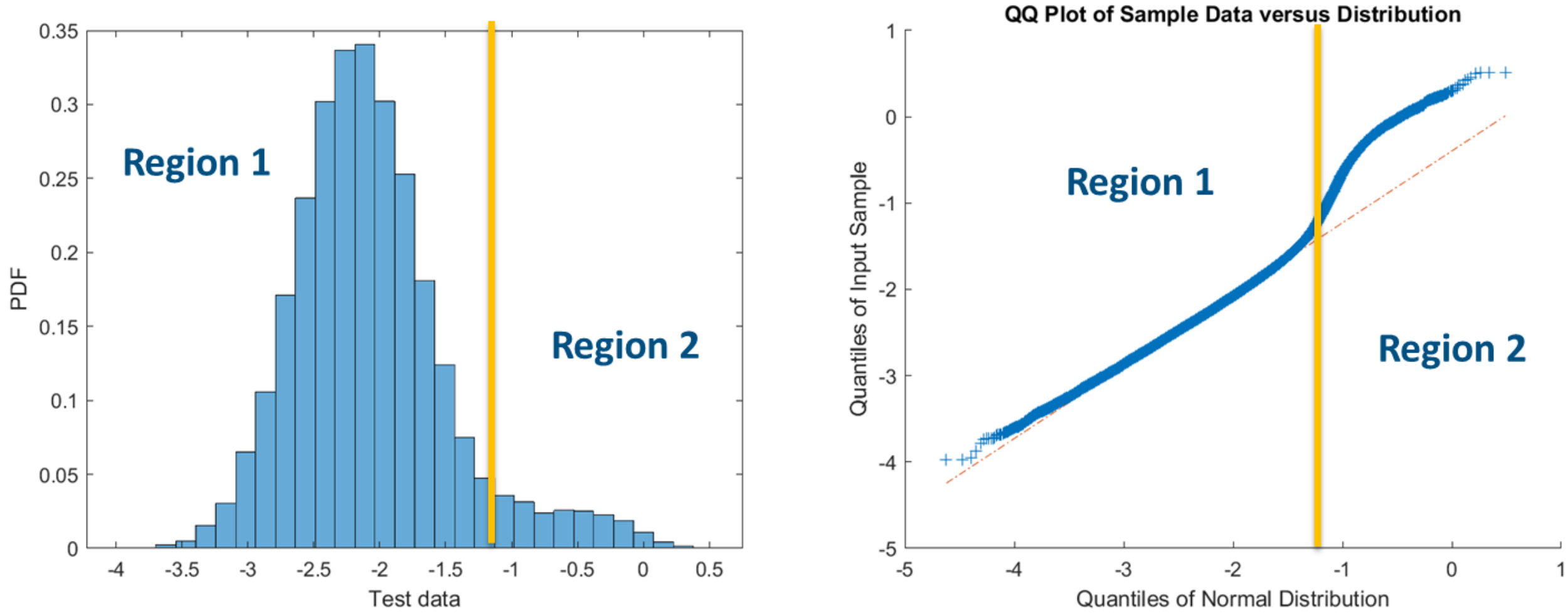

**Figure 3:** PDF for the spectral slope in the polar region (left) and the corresponding quantile-quantile (q-q) plot if it is assumed that the dependent variable can be represented by the normal distribution.

The data in the equatorial region strongly resembles a Gamma distribution, but with the distribution mirrored to give exponential decay towards lower values. A transformation of $-p - min(p) + 0.01$ was applied to these data to mirror these datapoints and shift the distribution to positive values. Once this transformation was applied, then the Gamma distribution (without a further transformation of the data) was selected. The distribution of the data in the midlatitude region appears to be a composite of the distributions shown in both the auroral and equatorial regions. The distribution chosen was a normal distribution (without a transformation), but this was not truly representative of the dataset.

### 3.3. Choice of link function

Selection of the link function followed the methodology established in Swarm-VIP. In that earlier project, for dependent variables represented by the Gamma distribution, three standard link functions (the identity, inverse and logarithmic) were tested. Their suitability was assessed by scoring the statistical significance of each explanatory variable across all regions, following the scheme reported in Wood et al. (2024). The *log link* function consistently produced the strongest and most stable relationships. It was therefore used throughout Swarm-VIP-Dynamic for Gamma distributed responses. For $p$ variable modelled with a normal distribution, the *identity link* was used.

### 3.4. Model fitting and optimisation procedure

The model fitting procedure is also derived from the approach introduced in Swarm-VIP (Wood et al., 2024). There are two ways to fit a GLM (McCullagh and Nelder, 1983). Terms can be added one at a time to build a model (e.g. Wood et al., 2024) or a model can be constructed containing all explanatory variables and these can be removed until only those which are significant remain (e.g. Dorrian et al., 2019). These two methodologies can give different models. Suppose that dependent variable X is a related to explanatory variables A, B and C. Explanatory variable A may explain more of the variation in X than B or C on their own, but B and C in combination may explain more of the variation in X than A. A model fitted by adding terms one at a time will have the form X~A+B or X~A+C, whereas a model fitted by removing terms sequentially will have the form X~B+C. The large number of explanatory variables used in the present project meant that the only practical option was to add terms one at a time.

Model construction follows a structured, correlation-based selection procedure. Correlations were found between each explanatory variable and each dependent variable in turn using the training dataset (see Table S2 in Supplementary materials). The explanatory variable with the highest correlation to the dependent variable was added to the model. The explanatory variable with the next highest correlation was added to the dependent variable, provided it does not exceed a |0.25| correlation threshold relative to any variables already in the model. It is a requirement of a GLM that the explanatory variables are not correlated with one another, otherwise different explanatory variables are representing the same variations in the dataset, and this can result in overfitting of the model. The threshold of |0.25| was introduced in Swarm-VIP to avoid multicollinearity and was retained here. Thresholds of |0.30| and |0.35| were also trailed, but made very little difference to the model formulation and so are not reported here.

Once the provisional model was assembled, the optimization procedure was undertaken to determine whether all of the terms were justified and GLM parameter estimates are fitted and all terms failing the 5 % level of significance threshold are removed. This two-step significance requirement reduces the chance of spurious terms in the model, lowering the effective error rate from 5 % to 0.25 %. Following optimisation, the model was applied to the evaluation dataset, and final performance was assessed using standard goodness-of-fit metrics.

Hemispheric differences identified in Swarm-VIP play a central role in the Swarm-VIP-Dynamic framework. For the polar, auroral and midlatitude models, the same provisional model was created using the training dataset for both hemispheres combined. Optimisation and evaluation were performed separately for the northern and southern hemispheres. In the equatorial region, this separation is not applied in order to avoid creating a discontinuity across the equatorial boundary

### 3.5. Model evaluation

As mentioned above, the re-fitted by using optimisation dataset model was used to predict the data in the evaluation dataset. The performance of the model was then assessed against four goodness-of-fit statistics. The goodness-of-fit statistics selected are given below.

- Accuracy: Assessed using the root mean square error (RMSE) on a relative scale (rRMSE). This is the RMSE divided by the median of the observations. Ideally, RMSRE should be close to zero. If RMSRE = 1 then the differences between the model predictions and the observations are of a similar size to the observations.
- Bias: Assessed using the mean error (ME), the mean of the predicted values less the mean of the observed values. Ideally, this should close to zero. If it is positive, then the model consistently overpredicts the observations. If it is negative, then the model consistently underpredicts the observations.
- Precision: Assessed using the ratio of the standard deviation of the predictions to the standard deviation of the observations. Ideally, this should be close to 1. If it is greater than 1, then there is too much variability in the model (the predictions are too noisy). If it is less than 1, then there is not enough variability in the model (the model is overfitted).
- Association: Assessed using the Pearson correlation coefficient. Ideally, this should be close to 1.

These measures follow the definitions and rationale given in Wood et al. (2024) and Liemohn et al. (2021).

## 4. Results

The full modelling approach described in Section 3 was then applied. Following the procedure established in Swarm-VIP and refined here, models were constructed separately for each dependent variable (electron density, |Grad_Ne@100km|, RODI10s, RODI1s_FP, and spectral slope, *p*) in each latitudinal region (polar, auroral, midlatitude and equatorial). Before constructing the full multi-term GLMs, a set of single term models was fitted to provide a baseline measure of how much variability each individual explanatory variable could explain. For that each explanatory variable is tried one at a time meaning that each dependent variable in each latitudinal region, applying the same dependent variable transformation identified in Table 3.

Prior to fitting a multi-term model, single term models were created. The purpose of these models was twofold. Firstly, to determine the effect of each of the underlying driving processes on the system, and secondly to determine how much of the ionospheric variability could be explained by a single driver. For each proxy, the absolute value of the correlations was averaged across all latitudinal regions and dependent variables. The explanatory variables were categorised based upon the type of physical phenomena which they act as a proxy for (e.g., 'solar', 'geomagnetic' etc.), and within each category the explanatory variable with the largest mean correlation was identified as the most representative for that driver type to explain the variability of the dependent variables. These results are summarised in Table S3 in Supplementary material, along with the median, 10$^{th}$ and 90$^{th}$ percentiles of each dependent variable. All these correlations were less than 1, showing that the single term model does not perfectly capture the underlying trend. This is hardly surprising, and such a result is expected for every model ever developed in any system. To determine how much of an effect each explanatory variable had upon the prediction of the dependent variable, the correlation between the dependent variable and *the prediction of the dependent variable* from each of these models was found using the training dataset. The 10$^{th}$ and 90$^{th}$ percentiles of each explanatory variable were used to fed into the model to create predictions based upon these percentiles. The resulting *predicted range* of the dependent variable was compared to the *observed range* of the dependent variable. These were used as a proxy for the range of this variable. Then the *ratio* of the predicted range to the observed range was found and in every case it was less than 1. Goodness-of-fit statistics have not been created for single term models. This is because the purpose is to show whether a model can capture the full range of values, not to evaluate how well it performs. In all cases, the predicted range was substantially smaller than the observed range, confirming that no single proxy fully describes the physical processes controlling ionospheric variability. This result is fully consistent with the expectation that ionospheric plasma is simultaneously driven by solar radiation, geomagnetic forcing, background neutral winds, and small-scale instabilities, none of which can be represented adequately by a single variable. Therefore, a multi-term model, which includes proxies for multiple physical processes, is needed.

Multi-term models were then fitted, using the method outlined in Sect. 3. The choice of terms used to formulate the model was informed by the correlation between the explanatory variable and the *transformed* dependent variable shown in Table S4 in Supplementary materials. Three versions of the models were fitted. Explanatory variables and parameter estimates for all three versions are provided in text form (see Supplementary materials) and the goodness-of-fit statistics (based on Liemohn et al., 2021) are shown in Tables 4.

**Version 3.1:** This model is fitted using the method stated in Section 3 considering all of the available explanatory variables, with the exceptions of AE, AL and AU, as these indices were not available after 31$^{st}$ December 2019 at the point at which the databases were constructed (autumn 2024). Version 3.1 therefore represents the highest-fidelity scientific model, using the fullest set of physical drivers. These were intended to be the definitive models for this project.

**Version 3.2:** The models in version 3.1 include measurements from Swarm (e.g., FAC, Ionospheric Radial Current (IRC), DNSxACC, DNSxPOD), which are only valid at that location. This is a challenge for validation activities. Therefore, models have been produced which exclude these explanatory variables to ensure that the models can be validated using data from other missions or ground-based systems. As expected, performance is slightly reduced relative to Version 3.1, but the models remain physically coherent and scientifically interpretable.

**Version 3.3:** This is the model with the potential to be run in near real-time. The only explanatory variables that cannot be used in the model with the potential to be run in near real time (version 3.3) are those which are measured directly from Swarm. DNSxACC, density_DNSxPOD or F10.7_eff were replaced with F107O as this is highly correlated and represents similar underlying physical process; Swarm-based current proxies were removed or replaced), and altitude-dependent terms were removed. Any measure of the ionospheric current systems (terms IRC_FACTMS or FAC_FACTMS in the models) was removed, if a geomagnetic index is also included in the models. If a geomagnetic index is not included, then replace these with *Kp* (polar, auroral and midlatitude models) or *Dst* at (equatorial models). *Kp* and *Dst* have been chosen as they are widely used by operators, but indices with a higher time resolution could have chosen. Models using these higher resolution indices were trialled and there was very little change in the goodness-of-fit statistics (not shown). Altitude-dependent terms were removed (height, radius or altitude). As an operator will know the altitude of their satellite, the measure of altitude in could be left in. But models which permitted a measure of altitude to be used as an explanatory variable were trialled, and there was very little change in the goodness-of-fit statistics (not shown), so the measure of altitude was not included in the final models.

To implement these changes, these changes were made at the optimisation step. The optimisation process was then re-run and any of the other terms in the model which were no-longer statistically significant were removed. This method was chosen to give a good level of consistency between these models and version 3.1.

The equatorial region receives special treatment due to the complexity of the EIA. In current project, a new explanatory variable of Lat_Diff that defined as the difference between geographic and quasi-dipole geomagnetic latitude was trialled to account for longitudinal structure. Furthermore, refinements based on climatological EIA crest locations were implemented, producing the updated versions of each version of equatorial model. This function was $|9.5 - |\mathrm{MLAT}||$ for the electron density and $|12 - |\mathrm{MLAT}||$ for other dependent variables. In essence, the position of the maximum value of the electron density was taken to be at 9.5 MLAT and the position of the maximum gradient in the electron density was taken to be at 12 MLAT.

Across all regions, electron density models remain the strongest modelled quantity, with Version 3.1 achieving high precision and correlation. At smaller spatial scales (e.g., RODI1s_FP), performance degrades systematically, consistent with the absence of proxies describing instability growth or forcing by wave activity from below. Removing Swarm-specific proxies (Version 3.2) reduces performance moderately, while the transition to real-time proxies (Version 3.3) leads to the expected decline but retains the essential structure and association of Version 3.1.

**Table 4:** Goodness-of-fit statistics for the modelling.

<table>
<tr><th rowspan="2">Dependent Variable</th><th rowspan="2">Region</th><th rowspan="2">Hemisphere</th><th colspan="4">Model version 3.1</th><th colspan="4">Model version 3.2</th><th colspan="4">Model version 3.3</th></tr>
<tr><th>rRMSE</th><th>rME</th><th>Precision</th><th>Correlation</th><th>rRMSE</th><th>rME</th><th>Precision</th><th>Correlation</th><th>rRMSE</th><th>rME</th><th>Precision</th><th>Correlation</th></tr>
<tr><td rowspan="7">Electron Density</td><td rowspan="2">Polar</td><td>NH</td><td>0.43</td><td>0.08</td><td>1.21</td><td>0.84</td><td>0.43</td><td>0.08</td><td>1.21</td><td>0.84</td><td>0.52</td><td>0.09</td><td>1.24</td><td>0.76</td></tr>
<tr><td>SH</td><td>0.48</td><td>0.04</td><td>0.96</td><td>0.78</td><td>0.48</td><td>0.04</td><td>0.96</td><td>0.78</td><td>0.52</td><td>0.00</td><td>0.84</td><td>0.70</td></tr>
<tr><td rowspan="2">Auroral</td><td>NH</td><td>0.32</td><td>0.05</td><td>1.11</td><td>0.85</td><td>0.32</td><td>0.05</td><td>1.11</td><td>0.85</td><td>0.39</td><td>0.06</td><td>1.13</td><td>0.79</td></tr>
<tr><td>SH</td><td>0.40</td><td>-0.01</td><td>0.79</td><td>0.74</td><td>0.40</td><td>-0.01</td><td>0.79</td><td>0.74</td><td>0.46</td><td>0.01</td><td>0.87</td><td>0.67</td></tr>
<tr><td rowspan="2">Midlatitude</td><td>NH</td><td>0.33</td><td>0.02</td><td>0.95</td><td>0.79</td><td>0.30</td><td>-0.02</td><td>0.71</td><td>0.83</td><td>0.32</td><td>-0.01</td><td>0.73</td><td>0.78</td></tr>
<tr><td>SH</td><td>0.38</td><td>0.00</td><td>0.94</td><td>0.72</td><td>0.35</td><td>-0.03</td><td>0.78</td><td>0.74</td><td>0.39</td><td>-0.01</td><td>0.86</td><td>0.68</td></tr>
<tr><td>Equatorial</td><td>Both</td><td>1.66</td><td>-0.16</td><td>0.62</td><td>0.83</td><td>1.66</td><td>-0.16</td><td>0.62</td><td>0.83</td><td>2.40</td><td>-0.04</td><td>0.93</td><td>0.60</td></tr>
<tr><td rowspan="7">|Grad_Ne@100km|</td><td rowspan="2">Polar</td><td>NH</td><td>0.35</td><td>0.01</td><td>0.57</td><td>0.54</td><td>0.35</td><td>0.01</td><td>0.57</td><td>0.54</td><td>0.36</td><td>0.02</td><td>0.65</td><td>0.50</td></tr>
<tr><td>SH</td><td>0.37</td><td>0.00</td><td>0.58</td><td>0.55</td><td>0.37</td><td>0.00</td><td>0.58</td><td>0.55</td><td>0.38</td><td>0.01</td><td>0.50</td><td>0.51</td></tr>
<tr><td rowspan="2">Auroral</td><td>NH</td><td>0.32</td><td>0.02</td><td>0.52</td><td>0.44</td><td>0.32</td><td>0.02</td><td>0.52</td><td>0.44</td><td>0.33</td><td>0.02</td><td>0.49</td><td>0.42</td></tr>
<tr><td>SH</td><td>0.36</td><td>-0.01</td><td>0.42</td><td>0.45</td><td>0.36</td><td>-0.01</td><td>0.42</td><td>0.45</td><td>0.36</td><td>0.00</td><td>0.42</td><td>0.43</td></tr>
<tr><td rowspan="2">Midlatitude</td><td>NH</td><td>0.21</td><td>-0.04</td><td>0.09</td><td>0.06</td><td>0.22</td><td>-0.03</td><td>0.17</td><td>0.30</td><td>0.22</td><td>-0.02</td><td>0.17</td><td>0.29</td></tr>
<tr><td>SH</td><td>0.22</td><td>0.00</td><td>0.30</td><td>0.29</td><td>0.23</td><td>0.00</td><td>0.32</td><td>0.27</td><td>0.23</td><td>0.00</td><td>0.33</td><td>0.27</td></tr>
<tr><td>Equatorial</td><td>Both</td><td>0.37</td><td>0.03</td><td>0.64</td><td>0.60</td><td>0.37</td><td>-0.01</td><td>0.51</td><td>0.58</td><td>0.38</td><td>0.00</td><td>0.51</td><td>0.56</td></tr>
<tr><td rowspan="7">|RODI10s|</td><td rowspan="2">Polar</td><td>NH</td><td>0.40</td><td>0.02</td><td>0.65</td><td>0.60</td><td colspan="4" rowspan="7">Dependent variable not used for valdiation and performance assessment activities, so version 3.2 of the model is not required.</td><td>0.43</td><td>0.03</td><td>0.75</td><td>0.56</td></tr>
<tr><td>SH</td><td>0.45</td><td>-0.01</td><td>0.63</td><td>0.60</td><td>0.47</td><td>0.01</td><td>0.54</td><td>0.55</td></tr>
<tr><td rowspan="2">Auroral</td><td>NH</td><td>0.37</td><td>0.03</td><td>0.64</td><td>0.52</td><td>0.38</td><td>0.04</td><td>0.61</td><td>0.49</td></tr>
<tr><td>SH</td><td>0.42</td><td>-0.03</td><td>0.44</td><td>0.51</td><td>0.43</td><td>-0.02</td><td>0.45</td><td>0.48</td></tr>
<tr><td rowspan="2">Midlatitude</td><td>NH</td><td>0.43</td><td>0.00</td><td>0.40</td><td>0.40</td><td>0.43</td><td>-0.03</td><td>0.35</td><td>0.42</td></tr>
<tr><td>SH</td><td>0.45</td><td>-0.03</td><td>0.40</td><td>0.39</td><td>0.45</td><td>-0.01</td><td>0.41</td><td>0.38</td></tr>
<tr><td>Equatorial</td><td>Both</td><td>0.49</td><td>0.00</td><td>0.33</td><td>0.38</td><td>0.50</td><td>-0.01</td><td>0.33</td><td>0.35</td></tr>
<tr><td rowspan="7">|RODI1s_FP|</td><td rowspan="2">Polar</td><td>NH</td><td>0.29</td><td>0.02</td><td>0.61</td><td>0.37</td><td colspan="4" rowspan="7">Dependent variable not used for valdiation and performance assessment activities, so version 3.2 of the model is not required.</td><td>0.29</td><td>0.23</td><td>1.68</td><td>0.33</td></tr>
<tr><td>SH</td><td>0.33</td><td>0.03</td><td>0.52</td><td>0.44</td><td>0.33</td><td>0.18</td><td>0.95</td><td>0.44</td></tr>
<tr><td rowspan="2">Auroral</td><td>NH</td><td>0.29</td><td>0.13</td><td>1.07</td><td>0.32</td><td>0.30</td><td>0.19</td><td>1.27</td><td>0.29</td></tr>
<tr><td>SH</td><td>0.34</td><td>0.11</td><td>0.69</td><td>0.42</td><td>0.34</td><td>0.13</td><td>0.79</td><td>0.41</td></tr>
<tr><td rowspan="2">Midlatitude</td><td>NH</td><td>0.16</td><td>0.04</td><td>0.54</td><td>0.30</td><td>0.16</td><td>0.03</td><td>0.43</td><td>0.27</td></tr>
<tr><td>SH</td><td>0.20</td><td>0.05</td><td>0.62</td><td>0.35</td><td>0.20</td><td>0.04</td><td>0.53</td><td>0.33</td></tr>
<tr><td>Equatorial</td><td>Both</td><td>0.21</td><td>0.10</td><td>0.85</td><td>0.21</td><td>0.96</td><td>-0.93</td><td>0.08</td><td>0.21</td></tr>
<tr><td rowspan="7">Slope</td><td rowspan="2">Polar</td><td>NH</td><td>0.38</td><td>0.16</td><td>0.64</td><td>0.58</td><td>0.31</td><td>-0.07</td><td>0.68</td><td>0.55</td><td>0.33</td><td>-0.03</td><td>0.68</td><td>0.55</td></tr>
<tr><td>SH</td><td>0.31</td><td>0.05</td><td>0.64</td><td>0.60</td><td>0.28</td><td>-0.05</td><td>0.60</td><td>0.60</td><td>0.31</td><td>0.05</td><td>0.60</td><td>0.60</td></tr>
<tr><td rowspan="2">Auroral</td><td>NH</td><td>0.36</td><td>0.13</td><td>0.35</td><td>0.31</td><td>0.29</td><td>-0.05</td><td>0.32</td><td>0.34</td><td>0.34</td><td>0.08</td><td>0.32</td><td>0.34</td></tr>
<tr><td>SH</td><td>0.29</td><td>-0.01</td><td>0.43</td><td>0.45</td><td>0.29</td><td>-0.02</td><td>0.42</td><td>0.46</td><td>0.29</td><td>-0.01</td><td>0.42</td><td>0.46</td></tr>
<tr><td rowspan="2">Midlatitude</td><td>NH</td><td>0.22</td><td>0.08</td><td>1.70</td><td>0.40</td><td>0.19</td><td>-0.03</td><td>0.47</td><td>0.40</td><td>0.21</td><td>0.04</td><td>0.47</td><td>0.40</td></tr>
<tr><td>SH</td><td>0.20</td><td>0.03</td><td>0.58</td><td>0.46</td><td>0.18</td><td>-0.03</td><td>0.57</td><td>0.47</td><td>0.20</td><td>0.04</td><td>0.57</td><td>0.47</td></tr>
<tr><td>Equatorial</td><td>Both</td><td>0.30</td><td>-0.07</td><td>0.37</td><td>0.23</td><td>0.79</td><td>-0.69</td><td>0.18</td><td>0.13</td><td>0.30</td><td>-0.04</td><td>0.40</td><td>0.12</td></tr>
</table>

## 5. Discussion

The Swarm-VIP-Dynamic models represent a substantial advance over the original Swarm-VIP framework, both in terms of data coverage and methodological flexibility. By exploiting an expanded Swarm dataset extending into the ascending and maximum phases of solar cycle 25, and by allowing region-specific model structures, the new models achieve improved representation of the topside ionosphere across a broad range of latitudes, activity levels, and spatial scales. Nevertheless, as with any statistical model of a complex geophysical system, their strengths and limitations must be interpreted carefully in the context of the underlying physics and the availability of suitable proxies.

For electron density (see Table 4), the Version 3.1 models perform particularly well at polar, auroral and midlatitudes. The rRMSE, rME, precision and correlation are measures of accuracy, bias, precision and association respectively. The goodness-of-fit statistics indicate strong association, low bias, and precision values close to their best possible values of 1, 0 and 1 respectively, demonstrating that the models capture the underlying trend in the observations (as shown by the measure of association), the variability observed in the observations (as shown by the precision) and do not consistently overpredict or underpredict (as shown by the bias). The statistic for the accuracy (rRMSE) would, ideally, be a little closer to zero, but it is not unreasonable. This would improve if the measure of association was closer to the ideal value of 0. It may also be that, while the overall variability of the observations is well captured by the model, there can be significant differences for particular observations. This suggests that targeted case-study analyses with examining periods of strong geomagnetic forcing or particular heliogeophysical conditions or unusual thermospheric conditions could provide additional insight into when and why the models deviate from observations and where the model performance is better or worse than at other times. Such investigations lie beyond the scope of the present work. A case studies during geomagnetically disturbed conditions are reported in the companion paper (Urbář et al., 2026) and further comparisons represent a natural direction for future studies.

For mid-latitude model the possibility that the range of latitudes permitted in the dataset was too great was considered (see Section 2 Dataset). Points were selected to be part of this dataset if the IPIR region flag placed them in this region and the magnetic latitude was in the range 27° – 75° MLAT. If the poleward limit of this range meant that plasma structures from the auroral or polar regions were driving the mid-latitude models, then this could affect the performance of the models in this region. Tests were conducted with a more restricted range of latitudes and the models did not change significantly (not show here). Therefore, the definition of the mid-latitude region used when creating the databases (Section 2) was deemed to be appropriate.

At equatorial latitudes, the situation is more complex. Although the association remains relatively strong, accuracy, bias and precision indicate scope for improvement. It would be useful to see if there are particular physical phenomena which are either well represented or poorly represented in the model. Independent performance assessment confirms that the classical double-crest structure of the EIA is not adequately captured by the models. This comparison, and modifications to the model to better capture this variability, will be discussed in a subsequent paper. The introduction of additional explanatory variables based on the latitudinal distance to the climatological EIA crest locations improve the morphological realism of the equatorial models, but does not fully resolve all discrepancies. Most notably, this model will still not capture the variability resulting from Equatorial Plasma Bubbles (EPBs) as these are driven by an instability process. This outcome highlights the intrinsic difficulty of statistically modelling equatorial electrodynamics, where longitudinal structure, enhanced plasma variability after sunset, and plasma instabilities play a dominant role.

A clear and physically meaningful pattern emerges when examining models of ionospheric variability (|Grad_Ne@100km|, RODI10s, RODI1s_FP and spectral slope, p). Across all regions, goodness-of-fit statistics degrade systematically as the characteristic scale size decreases (lower sections of Table 4). A reduction in the precision indicate that the models fail to reproduce the full amplitude of observed variability, while reduced correlations imply that only part of the observed trends are captured. This scale-dependent deterioration strongly suggests the presence of missing physical process that become increasingly important at smaller scales.

One likely missing driver is the instability growth rate, which directly governs the formation and evolution of plasma irregularities (as the goodness-of-fit statistics become progressively worse at smaller scales). For example, the Gradient Drift Instability commonly occurs at polar latitudes (Tsunoda, 1988; Sojka et al, 1998). To calculate this growth rate, it is necessary to know both the relative velocity of the ions and the neutral species, and the background electron density. These quantities are not routinely available from Swarm observations in low-Earth orbit, making it impractical to include an explicit instability-growth proxy in the present models. The absence of such terms naturally limits the ability of any statistical model to capture fine-scale variability, regardless of the sophistication of the fitting procedure. Similarly, at equatorial latitudes, the GLMs cannot capture EPBs. This is a clear limitation of the models, and a likely reason for the reduced performance of the models in the equatorial region, as shown in the goodness-of-fit statistics (Table 4).

At midlatitudes, recent study by Boyde et al. (2025) and Wood et al. (2026), each using over 2,700 hours of Low Frequency Array (LOFAR) data, attributed a significant proportion of ionospheric structures to driving from below by quasi-upward propagating atmospheric gravity waves (AGWs). Boyde et al. (2025) identified travelling ionospheric disturbances and determined their propagation direction. It was shown that these were, primarily, in the anti-wind direction, and indicate that they were driven by AGWs, while Wood et al. (2026) showed a statistical link between ionospheric structures and thunderstorms which can launch AGWs by. These finding underscores the relative importance of driving from below at midlatitudes, where lower-atmospheric wave activity can dominate ionospheric variability. The lack of a proxy for such wave activity in the current models likely contributes to their reduced precision at these scales. The QUID-REGIS project has developed indices describing AGWs activity over Europe (QUID-REGIS paper by Mackovjak et al. (2026)), and a future collaboration could test whether incorporating these indices improves midlatitude model performance.

If this approach is successful then it could, potentially, be extended to auroral latitudes, where the polar vortex can also launch AGWs (Sato & Yoshiki, 2008). Recent work linking the Northern Annular Mode (NAM) index to measure of the strength of the polar vortex (Gerber and Martineau, 2018; Kumar et al., 2025) suggests that stratifying models by vortex strength could provide an additional pathway to improved representation of polar-cap structuring. While such extensions are beyond the scope of the present study, they point to sensible, physically motivated avenues for future development.

The comparison between model versions 3.1, 3.2 and 3.3 illustrates the trade-off between physical accuracy and operational feasibility. As expected, goodness-of-fit statistics degrade progressively as more explanatory variables are removed or replaced by real-time proxies. This behaviour confirms that accurate specification of the ionosphere and the variability of this plasma requires representing the key physical processes and multiple physical processes simultaneously in the models. As was mentioned before, it is expected that the adjusted value that provides the best fit to the observed average thermosphere state may also provide a better representation of the ionosphere state as we see in the better performance for models Version 3.1 compared to models Versions 3.3 (F10.7_eff proxy as *f10_7eff_swarmc* for 3.1 and *F107O* for 3.3, respectively). Nevertheless, the Version 3.3 models retain meaningful skill despite relying exclusively on operational inputs, demonstrating that Swarm-based statistical modelling can contribute to near real-time space- weather applications, even if proxies for all of the underlying processes are not available in real time.

More generally, the results emphasise a fundamental limitation of statistical modelling where explanatory variables are proxies and not the physical processes themselves. Even when statistically significant, proxies cannot perfectly represent the underlying dynamics, particularly when the system exhibits stochastic behaviour or responds nonlinearly to forcing. The deterministic GLMs developed here therefore cannot be expected to capture rare extreme events or rapid, state-dependent changes if the changes occur more rapidly than the temporal resolution of the proxies used to represent these processes. This limitation is also reflected in the reduced precision and correlation at small scales.

## 6. Conclusion

This paper presents a suite of statistical models describing the plasma state and variability of the topside ionosphere, developed applying the technique of Generalised Linear Modelling (Section 3) and based primarily on in-situ observations from the ESA Swarm mission (Section 2). Building on the foundations established in the Swarm-VIP project, the present work extends both the methodological framework and the observational basis, enabling improved representation of ionospheric dynamics across a wider range of latitudes, spatial scales and solar-cycle conditions.

The Swarm-VIP-Dynamic models have been created for five dependent variables, namely the electron density, |Grad_Ne@100 km|, RODI10s, RODI1s_FP and the spectral slope, p. Each of these capture different aspects of ionospheric plasma structure and variability. Separate models were developed for polar, auroral, midlatitude and equatorial regions to account for regional differences in the dominant driving processes. For all regions (except equatorial) separate models were created for Northern and Southern hemispheres. The set of distributions was trialled for the dependent variables (Table 1) and chosen transformations (Table 2) after manual check-up of quantile-quantile plots applied to the dependent variables to represent the ionospheric plasma and the variability in this plasma, together with the distributions chosen in the second round of modelling. The modelling proceeded in two rounds, the second of which consolidated the database to 31 May 2024 (including the May 2024 superstorm), harmonised cleaning, re-validated distributions and transformations, reorganised model formulation by region and hemisphere, and produced three model families (see Supplementary materials section Explanatory variables and parameter estimates for model formulation). Version 3.1 represents the most complete scientific formulation, incorporating all statistically significant explanatory variables. Version 3.2 models are based on Version 3.1 and removed Swarm-dependent inputs (which are only valid at Swarm location) to facilitate validation activities. Version 3.3 presents models with the potential to be run in near real-time and replaces non-operational proxies with real-time alternatives, providing an assessment of what can realistically be achieved in an operational context. Each of these models has been optimised and evaluated (results are shown in Table 3).

Across all regions, the electron-density models demonstrate the strongest performance, particularly at polar, auroral and midlatitudes, where bias is small, association is high and precision approaches their best possible values (Table 4). These results confirm that the dominant large-scale drivers of the topside ionosphere are well represented by the selected heliogeophysical proxies when sufficient observational coverage is available (10 years of dataset from Swarm). Performance degrades systematically for measures of ionospheric variability as the characteristic spatial scale decreases, reflecting the increasing importance of physical processes that are not explicitly represented in the current proxy set. In the equatorial region, additional refinements were required to improve the representation of the Equatorial Ionisation Anomaly (introduced an additional explanatory variable, based upon the latitudinal distance from the average statistical location of the crests of the EIA), highlighting the complexity of low-latitude electrodynamics and the limitations of purely statistical approaches in this regime.

This work has demonstrated that no single proxy is capable of capturing the full observed variability of ionospheric plasma (see Table S3). Single-term models consistently underestimate the observed range of all dependent variables, confirming that ionospheric structure arises from the combined influence of multiple drivers acting simultaneously. The multi-term GLMs developed here therefore represent a physically motivated compromise between completeness, rather than an attempt to produce a perfect fit to the observations.

The limitations of the Swarm-VIP-Dynamic models are physically interpretable. The reduced precision and association at small spatial scales (lower sections of Table 4) point to missing drivers such as instability growth rates and lower-atmospheric wave forcing, which cannot be captured using currently available proxies. Similarly, the deterministic nature of the GLMs excludes representation of stochastic variability. These statistical models are only valid for the range of conditions contained within the training dataset, and so cannot be used to reliably model rare extreme events which were not captured by the training dataset. These limitations are not methodological shortcomings but reflect the intrinsic complexity of the ionosphere and the constraints imposed by observational availability. The results therefore provide clear guidance on where future improvements are most likely to be realised through the inclusion of instability-related quantities, and proxies for AGWs activity.

Despite these limitations, the near real-time models (Version 3.3) demonstrate that meaningful ionospheric specification is possible using only operational inputs. Although performance is reduced relative to the fully-proxied models (Version 3.1), the essential structure and trends of the ionosphere are retained, underscoring the potential value of Swarm-based statistical modelling for space-weather monitoring and nowcasting applications. The Swarm-VIP-Dynamic framework thus provides a practical bridge between scientific investigation and operational readiness.

Finally, while this paper has focused on model development and interpretation, a detailed assessment of model performance (Version 3.3), validation against independent datasets, and comparison with physics-based models is presented in a companion paper by Urbar et al. (2026).

**Acknowledgements**

This research is a part of the 4DSpace Strategic Research Initiative at the University of Oslo. For the provision of F10.7 dataset, we thank the Solar Radio Monitoring Program (https://www.spaceweather.gc.ca/forecast-prevision/solar-solaire/solarflux/sx-en.php) of the National Research Council and Natural Resources Canada.

**Funding Information**

This work is funded under the European Space Agency (ESA) Contract 4000143413/23/I-EB (Swarm-VIP-Dynamic) within the ESA Solid and Magnetic Science Cluster - 4D Ionosphere framework. YJ, DK and WJM acknowledge funding from the European Research Council (ERC) under the European Union's Horizon 2020 research and innovation programme (ERC Consolidator Grant 866357, POLAR-4DSpace).

**Data Availability Statement**

The Swarm data products are available at ftp://swarm-diss.eo.esa.int , the OMNI resolution data are available at https://spdf.gsfc.nasa.gov/pub/data/omni/ , the Service International des Indices Géomagnétiques data are available at https://isgi.unistra.fr/geomagnetic_indices.php and the Laboratory for Atmospheric and Space Physics data are available at https://lasp.colorado.edu. The processed MUSIC data in this paper is now available in CDF format through the Swarm dissemination server at https://swarm-diss.eo.esa.int/#swarm/Advanced/Plasma_Data/TDS_EFI_MUS_FP. The Python package and source code developed for evaluating the Swarm-VIP-Dynamic models are publicly available through the Royal Netherlands Meteorological Institute (KNMI) Open Source Software repository at https://gitlab.com/KNMI-OSS/spaceweather/libs/swarm-vip-dynamic-models.

## Annex A
## Complementary Data

Proxies for the heliogeophysical processes are taken from multiple data sources and associated with the Swarm data. These are summarised in the tables shown below. In all cases, these data have undergone significant processing by the data provider to identify and remove anomalous values. In some cases, such values were replaced with filler values, marked as NA or marked as NaN. In the Swarm-VIP-Dynamic dataset all such values are marked as NaN.
Initially, the geomagnetic indices AE, AL and AU were trialled as explanatory variables and were included as terms in a subset of the models. These were later discarded because their availability in the chosen repository ended in 2019, which prevented meaningful optimisation and evaluation during periods of high solar and geomagnetic activity. Therefore, these indices were not trailed as explanatory variables in the final versions of the models. Beyond standard sets, the second round added Hp30/Hp60 (Yamasaki et al., 2022) and the F30 cm (Dudok de Wit et al., 2014) /F10.7 (Tapping, 2013) solar radio flux (with 27- & 81-day centred and trailing averages for a broader representation of both instantaneous solar drivers and longer-term solar-cycle modulation).

**Statistics from solar wind parameters:** The values present within the Swarm-VIP-Dynamic databases are the average (mean) values and the standard deviations of these data calculated in a two-hour window leading up to on the observation to be updated.
**Calculated values:** Several values are calculated from the proxies for the heliogeophysical processes. These are:

- **Clock Angle:** The clock angle gives the relative importance of the *y*- and *z*-components of the IMF. It is defined as:

$$\arctan \frac{|B_y|}{B_z}$$

  A clock angle of 0º is purely IMF $B_z$ positive with a $B_y$ component of zero, 180º is purely IMF $B_z$ negative with a $B_y$ component of zero and 90º is completely dominated by $B_y$ with a $B_z$ of zero.
- **Solar Wind Coupling Function: Newell et al. (2007).** This is given by:

$$v^{4/3} \cdot B_T^{2/3} \cdot sin^{8/3}\left(\frac{\theta_c}{2}\right)$$

  v is the solar wind velocity, $B_T$ is the magnitude of the IMF and $\theta_c$ is the clock angle.
- **Solar wind coupling function: ELYA. (Elliot, personal communication).** A project which varied the powers in the solar wind coupling function presented by Newell et al. (2007) found that the most statistically signicant relationship to the structuring ratio observed by the EISCAT radars was given by:

$$vB_T^{\frac{1}{2}}sin^2\left(\frac{\theta_C}{2}\right)$$

  v is the solar wind velocity, $B_T$ is the magnitude of the IMF and $\theta_c$ is the clock angle.
- **Solar wind coupling function: Akasofu's ε parameter (Akasofu et al., 1981).** This is proportional to:

$$vB_T{}^2sin^4\left(\frac{\theta_C}{2}\right)$$

Sometimes this is expressed as $\varepsilon = vB_T{}^2 sin^4\left(\frac{\theta_C}{2}\right) l_0^2$ where $l_0$ is an empirically determined scale factor with units of length. As we are interested in the association between $\varepsilon$ and the dependent variable, the numerical value of $\varepsilon$ is irrelevant, only the variation matters. Therefore, the scale factor $l_0$ has not been used.

**Operating Missions as Nodes on the Internet (OMNI) database**

The Operating Missions as Nodes on the Internet (OMNI) database (Papitashvili et al., 2020) summarises the solar wind conditions and the geomagnetic indicies. It is available at https://spdf.gsfc.nasa.gov/pub/data/omni/. There are two versions, a high-resolution version at 1-minute resolution and a low-resolution version at 1-hour resolution.

**Geomagnetic indicies from the International des Indices Géomagnétiques**

The International des Indices Géomagnétiques (IAGA) publish a number of geomagnetic indicies, a subset of which are used within Swarm-VIP-Dynamic. These are available from: https://iaga-aiga.org/

**Geomagnetic indices from the GFZ German Research Centre for Geoscience**

The GFZ German Research Centre for Geosciences publish a number of geomagnetic indices, a subset of which are used within Swarm-VIP-Dynamic. These are available from: https://kp.gfz-potsdam.de/en/hp30-hp60

**Laboratory for Atmospheric and Space Physics**

The Laboratory for Atmospheric and Space Physics at the University of Colorado publish the F10.7 cm solar radio flux which can be used as a proxy for the solar activity (Tapping, 2013). These data are available at https://lasp.colorado.edu/lisird/.

**Table A1:** Variable Names Used in Databases and Modelling

| Variable Name | Description | Source |
|---|---|---|
| Year_IPIR | Year | IPIR |
| Day_of_year_IPIR | Day of year | IPIR |
| Hour_IPIR | Hour | IPIR |
| Minute_IPIR | Minutes | IPIR |
| Second_IPIR | Seconds | IPIR |
| Ne_IPIR | Plasma density; directly copied from the Langmuir probe files | IPIR |
| Grad_Ne_at_100km_IPIR | The electron density gradient in a running window calculated via linear regression over 27 data points for the 2 Hz electron density data | IPIR |
| Grad_Ne_at_50km_IPIR | The electron density gradient in a running window calculated via linear regression over 13 data points for the 2 Hz electron density data | IPIR |
| Grad_Ne_at_20km_IPIR | The electron density gradient in a running window calculated via linear regression over 5 data points for the 2 Hz electron density data | IPIR |
| ROD_IPIR | Rate Of change of Density | IPIR |
| RODI10s_IPIR | Rate Of change of Density Index (RODI) is the standard deviation of ROD over 10 seconds | IPIR |
| RODI20s_IPIR | Rate Of Density Index (RODI) is the standard deviation of ROD over 20 seconds | IPIR |
| Latitude_IPIR | Position in ITRF – Latitude<br>It is always the absolute value of this value which is used in the models | IPIR |
| Longitude_IPIR | Position in ITRF – Longitude | IPIR |
| CoMLATQD | MLAT Quasi-Dipole Co-ordinates<br>It is always the absolute value of this value which is used in the models | Calculated using codes developed in Swarm-VIP-Dynamic |
| CoMLONQD | MLON Quasi-Dipole Co-ordinates | Calculated using codes developed in Swarm-VIP-Dynamic |
| Radius_IPIR | Position in ITRF – Radius | IPIR |
| ST_LPEXT | Apparent Solar Time | LP_Extended |
| LT_fn | Sine function based on apparent solar time, going from -1 at midnight to +1 at midday | LP_Extended |
| CoMLT | MLT | Calculated using codes developed in Swarm-VIP-Dynamic |
| CorrMLTfn | Sine function based on MLT, going from -1 at midnight to +1 at midday | Calculated using codes developed in Swarm-VIP-Dynamic |
| CoSZA | Solar Zenith Angle | Calculated using codes developed in Swarm-VIP-Dynamic |
| density_DNSPOD | Density derived from GPS accelerations | DNS_POD_2_ |
| density_DNSACC | Thermospheric neutral density | DNSxACC_2_ |
| R | Sunspot number | OMNI low resolution data |
| R1Av | 27-day mean  of the sunspot number, centred on the day to be updated | Calculated from relevant parameters in the OMNI low resolution data |
| R2Av | 81-day mean of the sunspot number, centred on the day to be updated | Calculated from relevant parameters in the OMNI low resolution data |
| R1Med | 27-day median  of the sunspot number, centred on the day to be updated | Calculated from relevant parameters in the OMNI low resolution data |
| R2Med | 81-day median of the sunspot number, centred on the day to be updated | Calculated from relevant parameters in the OMNI low resolution data |
| F107O | F10.7 cm solar radio flux | https://lasp.colorado.edu/lisird/data/penticton_radio_flux/ |
| F107O1Av | 27 day mean value, centred on the day to be updated | https://lasp.colorado.edu/lisird/data/penticton_radio_flux/ |
| F107O2Av | 81 day mean value, centred on the day to be updated | https://lasp.colorado.edu/lisird/data/penticton_radio_flux/ |
| F107O1Med | 27 day median value, centred on the day to be updated | https://lasp.colorado.edu/lisird/data/penticton_radio_flux/ |
| F107O2Med | 81 day median value, centred on the day to be updated | https://lasp.colorado.edu/lisird/data/penticton_radio_flux/ |
| F107O1AvRA | 27 day mean value, leading up to the day to be updated | https://lasp.colorado.edu/lisird/data/penticton_radio_flux/ |
| F107O2AvRA | 81 day mean value, leading up to the day to be updated | https://lasp.colorado.edu/lisird/data/penticton_radio_flux/ |
| F107O1MedRA | 27 day median value, leading up to the day to be updated | https://lasp.colorado.edu/lisird/data/penticton_radio_flux/ |
| F107O2MedRA | 81 day median value, leading up to the day to be updated | https://lasp.colorado.edu/lisird/data/penticton_radio_flux/ |
| F30cm | F30 cm solar radio flux | https://solar.nro.nao.ac.jp |
| f30_mean27c | 27 day mean value, centred on the day to be updated | https://solar.nro.nao.ac.jp |
| f30_mean81c | 81 day mean value, centred on the day to be updated | https://solar.nro.nao.ac.jp |
| f30_med27c | 27 day median value, centred on the day to be updated | https://solar.nro.nao.ac.jp |
| f30_med81c | 81 day median value, centred on the day to be updated | https://solar.nro.nao.ac.jp |
| f30_mean27r | 27 day mean value, leading up to the day to be updated | https://solar.nro.nao.ac.jp |
| f30_mean81r | 81 day mean value, leading up to the day to be updated | https://solar.nro.nao.ac.jp |
| f30_med27r | 27 day median value, leading up to the day to be updated | https://solar.nro.nao.ac.jp |
| f30_med81r | 81 day median value, leading up to the day to be updated | https://solar.nro.nao.ac.jp |
| f10_7eff_swarmc | Proxy from Swarm C | Calculated using codes developed in Swarm-VIP-Dynamic |
| f10_7eff_mean27c | 27 day mean value, centred on the day to be updated | Calculated using codes developed in Swarm-VIP-Dynamic |
| f10_7eff_mean81c | 81 day mean value, centred on the day to be updated | Calculated using codes developed in Swarm-VIP-Dynamic |
| f10_7eff_med27c | 27 day median value, centred on the day to be updated | Calculated using codes developed in Swarm-VIP-Dynamic |
| f10_7eff_med81c | 81 day median value, centred on the day to be updated | Calculated using codes developed in Swarm-VIP-Dynamic |
| f10_7eff_mean27r | 27 day mean value, leading up to the day to be updated | Calculated using codes developed in Swarm-VIP-Dynamic |
| f10_7eff_mean81r | 81 day mean value, leading up to the day to be updated | Calculated using codes developed in Swarm-VIP-Dynamic |
| f10_7eff_med27r | 27 day median value, leading up to the day to be updated | Calculated using codes developed in Swarm-VIP-Dynamic |
| f10_7eff_med81r | 81 day median value, leading up to the day to be updated | Calculated using codes developed in Swarm-VIP-Dynamic |
| DOY_fn | A sine function that takes a value of -1 on $21^{st}$ December and +1 on $21^{st}$ June | Calculated using codes developed in Swarm-VIP-Dynamic |
| DOY_fn_2 | A sine function that takes a value of -1 on $21^{st}$ December and on $21^{st}$ June, with values of +1 on $21^{st}$ March and $21^{st}$ September. | Calculated using codes developed in Swarm-VIP-Dynamic |
| DOY_fn_3 | A sine function that takes a value of -1 at midwinter and +1 at midsummer, regardless of hemisphere | Calculated using codes developed in Swarm-VIP-Dynamic |
| IRC_FACTMS | Ionospheric radial current (IRC)<br>It is always the absolute value of this value which is used in the models | FAC_TMS_2F |
| FAC_FACTMS | Field-aligned current (FAC)<br>It is always the absolute value of this value which is used in the models | FAC_TMS_2F |
| Bx.Mean | As Akasofu, but for the x-component of the Interplanetary Magnetic Field (in nT) | Calculated from relevant parameters in the OMNI high resolution data |
| Bx.stdev | As Akasofu_sd, but for the x-component of the Interplanetary Magnetic Field (in nT) | Calculated from relevant parameters in the OMNI high resolution data |
| By.Mean | As Akasofu, but for the y-component of the Interplanetary Magnetic Field (in nT) | Calculated from relevant parameters in the OMNI high resolution data |
| By.stdev | As Akasofu_sd, but for the y-component of the Interplanetary Magnetic Field (in nT) | Calculated from relevant parameters in the OMNI high resolution data |
| AbsBy.Mean | As Akasofu, but for the absolute value of the y-component of the Interplanetary Magnetic Field (in nT) | Calculated from relevant parameters in the OMNI high resolution data |
| AbsBy.stdev | As Akasofu_sd, but for the absolute value of the y-component of the Interplanetary Magnetic Field (in nT) | Calculated from relevant parameters in the OMNI high resolution data |
| Bz.Mean | As Akasofu, but for the z-component of the Interplanetary Magnetic Field (in nT) | Calculated from relevant parameters in the OMNI high resolution data |
| Bz.stdev | As Akasofu_sd, but for the z-component of the Interplanetary Magnetic Field (in nT) | Calculated from relevant parameters in the OMNI high resolution data |
| Bt.Mean | As Akasofu, but for the Interplanetary Magnetic Field (in nT) | Calculated from relevant parameters in the OMNI high resolution data |
| Bt.stdev | As Akasofu_sd, but for the Interplanetary Magnetic Field (in nT) | Calculated from relevant parameters in the OMNI high resolution data |
| Clock.Mean | As Akasofu, but for the clock angle of the Interplanetary Magnetic Field (in nT) | Calculated from relevant parameters in the OMNI high resolution data |
| Clock.stdev | As Akasofu_sd, but for the clock angle of the Interplanetary Magnetic Field (in nT) | Calculated from relevant parameters in the OMNI high resolution data |
| swvel.Mean | As Akasofu, but for the solar wind velocity (in km $s^{-1}$) | Calculated from relevant parameters in the OMNI high resolution data |
| swvel.stdev | As Akasofu_sd, but for the solar wind velocity (in km $s^{-1}$) | Calculated from relevant parameters in the OMNI high resolution data |
| swden.Mean | As Akasofu, but for the solar wind density (in n $cc^{-3}$) | Calculated from relevant parameters in the OMNI high resolution data |
| swden.stdev | As Akasofu_sd, but for the solar wind density (in n $cc^{-3}$) | Calculated from relevant parameters in the OMNI high resolution data |

| swpress.Mean | As Akasofu, but for the solar wind pressure (in nPa) | Calculated from relevant parameters in the OMNI high resolution data |
|---|---|---|
| swpress.stdev | As Akasofu_sd, but for the solar wind pressure (in nPa) | Calculated from relevant parameters in the OMNI high resolution data |
| Newell.Mean | As Akasofu, but for the Newell solar wind coupling function | Calculated from relevant parameters in the OMNI high resolution data |
| Newell.stdev | As Akasofu_sd, but for the Newell solar wind coupling function | Calculated from relevant parameters in the OMNI high resolution data |
| Elya.Mean | As Akasofu, but for the Elya solar wind coupling function | Calculated from relevant parameters in the OMNI high resolution data |
| Elya.stdev | As Akasofu_sd, but for the Elya solar wind coupling function | Calculated from relevant parameters in the OMNI high resolution data |
| Akasofu.Mean | The average value of the Akasofu solar wind coupling function, across a two-hour window, starting two hours before the observation to be updated | Calculated from relevant parameters in the OMNI high resolution data |
| Akasofu.stdev | The standard deviation of the Akasofu solar wind coupling function, across a two-hour window, starting two hours before the observation to be updated | Calculated from relevant parameters in the OMNI high resolution data |
| ief.Mean | As Akasofu, but for the Interplanetary Electric Field (in mV $m^{-1}$) | OMNI high resolution data |
| ief.stdev | As Akasofu_sd, but for the Interplanetary Electric Field (in mV $m^{-1}$) | OMNI high resolution data |
| Kp | Kp index | OMNI low resolution data |
| Ap | Ap index (in nT) | OMNI low resolution data |
| AE | AE index (in nT) | OMNI low resolution data |
| AL | AL index (in nT) | OMNI low resolution data |
| AU | AU index (in nT) | OMNI low resolution data |
| aa | aa index (in nT) | https://isgi.unistra.fr/ |
| am | aa index (in nT) | https://isgi.unistra.fr/ |
| SYM_D | SYM_D index (in nT) | OMNI high resolution data |
| SYM_H | SYM_H index (in nT) | OMNI high resolution data |
| ASY_D | ASY_D index (in nT) | OMNI high resolution data |
| ASY_H | ASY_H index (in nT) | OMNI high resolution data |
| PCN | PCN index | OMNI high resolution data |
| Dst | Dst index (in nT) | OMNI low resolution data |
| Hp30 | Hp30 index | GFZ German Research Centre for Geosciences |
| Hp60 | Hp60 index | GFZ German Research Centre for Geosciences |
| tvt | Subset (0: Training, 1: Optimisation, 2: Evaluation) | NA |
| LT_fn_comp | Sine function based on apparent solar time, 90 degrees out of phase with LT_fn | LP_Extended |
| MLT_fn_comp | Sine function based on magnetic time, 90 degrees out of phase with MLT_fn | Calculated using codes developed in Swarm-VIP-Dynamic |
| Lat_Diff | The difference in the geographic and geomagnetic latitudes | Latitude from the IDPxIRR_2F data product – the quasi-diplole geomagnetic latitude calculated by the Swarm-VIP-Dynamic team |

To represent Equatorial Ionisation Anomaly (EIA), a Lat_Diff parameter was used as a function to represent the crests in EIA. This function was:
$\left|9.5 - |\mathrm{MLAT}|\right|$ for the electron density and
$\left|12 - |\mathrm{MLAT}|\right|$ for other dependent variables.

**Supplementary materials**

for

**Statistical Models of Ionospheric Variability and Irregularities in the Topside Ionosphere Based on the Swarm Satellite Data**

Daria Kotova[(1)], Alan Wood[(2)], Eelco Doornbos[(3)], Jaroslav Urbář[(4)], Luca Spogli[(5)], Yaqi Jin[(1)], Lucilla Alfonsi[(5)], Gareth Dorrian[(6)], Mainul Hoque[(7)], Kasper van Dam[(3)], Elisabetta Iorfida[(8)], and Wojciech J. Miloch[(1)]

(1) Department of Physics, University of Oslo, Norway Met Office, Exeter, UK
(2) Met Office, Exeter, UK Department of Physics, University of Oslo, Norway
(3) The Royal Netherlands Meteorological Institute (KNMI), The Netherlands
(4) Institute of Atmospheric Physics CAS, Czech Republic
(5) Istituto Nazionale di Geofisica e Vulcanologia, Italy
(6) Space Environment and Radio Engineering (SERENE) group, University of Birmingham, UK
(7) German Aerospace Center (DLR), Germany
(8) European Space Agency (ESA), Noordwijk, The Netherlands

*Corresponding author: Daria Kotova (daria.kotova@fys.uio.no)

Contents of this file: supporting the manuscript text Tables S1-S4 and description of Explanatory Variables & Parameter Estimates for three versions of models.

## Cleaning of databases

The databases were cleaned based on the flags in the various Swarm data products, as shown in Table S1.

**Table S1:** Details of cleaning databases for Swarm data products.

| Data Product | Field to check | Criteria for change | Fields changed to NaN | Notes |
|---|---|---|---|---|
| IPDxIRR_2F | Ne_quality_flag_IPIR | 40000 (anomalous data) | IPIR: Ne, ROD, delta_Ne10s, delta_Ne20s & delta_Ne40s | |
| IPDxIRR_2F | Ne_quality_flag_IPIR | 99999 (filler value) | IPIR: Ne, ROD, delta_Ne10s, delta_Ne20s & delta_Ne40s | |
| IPDxIRR_2F | In addition, in the region of low and mid-latitudes around 09 and 15 magnetic local time (MLT), there have been observed artifacts in the Langmuir probe data from all Swarm satellites. Details on their morphology and occurrence have been given in supplementary material to Jin et al. (2020). To | | Whole IPDxIRR_2F product | |
| | eliminate this artifact in the Langmuir probe data, a filter based on comparison with data obtained with the onboard GPS receiver was used. The following data filtering was used for two time intervals: (1) 8.1 MLT < $t$ < 9.3 MLT; (2) 14.3 MLT < $t$ < 16.0 MLT: if RODI data was larger than 1000 $cm^{-3}$/s and the rate of change of TEC index (ROTI) was less than 0.02 (0.01) TECU/s, then such data was discarded. A level of 0.02 TECU/s corresponds to the usual distribution of ROTI in 10 s near noise for a high level of solar activity and 0.01 TECU/s for a low one.<br><br>(1) high solar activity (HSA) from August 2014 to July 2015 and (2) low solar activity (LSA) from January to December 2018.<br><br>This definition of HSA ranged from 89 – 246.9 sfu with a median of 132 sfu and a mean of 135.1 sfu<br><br>This definition of LSA ranged from 65.8 - 82.1 sfu with a median of 69.5 sfu and a mean of 69.9 sfu<br><br>Therefore, a value of 85.5 sfu was used as the break point between HSA and LSA | | | |
| IPDxIRR_2F | ROD_IPIR, RODI10s_ IPIR and RODI20s_ IPIR | If all of these equal zero, then these are replaced by NaN. | ROD_IPIR, RODI10s_ IPIR and RODI20s_ IPIR | 1 |
| 2 Hz Langmuir Probe extended Dataset | SZA_LPEXT, ST_LPEXT, Diplat_LPEXT and Diplon_LPEXT | If these are all 0 (filler values of zero used in this product), then replace the product with NaN). | Co-ordinates from whole LP product | 2 |
| FAC_TMS_2F | FAC_TMS | Flags_B_FACTMS = 255 | FAC and IRC | 3 |
| FAC_TMS_2F | FAC_TMS | Flags_q_FACTMS = 255 | FAC and IRC | 3 |
| DNSxPOD_2 | density_DNSPOD | Negative | density_DNSPOD | |
| DNSxPOD_2 | density_DNSPOD | Greater than 1 (to remove filler values of $9.99x10^{32}$) | density_DNSPOD | |
| DNSxPOD_2 | validity_flag_DNSPOD | 1 (anomalous data) | density_DNSPOD | |
| DNSxACC_2_ | density_DNSACC | Negative | density_DNSACC | |
| DNSxACC_2_ | density_DNSACC | Greater than 1 (to remove filler values of $9.99x10^{32}$) | density_DNSACC | 4 |
| ACCxCAL_2_ | flag_val_ACCCAL | 1 (anomalous data) | density_DNSACC | 4 |
| ACCxCAL_2_ | flag_thr_ACCCAL | 1 (thrustors on) | density_DNSACC | 4 |

**Notes**

[1] The rationale for this is that it was unlikely that all of these would be exactly zero, and it was judged to be an anomaly.

[2] These data are from the Swarm 2 Hz Langmuir Probe extended dataset, which is used to get the magnetic co-ordinates of the satellite and the solar zenith angle (SZA). However, this data product uses zero, and is used to both represent real zeros and times of no data. Therefore, if the specified fields were all zero, this condition is not physically possible. In this case, the zeros were interpreted as meaning 'no data' and the entire data product was replaced by NaN.

[3] The use of these flags in this manner was at the recommendation of Enkelejda Qamili. This condition never occurred in the database.

[4] The DNSxACC_2_ product does not have quality flags. It was decided to use the flags from the ACCxCAL_2_ to identify anomalous points in DNSxACC_2_

**Table S2:** Correlations between explanatory variables for the training dataset in all geographic regions. Correlations of more than |0.25| are shaded in red. From the Excel file *Table S2_correlations-analysis.xlsx*

**Table S3:** An assessment of how well GLMs based upon one explanatory variable (which contain one proxy for one physical process) explain the variability exhibited by each of the dependent variables trialled within the Swarm-VIP-Dynamic project in each latitudinal region. In each case, the median value, 10th percentile and 90th percentile of the dependent variable is shown, and a proxy for the range is calculated from the difference between the 90th percentile and the 10th percentile of this variable. The 10th and 90th percentiles of the explanatory variables were used to make predictions of the dependent variable, and the difference between these predictions gives the predicted range. The ratio of the predicted range to the observed range is also shown. From the Excel file *Table S3_model_summary_present.xlsx*

**Table S4:** Correlations for single-term models. Each dependent variable was tested for each latitudinal region with each explanatory variable. From the Excel file *Table S4_Model-single-term.xlsx*

# Explanatory Variables & Parameter Estimates: Version 3.1

This shows the output from R (statistical modelling software). These statistics are also available in the Excel spreadsheet Explanatory-Varibles-Parameter-Estimates.xlsx

**Form of the model**

The form of a generalised linear model is:

$$g(E(y)) = \beta_0 + \beta_1 \cdot x_1 + \cdots + \beta_n \cdot x_n$$

$E(y)$ is the expected value of the dependent variable $y$

$x_1 \cdots x_n$ are the independent, or explanatory, variables

$\beta_1 \cdots \beta_n$ are the parameter estimates for the model

**Transformation applied: All dependent variables except the one-dimensional spectral index p**

In our implementation, the dependent variable is transformed and the nth root is modelled. A logarithmic link function is used. Therefore, the form of the equation is:

$$\sqrt[n]{E(y)} = exp(\beta_0 + \beta_1 \cdot x_1 + \cdots + \beta_n \cdot x_n)$$

**Transformation applied: One-dimensional spectral index p**

Different transforms were applied for this dependent variable. In the polar, auroral and midlatitude regions, a normal distribution was used. The form of the equation is:

$$E(y) - \min(E(y)) + 0.01 = exp(\beta_0 + \beta_1 \cdot x_1 + \cdots + \beta_n \cdot x_n)$$

In the equatorial regions, a gamma distribution with a logarithmic link function was used. Therefore, the form of the equation is:

$$-E(y) + \max(E(y)) + 0.01 = exp(\beta_0 + \beta_1 \cdot x_1 + \cdots + \beta_n \cdot x_n)$$

**Electron Density, Polar region, Northern Hemisphere**

```
                                   Estimate Std. Error t value Pr(>|t|)
(Intercept)                       3.964e+00  4.492e-02   88.26   <2e-16 ***
f10_7eff_swarmc[first_pt:last_pt] 9.824e-03  7.295e-05  134.66   <2e-16 ***
DOY_fn_3[first_pt:last_pt]        4.386e-01  4.015e-03  109.24   <2e-16 ***
LT_fn[first_pt:last_pt]           1.502e-01  3.656e-03   41.09   <2e-16 ***
DOY_fn_2[first_pt:last_pt]        7.784e-02  3.679e-03   21.16   <2e-16 ***
LT_fn_comp[first_pt:last_pt]      9.762e-02  3.559e-03   27.43   <2e-16 ***
abs(CoMLATQD[first_pt:last_pt])   5.927e-03  5.377e-04   11.02   <2e-16 ***
Hp30[first_pt:last_pt]            4.273e-02  1.864e-03   22.92   <2e-16 ***
swden.Mean[first_pt:last_pt]      8.417e-03  5.533e-04   15.21   <2e-16 ***
By.Mean[first_pt:last_pt]         1.183e-02  7.426e-04   15.94   <2e-16 ***
```

**Electron Density, Polar region, Southern Hemisphere**

```
                                    Estimate Std. Error t value Pr(>|t|)
(Intercept)                        3.703e+00  2.409e-02 153.670  < 2e-16 ***
f10_7eff_swarmc[first_pt:last_pt]  8.666e-03  3.969e-05 218.332  < 2e-16 ***
DOY_fn_3[first_pt:last_pt]         4.625e-01  2.185e-03 211.659  < 2e-16 ***
LT_fn[first_pt:last_pt]            1.855e-01  2.111e-03  87.863  < 2e-16 ***
DOY_fn_2[first_pt:last_pt]         1.014e-01  2.171e-03  46.686  < 2e-16 ***
LT_fn_comp[first_pt:last_pt]       1.181e-01  2.161e-03  54.665  < 2e-16 ***
abs(CoMLATQD[first_pt:last_pt])    1.013e-02  2.935e-04  34.512  < 2e-16 ***
Hp30[first_pt:last_pt]             2.793e-02  1.164e-03  23.989  < 2e-16 ***
SYM_D[first_pt:last_pt]            3.157e-03  4.507e-04   7.004 2.51e-12 ***
swden.Mean[first_pt:last_pt]       7.502e-03  3.256e-04  23.041  < 2e-16 ***
DOY_fn[first_pt:last_pt]          -5.402e-02  2.222e-03 -24.305  < 2e-16 ***
By.Mean[first_pt:last_pt]          4.422e-03  4.473e-04   9.886  < 2e-16 ***
```

```
Lat_Diff[first_pt:last_pt]        -6.488e-04  2.106e-04  -3.081  0.00207 **
```

**Electron Density, Auroral region, Northern Hemisphere**

```
                                   Estimate Std. Error t value Pr(>|t|)
(Intercept)                         4.630e+00  2.475e-02 187.074  < 2e-16 ***
f10_7eff_swarmc[first_pt:last_pt]  8.618e-03  5.401e-05 159.554  < 2e-16 ***
DOY_fn_3[first_pt:last_pt]          3.649e-01  2.608e-03 139.890  < 2e-16 ***
LT_fn[first_pt:last_pt]             1.640e-01  2.495e-03  65.719  < 2e-16 ***
LT_fn_comp[first_pt:last_pt]        6.982e-02  2.328e-03  29.990  < 2e-16 ***
DOY_fn_2[first_pt:last_pt]          3.523e-02  2.471e-03  14.256  < 2e-16 ***
Bt.Mean[first_pt:last_pt]           7.802e-03  7.104e-04  10.982  < 2e-16 ***
abs(CoMLATQD[first_pt:last_pt])     1.477e-03  2.714e-04   5.443 5.28e-08 ***
Clock.Mean[first_pt:last_pt]       -2.778e-04  4.858e-05  -5.719 1.08e-08 ***
swvel.Mean[first_pt:last_pt]       -2.036e-04  1.822e-05 -11.172  < 2e-16 ***
SYM_D[first_pt:last_pt]            -2.060e-03  5.824e-04  -3.537 0.000405 ***
```

**Electron Density, Auroral region, Southern Hemisphere**

```
                                   Estimate Std. Error t value Pr(>|t|)
(Intercept)                         4.823e+00  2.952e-02 163.383  < 2e-16 ***
f10_7eff_swarmc[first_pt:last_pt]  6.113e-03  6.222e-05  98.248  < 2e-16 ***
DOY_fn_3[first_pt:last_pt]          5.481e-01  3.236e-03 169.396  < 2e-16 ***
LT_fn[first_pt:last_pt]             2.410e-01  2.881e-03  83.654  < 2e-16 ***
LT_fn_comp[first_pt:last_pt]        6.843e-02  2.779e-03  24.622  < 2e-16 ***
DOY_fn_2[first_pt:last_pt]          1.086e-01  2.965e-03  36.638  < 2e-16 ***
Bt.Mean[first_pt:last_pt]           1.888e-03  8.186e-04   2.306 0.021120 *
abs(CoMLATQD[first_pt:last_pt])     8.408e-04  3.348e-04   2.511 0.012030 *
Clock.Mean[first_pt:last_pt]       -4.187e-04  5.765e-05  -7.263 3.88e-13 ***
swvel.Mean[first_pt:last_pt]       -2.292e-04  2.110e-05 -10.864  < 2e-16 ***
SYM_D[first_pt:last_pt]             6.667e-03  6.551e-04  10.177  < 2e-16 ***
Bx.Mean[first_pt:last_pt]           2.436e-03  6.280e-04   3.879 0.000105 ***
```

**Electron Density, Midlatitude region, Northern Hemisphere**

```
                                  Estimate Std. Error  t value Pr(>|t|)
(Intercept)                        6.091e+00  7.283e-03  836.359  < 2e-16 ***
density_DNSPOD[first_pt:last_pt]  4.658e+11  4.105e+09  113.467  < 2e-16 ***
DOY_fn_3[first_pt:last_pt]         1.662e-01  2.299e-03   72.295  < 2e-16 ***
abs(CoMLATQD[first_pt:last_pt])  -1.473e-02  1.293e-04 -113.909  < 2e-16 ***
LT_fn_comp[first_pt:last_pt]       1.516e-02  2.173e-03    6.975 3.11e-12 ***
AbsBy.Mean[first_pt:last_pt]       6.682e-03  8.876e-04    7.528 5.26e-14 ***
swvel.stdev[first_pt:last_pt]    -2.720e-03  2.961e-04   -9.187  < 2e-16 ***
DOY_fn_2[first_pt:last_pt]         1.307e-02  2.278e-03    5.735 9.82e-09 ***
Bx.Mean[first_pt:last_pt]          3.084e-03  5.072e-04    6.081 1.21e-09 ***
```

**Electron Density, Midlatitude region, Southern Hemisphere**

```
                                  Estimate Std. Error  t value Pr(>|t|)
(Intercept)                        6.108e+00  7.653e-03  798.162  < 2e-16 ***
density_DNSPOD[first_pt:last_pt]  5.016e+11  4.184e+09  119.883  < 2e-16 ***
DOY_fn_3[first_pt:last_pt]         3.005e-01  2.549e-03  117.892  < 2e-16 ***
abs(CoMLATQD[first_pt:last_pt])  -1.608e-02  1.369e-04 -117.444  < 2e-16 ***
LT_fn_comp[first_pt:last_pt]       1.272e-02  2.336e-03    5.444 5.23e-08 ***
AbsBy.Mean[first_pt:last_pt]       6.040e-03  9.229e-04    6.545 6.02e-11 ***
DOY_fn_2[first_pt:last_pt]         2.255e-02  2.447e-03    9.214  < 2e-16 ***
Bz.Mean[first_pt:last_pt]         -1.649e-03  7.104e-04   -2.322   0.0202 *
SYM_D[first_pt:last_pt]            1.230e-02  5.960e-04   20.638  < 2e-16 ***
Bx.Mean[first_pt:last_pt]          3.420e-03  5.494e-04    6.225 4.86e-10 ***
```

**Electron Density, Equatorial region, Both Hemispheres (all terms considered)**

```
                                   Estimate Std. Error  t value Pr(>|t|)
(Intercept)                         1.146e+01  4.139e-03 2770.058  < 2e-16 ***
f10_7eff_swarmc[first_pt:last_pt]  1.267e-02  4.057e-05  312.170  < 2e-16 ***
LT_fn_comp[first_pt:last_pt]        5.971e-01  1.842e-03  324.210  < 2e-16 ***
LT_fn[first_pt:last_pt]             3.638e-01  1.809e-03  201.136  < 2e-16 ***
```

```
abs(CoMLATQD[first_pt:last_pt])   -1.982e-02  2.474e-04  -80.103  < 2e-16 ***
DOY_fn_3[first_pt:last_pt]         2.273e-01  1.778e-03  127.861  < 2e-16 ***
DOY_fn_2[first_pt:last_pt]         1.328e-01  1.951e-03   68.076  < 2e-16 ***
DOY_fn[first_pt:last_pt]          -1.609e-01  1.998e-03  -80.538  < 2e-16 ***
Kp[first_pt:last_pt]               1.614e-03  1.091e-04   14.801  < 2e-16 ***
Bx.Mean[first_pt:last_pt]          1.710e-03  4.326e-04    3.953 7.71e-05 ***
Lat_Diff[first_pt:last_pt]        -2.760e-03  1.847e-04  -14.942  < 2e-16 ***
EIA_fn[first_pt:last_pt]          -1.686e-02  3.714e-04  -45.410  < 2e-16 ***
```

**|Grad_Ne@100km|, Polar region, Northern Hemisphere**

```
                                    Estimate Std. Error t value Pr(>|t|)
(Intercept)                       -1.715e+00  2.168e-02  -79.09   <2e-16 ***
f10_7eff_swarmc[first_pt:last_pt]  3.951e-03  3.515e-05  112.40   <2e-16 ***
DOY_fn[first_pt:last_pt]          -6.292e-02  2.006e-03  -31.36   <2e-16 ***
Kp[first_pt:last_pt]               2.768e-03  1.061e-04   26.10   <2e-16 ***
LT_fn[first_pt:last_pt]            7.625e-02  1.880e-03   40.56   <2e-16 ***
DOY_fn_3[first_pt:last_pt]         1.199e-01  1.982e-03   60.51   <2e-16 ***
DOY_fn_2[first_pt:last_pt]         5.439e-02  1.932e-03   28.16   <2e-16 ***
abs(CoMLATQD[first_pt:last_pt])    6.814e-03  2.629e-04   25.92   <2e-16 ***
LT_fn_comp[first_pt:last_pt]       4.482e-02  1.916e-03   23.40   <2e-16 ***
swden.Mean[first_pt:last_pt]       3.628e-03  2.903e-04   12.49   <2e-16 ***
```

**|Grad_Ne@100km|, Polar region, Southern Hemisphere**

```
                                    Estimate Std. Error t value Pr(>|t|)
(Intercept)                       -1.752e+00  2.272e-02 -77.123   <2e-16 ***
f10_7eff_swarmc[first_pt:last_pt]  3.943e-03  3.742e-05 105.363   <2e-16 ***
DOY_fn[first_pt:last_pt]          -6.488e-02  2.095e-03 -30.972   <2e-16 ***
Kp[first_pt:last_pt]               2.842e-03  1.126e-04  25.237   <2e-16 ***
LT_fn[first_pt:last_pt]            7.896e-02  1.991e-03  39.659   <2e-16 ***
DOY_fn_3[first_pt:last_pt]         1.184e-01  2.060e-03  57.447   <2e-16 ***
DOY_fn_2[first_pt:last_pt]         5.248e-02  2.051e-03  25.582   <2e-16 ***
abs(CoMLATQD[first_pt:last_pt])    7.284e-03  2.768e-04  26.314   <2e-16 ***
LT_fn_comp[first_pt:last_pt]       5.118e-02  2.038e-03  25.114   <2e-16 ***
Lat_Diff[first_pt:last_pt]         4.700e-04  1.987e-04   2.366   0.0180 *
swden.Mean[first_pt:last_pt]       3.643e-03  3.073e-04  11.858   <2e-16 ***
SYM_D[first_pt:last_pt]            1.030e-03  4.253e-04   2.421   0.0155 *
```

**|Grad_Ne@100km|, Auroral region, Northern Hemisphere**

```
                                    Estimate Std. Error t value Pr(>|t|)
(Intercept)                       -1.684e+00  2.231e-02 -75.478  < 2e-16 ***
f10_7eff_swarmc[first_pt:last_pt]  3.588e-03  5.826e-05  61.589  < 2e-16 ***
DOY_fn_3[first_pt:last_pt]         8.253e-03  2.740e-03   3.012  0.00259 **
Bt.Mean[first_pt:last_pt]          1.086e-02  7.710e-04  14.086  < 2e-16 ***
LT_fn_comp[first_pt:last_pt]       3.919e-02  2.525e-03  15.518  < 2e-16 ***
abs(CoMLATQD[first_pt:last_pt])    7.838e-03  2.805e-04  27.944  < 2e-16 ***
DOY_fn_2[first_pt:last_pt]         3.266e-02  2.653e-03  12.313  < 2e-16 ***
Bz.Mean[first_pt:last_pt]         -7.242e-03  7.469e-04  -9.696  < 2e-16 ***
```

**|Grad_Ne@100km|, Auroral region, Southern Hemisphere**

```
                                    Estimate Std. Error t value Pr(>|t|)
(Intercept)                       -1.412e+00  2.555e-02 -55.277  < 2e-16 ***
f10_7eff_swarmc[first_pt:last_pt]  2.701e-03  6.173e-05  43.766  < 2e-16 ***
DOY_fn[first_pt:last_pt]          -1.643e-01  3.179e-03 -51.677  < 2e-16 ***
Bt.Mean[first_pt:last_pt]          5.756e-03  8.066e-04   7.137 9.76e-13 ***
LT_fn_comp[first_pt:last_pt]       3.800e-02  2.763e-03  13.755  < 2e-16 ***
abs(CoMLATQD[first_pt:last_pt])    5.493e-03  3.301e-04  16.641  < 2e-16 ***
DOY_fn_2[first_pt:last_pt]         3.637e-02  2.926e-03  12.430  < 2e-16 ***
LT_fn[first_pt:last_pt]            4.885e-02  2.863e-03  17.062  < 2e-16 ***
Bz.Mean[first_pt:last_pt]         -5.780e-03  7.834e-04  -7.378 1.65e-13 ***
```

**|Grad_Ne@100km|, Midlatitude region, Northern Hemisphere**

```
                                       Estimate Std. Error t value Pr(>|t|)
```

```
(Intercept)                          -2.210e+00  1.593e-01 -13.869  < 2e-16 ***
density_DNSACC[first_pt:last_pt]     -2.192e+10  8.854e+09  -2.475  0.01332 *
Newell.Mean[first_pt:last_pt]         5.676e-06  5.923e-07   9.582  < 2e-16 ***
Height_HIGHRESIPIR[first_pt:last_pt]  3.379e-03  3.594e-04   9.400  < 2e-16 ***
CoMLT[first_pt:last_pt]               8.815e-04  2.896e-04   3.044  0.00234 **
DOY_fn_2[first_pt:last_pt]           -8.225e-03  2.975e-03  -2.765  0.00571 **
```

**|Grad_Ne@100km|, Midlatitude region, Southern Hemisphere**

```
                                   Estimate Std. Error  t value Pr(>|t|)
(Intercept)                      -6.938e-01  3.063e-03 -226.482  < 2e-16 ***
density_DNSACC[first_pt:last_pt]  1.239e+11  3.054e+09   40.556  < 2e-16 ***
DOY_fn_3[first_pt:last_pt]        5.757e-02  1.976e-03   29.131  < 2e-16 ***
Newell.Mean[first_pt:last_pt]     2.260e-06  3.783e-07    5.973 2.35e-09 ***
Bt.stdev[first_pt:last_pt]        8.075e-03  2.667e-03    3.028  0.00247 **
CoMLT[first_pt:last_pt]          -1.120e-03  1.787e-04   -6.266 3.74e-10 ***
DOY_fn_2[first_pt:last_pt]       -1.851e-02  1.771e-03  -10.456  < 2e-16 ***
```

**|Grad_Ne@100km|, Equatorial region, Both Hemispheres (all terms considered)**

```
                                   Estimate Std. Error  t value Pr(>|t|)
(Intercept)                      -7.661e-01  2.525e-03 -303.400  < 2e-16 ***
density_DNSACC[first_pt:last_pt]  2.951e+11  1.640e+09  179.899  < 2e-16 ***
CoSZA[first_pt:last_pt]           6.664e-04  8.078e-06   82.498  < 2e-16 ***
DOY_fn[first_pt:last_pt]         -2.138e-02  1.229e-03  -17.395  < 2e-16 ***
DOY_fn_2[first_pt:last_pt]        2.620e-02  1.113e-03   23.554  < 2e-16 ***
DOY_fn_3[first_pt:last_pt]        4.020e-02  1.084e-03   37.091  < 2e-16 ***
Hp30[first_pt:last_pt]           -4.996e-03  6.152e-04   -8.121 4.66e-16 ***
Bx.Mean[first_pt:last_pt]         9.851e-04  2.451e-04    4.019 5.84e-05 ***
Lat_Diff[first_pt:last_pt]       -1.137e-03  1.079e-04  -10.536  < 2e-16 ***
swden.Mean[first_pt:last_pt]     -8.703e-04  1.615e-04   -5.389 7.10e-08 ***
EIA_fn[first_pt:last_pt]         -2.338e-02  1.551e-04 -150.805  < 2e-16 ***
```

**|RODI10s|, Polar region, Northern Hemisphere**

```
                                    Estimate Std. Error t value Pr(>|t|)
(Intercept)                        8.521e-01  2.411e-02  35.335   <2e-16 ***
f10_7eff_swarmc[first_pt:last_pt]  5.223e-03  3.896e-05 134.043   <2e-16 ***
DOY_fn[first_pt:last_pt]          -6.791e-02  2.222e-03 -30.563   <2e-16 ***
LT_fn[first_pt:last_pt]            9.619e-02  2.078e-03  46.299   <2e-16 ***
DOY_fn_3[first_pt:last_pt]         1.551e-01  2.190e-03  70.808   <2e-16 ***
Kp[first_pt:last_pt]               3.168e-03  1.201e-04  26.382   <2e-16 ***
abs(CoMLATQD[first_pt:last_pt])    9.366e-03  2.904e-04  32.248   <2e-16 ***
DOY_fn_2[first_pt:last_pt]         6.447e-02  2.135e-03  30.200   <2e-16 ***
LT_fn_comp[first_pt:last_pt]       5.825e-02  2.117e-03  27.520   <2e-16 ***
swden.Mean[first_pt:last_pt]       3.837e-03  3.209e-04  11.958   <2e-16 ***
Clock.stdev[first_pt:last_pt]      2.778e-04  1.140e-04   2.437   0.0148 *
```

**|RODI10s|, Polar region, Southern Hemisphere**

```
                                    Estimate Std. Error t value Pr(>|t|)
(Intercept)                        8.012e-01  2.533e-02  31.636  < 2e-16 ***
f10_7eff_swarmc[first_pt:last_pt]  5.211e-03  4.155e-05 125.415  < 2e-16 ***
DOY_fn[first_pt:last_pt]          -7.067e-02  2.325e-03 -30.399  < 2e-16 ***
LT_fn[first_pt:last_pt]            9.921e-02  2.204e-03  45.014  < 2e-16 ***
DOY_fn_3[first_pt:last_pt]         1.525e-01  2.281e-03  66.866  < 2e-16 ***
Kp[first_pt:last_pt]               3.260e-03  1.277e-04  25.525  < 2e-16 ***
abs(CoMLATQD[first_pt:last_pt])    9.999e-03  3.064e-04  32.631  < 2e-16 ***
DOY_fn_2[first_pt:last_pt]         6.175e-02  2.271e-03  27.193  < 2e-16 ***
LT_fn_comp[first_pt:last_pt]       6.677e-02  2.256e-03  29.598  < 2e-16 ***
Lat_Diff[first_pt:last_pt]         9.897e-04  2.199e-04   4.501 6.78e-06 ***
swden.Mean[first_pt:last_pt]       3.820e-03  3.402e-04  11.229  < 2e-16 ***
SYM_D[first_pt:last_pt]            1.205e-03  4.709e-04   2.559   0.0105 *
Clock.stdev[first_pt:last_pt]      2.935e-04  1.207e-04   2.430   0.0151 *
```

**|RODI10s|, Auroral region, Northern Hemisphere**

| | Estimate | Std. Error | t value | Pr(>\|t\|) | |
|---|---|---|---|---|---|
| (Intercept) | 6.312e-01 | 2.537e-02 | 24.878 | < 2e-16 | *** |
| f10_7eff_swarmc[first_pt:last_pt] | 4.771e-03 | 6.406e-05 | 74.477 | < 2e-16 | *** |
| DOY_fn[first_pt:last_pt] | 2.533e-02 | 3.037e-03 | 8.340 | < 2e-16 | *** |
| abs(CoMLATQD[first_pt:last_pt]) | 1.394e-02 | 3.173e-04 | 43.926 | < 2e-16 | *** |
| Bt.Mean[first_pt:last_pt] | 1.364e-02 | 8.675e-04 | 15.727 | < 2e-16 | *** |
| DOY_fn_2[first_pt:last_pt] | 3.828e-02 | 2.924e-03 | 13.095 | < 2e-16 | *** |
| LT_fn_comp[first_pt:last_pt] | 4.439e-02 | 2.764e-03 | 16.062 | < 2e-16 | *** |
| CorrMLTfn[first_pt:last_pt] | -2.553e-02 | 2.905e-03 | -8.789 | < 2e-16 | *** |
| Elya.stdev[first_pt:last_pt] | 6.357e-05 | 1.709e-05 | 3.720 | 0.000199 | *** |
| SYM_D[first_pt:last_pt] | -2.862e-03 | 6.907e-04 | -4.144 | 3.43e-05 | *** |
| ief.Mean[first_pt:last_pt] | 1.859e-02 | 1.897e-03 | 9.797 | < 2e-16 | *** |

**|RODI10s|, Auroral region, Southern Hemisphere**

| | Estimate | Std. Error | t value | Pr(>\|t\|) | |
|---|---|---|---|---|---|
| (Intercept) | 8.818e-01 | 2.949e-02 | 29.903 | <2e-16 | *** |
| f10_7eff_swarmc[first_pt:last_pt] | 3.623e-03 | 6.936e-05 | 52.228 | <2e-16 | *** |
| DOY_fn[first_pt:last_pt] | -1.862e-01 | 3.494e-03 | -53.289 | <2e-16 | *** |
| abs(CoMLATQD[first_pt:last_pt]) | 1.210e-02 | 3.818e-04 | 31.696 | <2e-16 | *** |
| Bt.Mean[first_pt:last_pt] | 9.035e-03 | 9.045e-04 | 9.989 | <2e-16 | *** |
| DOY_fn_2[first_pt:last_pt] | 4.424e-02 | 3.269e-03 | 13.535 | <2e-16 | *** |
| LT_fn_comp[first_pt:last_pt] | 4.618e-02 | 3.105e-03 | 14.874 | <2e-16 | *** |
| CorrMLTfn[first_pt:last_pt] | -2.818e-02 | 3.183e-03 | -8.854 | <2e-16 | *** |
| ief.Mean[first_pt:last_pt] | 1.901e-02 | 1.994e-03 | 9.536 | <2e-16 | *** |

**|RODI10s|, Midlatitude region, Northern Hemisphere**

| | Estimate | Std. Error | t value | Pr(>\|t\|) | |
|---|---|---|---|---|---|
| (Intercept) | -4.797e+01 | 3.011e+00 | -15.930 | < 2e-16 | *** |
| CorrMLTfn[first_pt:last_pt] | -1.883e-01 | 2.673e-03 | -70.460 | < 2e-16 | *** |
| f10_7eff_swarmc[first_pt:last_pt] | -6.349e-04 | 1.333e-04 | -4.762 | 1.92e-06 | *** |
| abs(CoMLATQD[first_pt:last_pt]) | 3.941e-03 | 1.836e-04 | 21.464 | < 2e-16 | *** |
| DOY_fn_3[first_pt:last_pt] | -1.259e-02 | 2.925e-03 | -4.303 | 1.69e-05 | *** |
| Radius_IPIR[first_pt:last_pt] | 7.266e-06 | 4.428e-07 | 16.408 | < 2e-16 | *** |
| SYM_H[first_pt:last_pt] | -1.241e-03 | 1.451e-04 | -8.556 | < 2e-16 | *** |
| MLT_fn_comp[first_pt:last_pt] | -6.265e-02 | 2.688e-03 | -23.304 | < 2e-16 | *** |
| Bz.stdev[first_pt:last_pt] | 1.538e-02 | 1.899e-03 | 8.099 | 5.71e-16 | *** |
| DOY_fn_2[first_pt:last_pt] | -2.900e-02 | 2.846e-03 | -10.190 | < 2e-16 | *** |

**|RODI10s|, Midlatitude region, Southern Hemisphere**

| | Estimate | Std. Error | t value | Pr(>\|t\|) | |
|---|---|---|---|---|---|
| (Intercept) | -1.236e+01 | 3.015e+00 | -4.100 | 4.14e-05 | *** |
| CorrMLTfn[first_pt:last_pt] | -1.727e-01 | 2.540e-03 | -67.974 | < 2e-16 | *** |
| f10_7eff_swarmc[first_pt:last_pt] | 2.223e-03 | 1.346e-04 | 16.514 | < 2e-16 | *** |
| abs(CoMLATQD[first_pt:last_pt]) | 4.141e-03 | 1.569e-04 | 26.400 | < 2e-16 | *** |
| DOY_fn_3[first_pt:last_pt] | 8.286e-02 | 3.036e-03 | 27.295 | < 2e-16 | *** |
| Radius_IPIR[first_pt:last_pt] | 2.002e-06 | 4.432e-07 | 4.517 | 6.28e-06 | *** |
| SYM_H[first_pt:last_pt] | -9.896e-04 | 1.382e-04 | -7.163 | 8.01e-13 | *** |
| MLT_fn_comp[first_pt:last_pt] | -3.564e-02 | 2.544e-03 | -14.010 | < 2e-16 | *** |
| Bz.stdev[first_pt:last_pt] | 1.494e-02 | 1.798e-03 | 8.306 | < 2e-16 | *** |
| DOY_fn_2[first_pt:last_pt] | -4.131e-02 | 2.695e-03 | -15.327 | < 2e-16 | *** |

**|RODI10s|, Equatorial region, Both Hemispheres(all terms considered)**

| | Estimate | Std. Error | t value | Pr(>\|t\|) | |
|---|---|---|---|---|---|
| (Intercept) | 1.803e+00 | 3.132e-03 | 575.757 | < 2e-16 | *** |
| density_DNSACC[first_pt:last_pt] | 1.621e+11 | 2.152e+09 | 75.326 | < 2e-16 | *** |
| ST_LPEXT[first_pt:last_pt] | 8.068e-03 | 1.515e-04 | 53.273 | < 2e-16 | *** |
| DOY_fn_3[first_pt:last_pt] | 5.228e-02 | 1.370e-03 | 38.147 | < 2e-16 | *** |
| DOY_fn[first_pt:last_pt] | -3.275e-02 | 1.615e-03 | -20.274 | < 2e-16 | *** |
| Hp30[first_pt:last_pt] | 3.919e-03 | 7.814e-04 | 5.015 | 5.30e-07 | *** |
| DOY_fn_2[first_pt:last_pt] | -2.340e-02 | 1.473e-03 | -15.884 | < 2e-16 | *** |
| Bx.Mean[first_pt:last_pt] | 1.333e-03 | 3.112e-04 | 4.283 | 1.85e-05 | *** |
| swden.Mean[first_pt:last_pt] | -5.391e-04 | 2.055e-04 | -2.624 | 0.0087 | ** |

```
Lat_Diff[first_pt:last_pt]        -2.043e-03  1.370e-04 -14.910  < 2e-16 ***
EIA_fn[first_pt:last_pt]          -1.767e-02  1.968e-04 -89.768  < 2e-16 ***
```

**|RODI1s FP|, Polar region, Northern Hemisphere**

```
                                   Estimate Std. Error t value Pr(>|t|)
(Intercept)                       1.330e+00  8.031e-02  16.565  < 2e-16 ***
density_DNSPOD[first_pt:last_pt]  4.394e+11  6.788e+09  64.728  < 2e-16 ***
DOY_fn[first_pt:last_pt]         -3.335e-02  2.382e-03 -14.004  < 2e-16 ***
DOY_fn_3[first_pt:last_pt]        1.101e-01  2.328e-03  47.278  < 2e-16 ***
DOY_fn_2[first_pt:last_pt]        4.036e-02  2.267e-03  17.805  < 2e-16 ***
CorrMLTfn[first_pt:last_pt]       6.509e-02  2.301e-03  28.288  < 2e-16 ***
Newell.stdev[first_pt:last_pt]    1.213e-05  1.061e-06  11.432  < 2e-16 ***
Height_HIGHRESIPIR[first_pt:last_pt] 2.776e-03  1.559e-04  17.814  < 2e-16 ***
abs(CoMLATQD[first_pt:last_pt])   3.463e-03  3.221e-04  10.752  < 2e-16 ***
swden.Mean[first_pt:last_pt]      4.579e-03  3.412e-04  13.420  < 2e-16 ***
LT_fn_comp[first_pt:last_pt]      2.209e-02  2.264e-03   9.758  < 2e-16 ***
Bx.Mean[first_pt:last_pt]        -2.662e-03  5.446e-04  -4.888 1.02e-06 ***
SYM_D[first_pt:last_pt]           2.170e-03  4.998e-04   4.343 1.41e-05 ***
```

**|RODI1s FP|, Polar region, Southern Hemisphere**

```
                                   Estimate Std. Error t value Pr(>|t|)
(Intercept)                       1.295e+00  8.291e-02  15.624  < 2e-16 ***
density_DNSPOD[first_pt:last_pt]  4.439e+11  7.045e+09  63.009  < 2e-16 ***
DOY_fn[first_pt:last_pt]         -3.847e-02  2.507e-03 -15.342  < 2e-16 ***
DOY_fn_3[first_pt:last_pt]        1.042e-01  2.442e-03  42.668  < 2e-16 ***
DOY_fn_2[first_pt:last_pt]        4.195e-02  2.439e-03  17.198  < 2e-16 ***
CorrMLTfn[first_pt:last_pt]       6.871e-02  2.479e-03  27.722  < 2e-16 ***
Newell.stdev[first_pt:last_pt]    1.186e-05  1.123e-06  10.562  < 2e-16 ***
Height_HIGHRESIPIR[first_pt:last_pt] 2.714e-03  1.635e-04  16.602  < 2e-16 ***
abs(CoMLATQD[first_pt:last_pt])   4.270e-03  3.377e-04  12.646  < 2e-16 ***
Lat_Diff[first_pt:last_pt]        1.204e-03  2.357e-04   5.109 3.26e-07 ***
swden.Mean[first_pt:last_pt]      4.569e-03  3.640e-04  12.550  < 2e-16 ***
LT_fn_comp[first_pt:last_pt]      2.857e-02  2.445e-03  11.688  < 2e-16 ***
Bx.Mean[first_pt:last_pt]        -1.950e-03  5.794e-04  -3.366 0.000763 ***
SYM_D[first_pt:last_pt]           2.168e-03  5.281e-04   4.104 4.07e-05 ***
```

**|RODI1s FP|, Auroral region, Northern Hemisphere**

```
                                   Estimate Std. Error t value Pr(>|t|)
(Intercept)                       1.447e+00  3.117e-02  46.437  < 2e-16 ***
f10_7eff_swarmc[first_pt:last_pt] 5.134e-03  8.919e-05  57.560  < 2e-16 ***
DOY_fn[first_pt:last_pt]          5.551e-02  3.431e-03  16.180  < 2e-16 ***
abs(CoMLATQD[first_pt:last_pt])   1.605e-02  3.956e-04  40.574  < 2e-16 ***
Bt.Mean[first_pt:last_pt]         1.568e-02  1.036e-03  15.137  < 2e-16 ***
DOY_fn_2[first_pt:last_pt]        3.917e-02  3.675e-03  10.660  < 2e-16 ***
LT_fn_comp[first_pt:last_pt]      2.021e-02  3.408e-03   5.931 3.09e-09 ***
Elya.stdev[first_pt:last_pt]      1.399e-04  2.043e-05   6.846 7.91e-12 ***
abs(IRC_FACTMS[first_pt:last_pt]) 2.888e-03  6.191e-04   4.665 3.11e-06 ***
ief.Mean[first_pt:last_pt]        1.766e-02  2.298e-03   7.686 1.62e-14 ***
```

**|RODI1s FP|, Auroral region, Southern Hemisphere**

```
                                   Estimate Std. Error t value Pr(>|t|)
(Intercept)                        1.536e+00  3.606e-02  42.599  < 2e-16 ***
f10_7eff_swarmc[first_pt:last_pt]  3.720e-03  9.659e-05  38.517  < 2e-16 ***
DOY_fn[first_pt:last_pt]          -1.504e-01  3.965e-03 -37.932  < 2e-16 ***
abs(CoMLATQD[first_pt:last_pt])    1.661e-02  4.642e-04  35.780  < 2e-16 ***
Bt.Mean[first_pt:last_pt]          1.222e-02  1.104e-03  11.068  < 2e-16 ***
DOY_fn_2[first_pt:last_pt]         5.894e-02  4.051e-03  14.550  < 2e-16 ***
LT_fn_comp[first_pt:last_pt]       2.515e-02  3.748e-03   6.710 2.02e-11 ***
CorrMLTfn[first_pt:last_pt]       -2.995e-02  3.822e-03  -7.837 4.95e-15 ***
Elya.stdev[first_pt:last_pt]       8.372e-05  2.208e-05   3.792  0.00015 ***
SYM_D[first_pt:last_pt]            2.535e-03  8.782e-04   2.887  0.00390 **
ief.Mean[first_pt:last_pt]         2.577e-02  2.457e-03  10.488  < 2e-16 ***
```

**|RODI1s FP|, Midlatitude region, Northern Hemisphere**

```
                                 Estimate Std. Error t value Pr(>|t|)    
(Intercept)                     6.766e-01  7.012e-02   9.650  < 2e-16 ***
abs(CoMLATQD[first_pt:last_pt])  3.567e-03  1.095e-04  32.571  < 2e-16 ***
Height_LPEXT[first_pt:last_pt]   4.069e-03  1.607e-04  25.321  < 2e-16 ***
CorrMLTfn[first_pt:last_pt]     -3.889e-02  1.705e-03 -22.815  < 2e-16 ***
DOY_fn_3[first_pt:last_pt]       9.808e-03  1.769e-03   5.544 2.99e-08 ***
Dst[first_pt:last_pt]           -1.331e-03  9.728e-05 -13.687  < 2e-16 ***
Bz.stdev[first_pt:last_pt]       8.268e-03  1.220e-03   6.774 1.29e-11 ***
MLT_fn_comp[first_pt:last_pt]   -4.429e-03  1.768e-03  -2.505   0.0123 *  
Bx.Mean[first_pt:last_pt]        1.159e-03  4.266e-04   2.717   0.0066 ** 
DOY_fn_2[first_pt:last_pt]       1.185e-02  1.901e-03   6.235 4.62e-10 ***
```

**|RODI1s FP|, Midlatitude region, Southern Hemisphere**

```
                                 Estimate Std. Error t value Pr(>|t|)    
(Intercept)                     -0.1663240  0.0730122  -2.278   0.0227 *  
abs(CoMLATQD[first_pt:last_pt])  0.0042616  0.0001371  31.088  < 2e-16 ***
Height_LPEXT[first_pt:last_pt]   0.0057363  0.0001668  34.383  < 2e-16 ***
CorrMLTfn[first_pt:last_pt]     -0.0598582  0.0019399 -30.857  < 2e-16 ***
DOY_fn_3[first_pt:last_pt]       0.0485200  0.0019689  24.644  < 2e-16 ***
Dst[first_pt:last_pt]           -0.0011284  0.0001087 -10.384  < 2e-16 ***
Bz.stdev[first_pt:last_pt]       0.0092548  0.0013824   6.695 2.21e-11 ***
MLT_fn_comp[first_pt:last_pt]   -0.0145006  0.0019777  -7.332 2.34e-13 ***
DOY_fn_2[first_pt:last_pt]       0.0170091  0.0021518   7.905 2.81e-15 ***
```

**RODI1s FP, Equatorial region, Both Hemispheres(all terms considered)**

```
                                  Estimate Std. Error t value Pr(>|t|)    
(Intercept)                      2.699e+00  3.035e-03 889.148  < 2e-16 ***
density_DNSPOD[first_pt:last_pt]  1.778e+11  2.155e+09  82.506  < 2e-16 ***
abs(CoMLATQD[first_pt:last_pt]) -5.839e-03  1.060e-04 -55.086  < 2e-16 ***
ST_LPEXT[first_pt:last_pt]       4.761e-03  1.128e-04  42.192  < 2e-16 ***
DOY_fn[first_pt:last_pt]        -9.046e-03  1.157e-03  -7.822 5.26e-15 ***
DOY_fn_3[first_pt:last_pt]       2.245e-02  1.042e-03  21.554  < 2e-16 ***
Kp[first_pt:last_pt]             5.455e-04  6.901e-05   7.905 2.69e-15 ***
DOY_fn_2[first_pt:last_pt]       4.427e-02  1.219e-03  36.302  < 2e-16 ***
Bx.Mean[first_pt:last_pt]        1.870e-03  2.685e-04   6.966 3.28e-12 ***
Lat_Diff[first_pt:last_pt]      -1.273e-03  1.108e-04 -11.494  < 2e-16 ***
swden.Mean[first_pt:last_pt]     1.143e-03  1.730e-04   6.605 4.00e-11 ***
Clock.stdev[first_pt:last_pt]   -3.378e-04  6.071e-05  -5.565 2.63e-08 ***
EIA_fn[first_pt:last_pt]        -4.174e-04  1.895e-04  -2.202   0.0276 *  
```

**Slope, Polar region, Northern Hemisphere**

```
                                  Estimate Std. Error  t value Pr(>|t|)    
(Intercept)                      3.709e+00  5.048e-02   73.473  < 2e-16 ***
DOY_fn_3[first_pt:last_pt]      -4.809e-01  4.424e-03 -108.717  < 2e-16 ***
density_DNSPOD[first_pt:last_pt] -3.194e+11  1.143e+10  -27.932  < 2e-16 ***
abs(CoMLATQD[first_pt:last_pt]) -1.524e-02  5.901e-04  -25.827  < 2e-16 ***
LT_fn[first_pt:last_pt]         -1.551e-01  4.150e-03  -37.384  < 2e-16 ***
swvel.Mean[first_pt:last_pt]    -6.501e-04  3.289e-05  -19.769  < 2e-16 ***
ief.Mean[first_pt:last_pt]      -4.504e-02  3.095e-03  -14.552  < 2e-16 ***
DOY_fn_2[first_pt:last_pt]      -5.274e-02  4.332e-03  -12.174  < 2e-16 ***
Bx.stdev[first_pt:last_pt]      -4.884e-02  3.668e-03  -13.315  < 2e-16 ***
DOY_fn[first_pt:last_pt]         3.515e-02  4.526e-03    7.766  8.3e-15 ***
LT_fn_comp[first_pt:last_pt]    -5.069e-02  4.294e-03  -11.805  < 2e-16 ***
```

**Slope, Polar region, Southern Hemisphere**

```
                                 Estimate Std. Error  t value Pr(>|t|)    
(Intercept)                      3.694e+00  5.345e-02   69.117  < 2e-16 ***
DOY_fn_3[first_pt:last_pt]      -4.750e-01  4.619e-03 -102.841  < 2e-16 ***
density_DNSPOD[first_pt:last_pt] -3.420e+11  1.192e+10  -28.694  < 2e-16 ***
abs(CoMLATQD[first_pt:last_pt]) -1.511e-02  6.281e-04  -24.057  < 2e-16 ***
LT_fn[first_pt:last_pt]         -1.592e-01  4.426e-03  -35.975  < 2e-16 ***
```

```
swvel.Mean[first_pt:last_pt]     -6.289e-04  3.481e-05  -18.065  < 2e-16 ***
ief.Mean[first_pt:last_pt]       -4.272e-02  3.228e-03  -13.235  < 2e-16 ***
DOY_fn_2[first_pt:last_pt]       -5.029e-02  4.638e-03  -10.842  < 2e-16 ***
Bx.stdev[first_pt:last_pt]       -4.622e-02  3.878e-03  -11.919  < 2e-16 ***
DOY_fn[first_pt:last_pt]          4.063e-02  4.743e-03    8.567  < 2e-16 ***
LT_fn_comp[first_pt:last_pt]     -5.920e-02  4.612e-03  -12.836  < 2e-16 ***
Lat_Diff[first_pt:last_pt]        1.764e-03  4.435e-04    3.978 6.97e-05 ***
```

**Slope, Auroral region, Northern Hemisphere**

```
                                       Estimate Std. Error t value Pr(>|t|)
(Intercept)                           1.069e+01  2.356e-01   45.35   <2e-16 ***
abs(CoMLATQD[first_pt:last_pt])      -1.766e-02  6.800e-04  -25.98   <2e-16 ***
DOY_fn[first_pt:last_pt]             -1.517e-01  5.919e-03  -25.63   <2e-16 ***
CorrMLTfn[first_pt:last_pt]           8.359e-02  6.113e-03   13.67   <2e-16 ***
Bz.stdev[first_pt:last_pt]           -4.276e-02  3.907e-03  -10.94   <2e-16 ***
swvel.Mean[first_pt:last_pt]         -5.053e-04  4.732e-05  -10.68   <2e-16 ***
ief.Mean[first_pt:last_pt]           -5.771e-02  3.995e-03  -14.45   <2e-16 ***
Height_HIGHRESIPIR[first_pt:last_pt] -1.625e-02  5.161e-04  -31.49   <2e-16 ***
```

**Slope, Auroral region, Southern Hemisphere**

```
                                       Estimate Std. Error t value Pr(>|t|)
(Intercept)                           8.005e+00  2.339e-01  34.228  < 2e-16 ***
DOY_fn_3[first_pt:last_pt]           -3.200e-01  6.462e-03 -49.522  < 2e-16 ***
abs(CoMLATQD[first_pt:last_pt])      -2.175e-02  7.683e-04 -28.307  < 2e-16 ***
CorrMLTfn[first_pt:last_pt]           1.430e-01  6.347e-03  22.537  < 2e-16 ***
Bz.stdev[first_pt:last_pt]           -3.771e-02  3.968e-03  -9.504  < 2e-16 ***
swvel.Mean[first_pt:last_pt]         -3.608e-04  4.883e-05  -7.390 1.55e-13 ***
ief.Mean[first_pt:last_pt]           -4.441e-02  4.097e-03 -10.839  < 2e-16 ***
Height_HIGHRESIPIR[first_pt:last_pt] -8.948e-03  4.908e-04 -18.230  < 2e-16 ***
By.Mean[first_pt:last_pt]            -6.538e-03  1.332e-03  -4.910 9.21e-07 ***
```

**Slope, Midlatitude region, Northern Hemisphere**

```
                                  Estimate Std. Error t value Pr(>|t|)
(Intercept)                      7.558e+01  3.848e+00  19.640  < 2e-16 ***
abs(CoMLATQD[first_pt:last_pt]) -1.893e-02  4.027e-04 -46.998  < 2e-16 ***
CorrMLTfn[first_pt:last_pt]      2.907e-01  6.249e-03  46.521  < 2e-16 ***
DOY_fn_3[first_pt:last_pt]      -1.006e-01  6.324e-03 -15.912  < 2e-16 ***
Dst[first_pt:last_pt]            5.827e-03  3.491e-04  16.689  < 2e-16 ***
MLT_fn_comp[first_pt:last_pt]    1.186e-01  6.478e-03  18.303  < 2e-16 ***
AbsBy.Mean[first_pt:last_pt]    -1.193e-02  2.524e-03  -4.726 2.31e-06 ***
Radius_IPIR[first_pt:last_pt]   -1.052e-05  5.645e-07 -18.635  < 2e-16 ***
```

**Slope, Midlatitude region, Southern Hemisphere**

```
                                    Estimate Std. Error t value Pr(>|t|)
(Intercept)                        -5.099e+01  8.575e+00  -5.946 2.79e-09 ***
abs(CoMLATQD[first_pt:last_pt])    -2.636e-02  3.682e-04 -71.577  < 2e-16 ***
CorrMLTfn[first_pt:last_pt]         2.925e-01  5.920e-03  49.417  < 2e-16 ***
f10_7eff_swarmc[first_pt:last_pt]  -5.917e-03  3.967e-04 -14.915  < 2e-16 ***
DOY_fn_3[first_pt:last_pt]         -1.232e-01  7.040e-03 -17.505  < 2e-16 ***
Dst[first_pt:last_pt]               5.379e-03  3.401e-04  15.815  < 2e-16 ***
MLT_fn_comp[first_pt:last_pt]       1.235e-01  6.010e-03  20.549  < 2e-16 ***
AbsBy.Mean[first_pt:last_pt]       -8.443e-03  2.384e-03  -3.542 0.000398 ***
Radius_IPIR[first_pt:last_pt]       8.175e-06  1.260e-06   6.486 9.02e-11 ***
```

**Slope, Equatorial region, Both Hemispheres (EIA fn was not significant, no new model fitted)**

```
                                    Estimate Std. Error t value Pr(>|t|)
(Intercept)                         6.633e-01  7.498e-03  88.453  < 2e-16 ***
density_DNSPOD[first_pt:last_pt]    1.323e+11  3.409e+09  38.805  < 2e-16 ***
DOY_fn_3[first_pt:last_pt]          8.816e-02  1.653e-03  53.319  < 2e-16 ***
DOY_fn_2[first_pt:last_pt]         -2.034e-02  1.908e-03 -10.658  < 2e-16 ***
AbsBy.Mean[first_pt:last_pt]        2.849e-03  7.082e-04   4.023 5.75e-05 ***
DOY_fn[first_pt:last_pt]            4.485e-02  1.794e-03  25.004  < 2e-16 ***
```

```
Clock.stdev[first_pt:last_pt]      -4.537e-04  9.627e-05  -4.713 2.45e-06 ***
abs(CoMLATQD[first_pt:last_pt])    -5.223e-03  2.815e-04 -18.557  < 2e-16 ***
abs(FAC_FACTMS[first_pt:last_pt]) -4.754e-04  1.648e-04  -2.884  0.00392 **
Lat_Diff[first_pt:last_pt]         -3.339e-03  1.772e-04 -18.845  < 2e-16 ***
```

## Explanatory Variables & Parameter Estimates: Version 3.2

This shows the output from R (statistical modelling software). These statistics are also available in the Excel spreadsheet Explanatory-Varibles-Parameter-Estimates.xlsx

**Transformation applied: All dependent variables except the one-dimensional spectral index p**

In our implementation, the dependent variable is transformed and the nth root is modelled. A logarithmic link function is used. Therefore, the form of the equation is:

$$\sqrt[n]{E(y)} = exp(\beta_0 + \beta_1 \cdot x_1 + \cdots + \beta_n \cdot x_n)$$

**Transformation applied: One-dimensional spectral index p**

Different transforms were applied for this dependent variable. In the polar, auroral and midlatitude regions, a normal distribution was used. The form of the equation is:

$$E(y) - \min(E(y)) + 0.01 = exp(\beta_0 + \beta_1 \cdot x_1 + \cdots + \beta_n \cdot x_n)$$

In the equatorial regions, a gamma distribution with a logarithmic link function was used. Therefore, the form of the equation is:

$$-E(y) + \max(E(y)) + 0.01 = exp(\beta_0 + \beta_1 \cdot x_1 + \cdots + \beta_n \cdot x_n)$$

```
Electron Density, Polar region, Northern Hemisphere
Model is unchanged from version 3.1 and is restated here for completeness.
                                   Estimate Std. Error t value Pr(>|t|)
(Intercept)                       3.964e+00  4.492e-02   88.26   <2e-16 ***
f10_7eff_swarmc[first_pt:last_pt] 9.824e-03  7.295e-05  134.66   <2e-16 ***
DOY_fn_3[first_pt:last_pt]        4.386e-01  4.015e-03  109.24   <2e-16 ***
LT_fn[first_pt:last_pt]           1.502e-01  3.656e-03   41.09   <2e-16 ***
DOY_fn_2[first_pt:last_pt]        7.784e-02  3.679e-03   21.16   <2e-16 ***
LT_fn_comp[first_pt:last_pt]      9.762e-02  3.559e-03   27.43   <2e-16 ***
abs(CoMLATQD[first_pt:last_pt])   5.927e-03  5.377e-04   11.02   <2e-16 ***
Hp30[first_pt:last_pt]            4.273e-02  1.864e-03   22.92   <2e-16 ***
swden.Mean[first_pt:last_pt]      8.417e-03  5.533e-04   15.21   <2e-16 ***
By.Mean[first_pt:last_pt]         1.183e-02  7.426e-04   15.94   <2e-16 ***
Electron Density, Polar region, Southern Hemisphere
Model is unchanged from version 3.1 and is restated here for completeness.
                                    Estimate Std. Error t value Pr(>|t|)
(Intercept)                        3.703e+00  2.409e-02 153.670  < 2e-16 ***
f10_7eff_swarmc[first_pt:last_pt]  8.666e-03  3.969e-05 218.332  < 2e-16 ***
DOY_fn_3[first_pt:last_pt]         4.625e-01  2.185e-03 211.659  < 2e-16 ***
LT_fn[first_pt:last_pt]            1.855e-01  2.111e-03  87.863  < 2e-16 ***
DOY_fn_2[first_pt:last_pt]         1.014e-01  2.171e-03  46.686  < 2e-16 ***
LT_fn_comp[first_pt:last_pt]       1.181e-01  2.161e-03  54.665  < 2e-16 ***
abs(CoMLATQD[first_pt:last_pt])    1.013e-02  2.935e-04  34.512  < 2e-16 ***
Hp30[first_pt:last_pt]             2.793e-02  1.164e-03  23.989  < 2e-16 ***
SYM_D[first_pt:last_pt]            3.157e-03  4.507e-04   7.004 2.51e-12 ***
swden.Mean[first_pt:last_pt]       7.502e-03  3.256e-04  23.041  < 2e-16 ***
DOY_fn[first_pt:last_pt]          -5.402e-02  2.222e-03 -24.305  < 2e-16 ***
By.Mean[first_pt:last_pt]          4.422e-03  4.473e-04   9.886  < 2e-16 ***
Lat_Diff[first_pt:last_pt]        -6.488e-04  2.106e-04  -3.081  0.00207 **
Electron Density, Auroral region, Northern Hemisphere
Model is unchanged from version 3.1 and is restated here for completeness.
                                    Estimate Std. Error t value Pr(>|t|)
(Intercept)                        4.630e+00  2.475e-02 187.074  < 2e-16 ***
f10_7eff_swarmc[first_pt:last_pt]  8.618e-03  5.401e-05 159.554  < 2e-16 ***
DOY_fn_3[first_pt:last_pt]         3.649e-01  2.608e-03 139.890  < 2e-16 ***
```

```
LT_fn[first_pt:last_pt]            1.640e-01  2.495e-03  65.719  < 2e-16 ***
LT_fn_comp[first_pt:last_pt]       6.982e-02  2.328e-03  29.990  < 2e-16 ***
DOY_fn_2[first_pt:last_pt]         3.523e-02  2.471e-03  14.256  < 2e-16 ***
Bt.Mean[first_pt:last_pt]          7.802e-03  7.104e-04  10.982  < 2e-16 ***
abs(CoMLATQD[first_pt:last_pt])    1.477e-03  2.714e-04   5.443 5.28e-08 ***
Clock.Mean[first_pt:last_pt]      -2.778e-04  4.858e-05  -5.719 1.08e-08 ***
swvel.Mean[first_pt:last_pt]      -2.036e-04  1.822e-05 -11.172  < 2e-16 ***
SYM_D[first_pt:last_pt]           -2.060e-03  5.824e-04  -3.537 0.000405 ***
```

**Electron Density, Auroral region, Southern Hemisphere**

Model is unchanged from version 3.1 and is restated here for completeness.

```
                                    Estimate Std. Error t value Pr(>|t|)
(Intercept)                        4.823e+00  2.952e-02 163.383  < 2e-16 ***
f10_7eff_swarmc[first_pt:last_pt]  6.113e-03  6.222e-05  98.248  < 2e-16 ***
DOY_fn_3[first_pt:last_pt]         5.481e-01  3.236e-03 169.396  < 2e-16 ***
LT_fn[first_pt:last_pt]            2.410e-01  2.881e-03  83.654  < 2e-16 ***
LT_fn_comp[first_pt:last_pt]       6.843e-02  2.779e-03  24.622  < 2e-16 ***
DOY_fn_2[first_pt:last_pt]         1.086e-01  2.965e-03  36.638  < 2e-16 ***
Bt.Mean[first_pt:last_pt]          1.888e-03  8.186e-04   2.306 0.021120 *
abs(CoMLATQD[first_pt:last_pt])    8.408e-04  3.348e-04   2.511 0.012030 *
Clock.Mean[first_pt:last_pt]      -4.187e-04  5.765e-05  -7.263 3.88e-13 ***
swvel.Mean[first_pt:last_pt]      -2.292e-04  2.110e-05 -10.864  < 2e-16 ***
SYM_D[first_pt:last_pt]            6.667e-03  6.551e-04  10.177  < 2e-16 ***
Bx.Mean[first_pt:last_pt]          2.436e-03  6.280e-04   3.879 0.000105 ***
```

**Electron Density, Midlatitude region, Northern Hemisphere**

Model has been replaced from version 3.1

```
                                    Estimate Std. Error  t value Pr(>|t|)
(Intercept)                        5.857e+00  8.048e-03  727.775  < 2e-16 ***
f10_7eff_swarmc[first_pt:last_pt]  5.263e-03  5.019e-05  104.860  < 2e-16 ***
DOY_fn_3[first_pt:last_pt]         1.601e-01  2.373e-03   67.476  < 2e-16 ***
abs(CoMLATQD[first_pt:last_pt])   -1.497e-02  1.328e-04 -112.771  < 2e-16 ***
LT_fn_comp[first_pt:last_pt]       5.436e-02  2.198e-03   24.729  < 2e-16 ***
AbsBy.Mean[first_pt:last_pt]       9.795e-03  9.084e-04   10.783  < 2e-16 ***
swvel.stdev[first_pt:last_pt]     -1.153e-03  3.016e-04   -3.822 0.000133 ***
DOY_fn_2[first_pt:last_pt]         2.495e-02  2.331e-03   10.703  < 2e-16 ***
Bx.Mean[first_pt:last_pt]          1.484e-03  5.195e-04    2.857 0.004276 **
```

**Electron Density, Midlatitude region, Southern Hemisphere**

Model has been replaced from version 3.1

```
                                    Estimate Std. Error  t value Pr(>|t|)
(Intercept)                        5.812e+00  8.128e-03  715.022  < 2e-16 ***
f10_7eff_swarmc[first_pt:last_pt]  6.290e-03  5.071e-05  124.055  < 2e-16 ***
DOY_fn_3[first_pt:last_pt]         3.337e-01  2.417e-03  138.063  < 2e-16 ***
abs(CoMLATQD[first_pt:last_pt])   -1.650e-02  1.337e-04 -123.427  < 2e-16 ***
LT_fn_comp[first_pt:last_pt]       6.670e-02  2.238e-03   29.802  < 2e-16 ***
AbsBy.Mean[first_pt:last_pt]       5.783e-03  9.012e-04    6.416 1.41e-10 ***
DOY_fn_2[first_pt:last_pt]         1.782e-02  2.391e-03    7.454 9.20e-14 ***
Bz.Mean[first_pt:last_pt]         -4.937e-03  6.936e-04   -7.118 1.11e-12 ***
SYM_D[first_pt:last_pt]            1.091e-02  5.835e-04   18.695  < 2e-16 ***
Bx.Mean[first_pt:last_pt]          2.488e-03  5.364e-04    4.637 3.54e-06 ***
```

**Electron Density, Equatorial region, Both Hemispheres**

Model is unchanged from version 3.1 and is restated here for completeness.

```
                                    Estimate Std. Error  t value Pr(>|t|)
(Intercept)                        1.146e+01  4.139e-03 2770.058  < 2e-16 ***
f10_7eff_swarmc[first_pt:last_pt]  1.267e-02  4.057e-05  312.170  < 2e-16 ***
LT_fn_comp[first_pt:last_pt]       5.971e-01  1.842e-03  324.210  < 2e-16 ***
LT_fn[first_pt:last_pt]            3.638e-01  1.809e-03  201.136  < 2e-16 ***
abs(CoMLATQD[first_pt:last_pt])   -1.982e-02  2.474e-04  -80.103  < 2e-16 ***
DOY_fn_3[first_pt:last_pt]         2.273e-01  1.778e-03  127.861  < 2e-16 ***
DOY_fn_2[first_pt:last_pt]         1.328e-01  1.951e-03   68.076  < 2e-16 ***
```

```
DOY_fn[first_pt:last_pt]          -1.609e-01  1.998e-03  -80.538  < 2e-16 ***
Kp[first_pt:last_pt]               1.614e-03  1.091e-04   14.801  < 2e-16 ***
Bx.Mean[first_pt:last_pt]          1.710e-03  4.326e-04    3.953 7.71e-05 ***
Lat_Diff[first_pt:last_pt]        -2.760e-03  1.847e-04  -14.942  < 2e-16 ***
EIA_fn[first_pt:last_pt]          -1.686e-02  3.714e-04  -45.410  < 2e-16 ***
```

**|Grad Ne@100km|, Polar region, Northern Hemisphere**

Model is unchanged from version 3.1 and is restated here for completeness.

```
                                    Estimate Std. Error t value Pr(>|t|)
(Intercept)                       -1.715e+00  2.168e-02  -79.09   <2e-16 ***
f10_7eff_swarmc[first_pt:last_pt]  3.951e-03  3.515e-05  112.40   <2e-16 ***
DOY_fn[first_pt:last_pt]          -6.292e-02  2.006e-03  -31.36   <2e-16 ***
Kp[first_pt:last_pt]               2.768e-03  1.061e-04   26.10   <2e-16 ***
LT_fn[first_pt:last_pt]            7.625e-02  1.880e-03   40.56   <2e-16 ***
DOY_fn_3[first_pt:last_pt]         1.199e-01  1.982e-03   60.51   <2e-16 ***
DOY_fn_2[first_pt:last_pt]         5.439e-02  1.932e-03   28.16   <2e-16 ***
abs(CoMLATQD[first_pt:last_pt])    6.814e-03  2.629e-04   25.92   <2e-16 ***
LT_fn_comp[first_pt:last_pt]       4.482e-02  1.916e-03   23.40   <2e-16 ***
swden.Mean[first_pt:last_pt]       3.628e-03  2.903e-04   12.49   <2e-16 ***
```

**|Grad Ne@100km|, Polar region, Southern Hemisphere**

Model is unchanged from version 3.1 and is restated here for completeness.

```
                                    Estimate Std. Error t value Pr(>|t|)
(Intercept)                       -1.752e+00  2.272e-02 -77.123   <2e-16 ***
f10_7eff_swarmc[first_pt:last_pt]  3.943e-03  3.742e-05 105.363   <2e-16 ***
DOY_fn[first_pt:last_pt]          -6.488e-02  2.095e-03 -30.972   <2e-16 ***
Kp[first_pt:last_pt]               2.842e-03  1.126e-04  25.237   <2e-16 ***
LT_fn[first_pt:last_pt]            7.896e-02  1.991e-03  39.659   <2e-16 ***
DOY_fn_3[first_pt:last_pt]         1.184e-01  2.060e-03  57.447   <2e-16 ***
DOY_fn_2[first_pt:last_pt]         5.248e-02  2.051e-03  25.582   <2e-16 ***
abs(CoMLATQD[first_pt:last_pt])    7.284e-03  2.768e-04  26.314   <2e-16 ***
LT_fn_comp[first_pt:last_pt]       5.118e-02  2.038e-03  25.114   <2e-16 ***
Lat_Diff[first_pt:last_pt]         4.700e-04  1.987e-04   2.366   0.0180 *
swden.Mean[first_pt:last_pt]       3.643e-03  3.073e-04  11.858   <2e-16 ***
SYM_D[first_pt:last_pt]            1.030e-03  4.253e-04   2.421   0.0155 *
```

**|Grad Ne@100km|, Auroral region, Northern Hemisphere**

Model is unchanged from version 3.1 and is restated here for completeness.

```
                                    Estimate Std. Error t value Pr(>|t|)
(Intercept)                       -1.684e+00  2.231e-02 -75.478  < 2e-16 ***
f10_7eff_swarmc[first_pt:last_pt]  3.588e-03  5.826e-05  61.589  < 2e-16 ***
DOY_fn_3[first_pt:last_pt]         8.253e-03  2.740e-03   3.012  0.00259 **
Bt.Mean[first_pt:last_pt]          1.086e-02  7.710e-04  14.086  < 2e-16 ***
LT_fn_comp[first_pt:last_pt]       3.919e-02  2.525e-03  15.518  < 2e-16 ***
abs(CoMLATQD[first_pt:last_pt])    7.838e-03  2.805e-04  27.944  < 2e-16 ***
DOY_fn_2[first_pt:last_pt]         3.266e-02  2.653e-03  12.313  < 2e-16 ***
Bz.Mean[first_pt:last_pt]         -7.242e-03  7.469e-04  -9.696  < 2e-16 ***
```

**|Grad Ne@100km|, Auroral region, Southern Hemisphere**

Model is unchanged from version 3.1 and is restated here for completeness.

```
                                    Estimate Std. Error t value Pr(>|t|)
(Intercept)                       -1.412e+00  2.555e-02 -55.277  < 2e-16 ***
f10_7eff_swarmc[first_pt:last_pt]  2.701e-03  6.173e-05  43.766  < 2e-16 ***
DOY_fn[first_pt:last_pt]          -1.643e-01  3.179e-03 -51.677  < 2e-16 ***
Bt.Mean[first_pt:last_pt]          5.756e-03  8.066e-04   7.137 9.76e-13 ***
LT_fn_comp[first_pt:last_pt]       3.800e-02  2.763e-03  13.755  < 2e-16 ***
abs(CoMLATQD[first_pt:last_pt])    5.493e-03  3.301e-04  16.641  < 2e-16 ***
DOY_fn_2[first_pt:last_pt]         3.637e-02  2.926e-03  12.430  < 2e-16 ***
LT_fn[first_pt:last_pt]            4.885e-02  2.863e-03  17.062  < 2e-16 ***
Bz.Mean[first_pt:last_pt]         -5.780e-03  7.834e-04  -7.378 1.65e-13 ***
```

**|Grad Ne@100km|, Midlatitude region, Northern Hemisphere**

Model has been replaced from version 3.1

```
                                    Estimate Std. Error  t value Pr(>|t|)    
(Intercept)                       -7.215e-01  3.707e-03 -194.636  < 2e-16 ***
f10_7eff_swarmc[first_pt:last_pt]  8.297e-04  3.472e-05   23.899  < 2e-16 ***
Newell.Mean[first_pt:last_pt]      4.160e-06  3.672e-07   11.330  < 2e-16 ***
CoMLT[first_pt:last_pt]           -2.140e-03  1.744e-04  -12.270  < 2e-16 ***
DOY_fn_2[first_pt:last_pt]        -8.828e-03  1.693e-03   -5.214 1.86e-07 ***
```

**|Grad Ne@100km|, Midlatitude region, Southern Hemisphere**

Model has been replaced from version 3.1

```
                                    Estimate Std. Error  t value Pr(>|t|)    
(Intercept)                       -7.993e-01  3.453e-03 -231.458  < 2e-16 ***
f10_7eff_swarmc[first_pt:last_pt]  1.745e-03  3.260e-05   53.528  < 2e-16 ***
DOY_fn_3[first_pt:last_pt]         5.955e-02  1.578e-03   37.747  < 2e-16 ***
Newell.Mean[first_pt:last_pt]      2.753e-06  3.405e-07    8.085 6.39e-16 ***
Bt.stdev[first_pt:last_pt]         1.044e-02  2.430e-03    4.298 1.73e-05 ***
CoMLT[first_pt:last_pt]           -4.749e-04  1.517e-04   -3.130  0.00175 ** 
DOY_fn_2[first_pt:last_pt]        -1.382e-02  1.561e-03   -8.849  < 2e-16 ***
```

**|Grad Ne@100km|, Equatorial region, Both Hemispheres**

Model has been replaced from version 3.1

```
                                    Estimate Std. Error  t value Pr(>|t|)    
(Intercept)                       -9.867e-01  2.754e-03 -358.257  < 2e-16 ***
f10_7eff_swarmc[first_pt:last_pt]  3.537e-03  2.161e-05  163.672  < 2e-16 ***
CoSZA[first_pt:last_pt]            9.160e-04  7.618e-06  120.250  < 2e-16 ***
DOY_fn[first_pt:last_pt]          -4.381e-02  1.068e-03  -41.029  < 2e-16 ***
DOY_fn_2[first_pt:last_pt]         2.959e-02  1.038e-03   28.515  < 2e-16 ***
DOY_fn_3[first_pt:last_pt]         3.993e-02  9.516e-04   41.957  < 2e-16 ***
Hp30[first_pt:last_pt]             4.341e-03  5.685e-04    7.637 2.23e-14 ***
Lat_Diff[first_pt:last_pt]        -6.050e-04  9.869e-05   -6.130 8.82e-10 ***
swden.Mean[first_pt:last_pt]      -1.378e-03  1.518e-04   -9.078  < 2e-16 ***
EIA_fn[first_pt:last_pt]          -2.253e-02  1.415e-04 -159.164  < 2e-16 ***
```

**|RODI10s|, Polar region, Northern Hemisphere**

Not required.

**|RODI10s|, Polar region, Southern Hemisphere**

Not required.

**|RODI10s|, Auroral region, Northern Hemisphere**

Not required.

**|RODI10s|, Auroral region, Southern Hemisphere**

Not required.

**|RODI10s|, Midlatitude region, Northern Hemisphere**

Not required.

**|RODI10s|, Midlatitude region, Southern Hemisphere**

Not required.

**|RODI10s|, Equatorial region, Both Hemispheres**

Not required.

**|RODI1s FP|, Polar region, Northern Hemisphere**

Not required.

**|RODI1s FP|, Polar region, Southern Hemisphere**

Not required.

**|RODI1s FP|, Auroral region, Northern Hemisphere**

Not required.

**|RODI1s FP|, Auroral region, Southern Hemisphere**

Not required.

**|RODI1s FP|, Midlatitude region, Northern Hemisphere**

Not required.

**|RODI1s FP|, Midlatitude region, Southern Hemisphere**

Not required.

**|RODI1s FP|, Equatorial region, Both Hemispheres**

Not required.

**Slope, Polar region, Northern Hemisphere**

Model has been replaced from version 3.1

```
                                     Estimate Std. Error t value Pr(>|t|)    
(Intercept)                         2.969e+00  1.053e-01  28.190  < 2e-16 ***
DOY_fn_3[first_pt:last_pt]         -4.652e-01  8.539e-03 -54.481  < 2e-16 ***
f10_7eff_swarmc[first_pt:last_pt] -5.151e-03  1.878e-04 -27.426  < 2e-16 ***
abs(CoMLATQD[first_pt:last_pt])    -6.420e-03  1.215e-03  -5.285 1.28e-07 ***
LT_fn[first_pt:last_pt]            -1.648e-01  8.349e-03 -19.735  < 2e-16 ***
swvel.Mean[first_pt:last_pt]       -7.124e-04  5.918e-05 -12.038  < 2e-16 ***
ief.Mean[first_pt:last_pt]         -3.820e-02  5.115e-03  -7.467 8.95e-14 ***
Bx.stdev[first_pt:last_pt]         -5.107e-02  6.627e-03  -7.706 1.42e-14 ***
LT_fn_comp[first_pt:last_pt]       -3.195e-02  7.900e-03  -4.045 5.28e-05 ***
```

**Slope, Polar region, Southern Hemisphere**

Model has been replaced from version 3.1

```
                                     Estimate Std. Error t value Pr(>|t|)    
(Intercept)                         3.706e+00  1.083e-01  34.208  < 2e-16 ***
DOY_fn_3[first_pt:last_pt]         -5.278e-01  1.045e-02 -50.505  < 2e-16 ***
f10_7eff_swarmc[first_pt:last_pt] -2.547e-03  2.414e-04 -10.553  < 2e-16 ***
abs(CoMLATQD[first_pt:last_pt])    -1.487e-02  1.261e-03 -11.790  < 2e-16 ***
LT_fn[first_pt:last_pt]            -2.013e-01  9.850e-03 -20.432  < 2e-16 ***
swvel.Mean[first_pt:last_pt]       -5.300e-04  7.097e-05  -7.469 9.19e-14 ***
ief.Mean[first_pt:last_pt]         -4.035e-02  6.103e-03  -6.611 4.14e-11 ***
DOY_fn_2[first_pt:last_pt]         -8.553e-02  1.060e-02  -8.068 8.52e-16 ***
Bx.stdev[first_pt:last_pt]         -4.323e-02  7.990e-03  -5.411 6.51e-08 ***
LT_fn_comp[first_pt:last_pt]       -1.030e-01  1.004e-02 -10.257  < 2e-16 ***
```

**Slope, Auroral region, Northern Hemisphere**

Model has been replaced from version 3.1

```
                                   Estimate Std. Error t value Pr(>|t|)    
(Intercept)                       3.4931104  0.0595671  58.642   <2e-16 ***
abs(CoMLATQD[first_pt:last_pt]) -0.0180200  0.0007011 -25.702   <2e-16 ***
DOY_fn[first_pt:last_pt]         -0.1478868  0.0061020 -24.236   <2e-16 ***
CorrMLTfn[first_pt:last_pt]       0.0801971  0.0063030  12.724   <2e-16 ***
Bz.stdev[first_pt:last_pt]       -0.0644200  0.0039642 -16.251   <2e-16 ***
swvel.Mean[first_pt:last_pt]     -0.0006698  0.0000485 -13.811   <2e-16 ***
ief.Mean[first_pt:last_pt]       -0.0371679  0.0040628  -9.148   <2e-16 ***
```

**Slope, Auroral region, Southern Hemisphere**

Model has been replaced from version 3.1

```
                                   Estimate Std. Error t value Pr(>|t|)    
(Intercept)                       3.901e+00  6.396e-02  60.986  < 2e-16 ***
DOY_fn_3[first_pt:last_pt]       -3.234e-01  6.530e-03 -49.525  < 2e-16 ***
abs(CoMLATQD[first_pt:last_pt]) -2.224e-02  7.763e-04 -28.642  < 2e-16 ***
CorrMLTfn[first_pt:last_pt]       1.405e-01  6.415e-03  21.894  < 2e-16 ***
Bz.stdev[first_pt:last_pt]       -4.984e-02  3.954e-03 -12.602  < 2e-16 ***
swvel.Mean[first_pt:last_pt]     -4.353e-04  4.919e-05  -8.850  < 2e-16 ***
ief.Mean[first_pt:last_pt]       -3.329e-02  4.096e-03  -8.126  4.8e-16 ***
By.Mean[first_pt:last_pt]        -6.564e-03  1.346e-03  -4.875  1.1e-06 ***
```

**Slope, Midlatitude region, Northern Hemisphere**

Model has been replaced from version 3.1

```
                                   Estimate Std. Error t value Pr(>|t|)    
(Intercept)                       3.8696576  0.0213824 180.974   <2e-16 ***
abs(CoMLATQD[first_pt:last_pt]) -0.0169025  0.0003913 -43.191   <2e-16 ***
CorrMLTfn[first_pt:last_pt]       0.2907435  0.0063058  46.107   <2e-16 ***
DOY_fn_3[first_pt:last_pt]       -0.0996266  0.0063813 -15.612   <2e-16 ***
Dst[first_pt:last_pt]             0.0071497  0.0003450  20.726   <2e-16 ***
MLT_fn_comp[first_pt:last_pt]     0.1191529  0.0065376  18.226   <2e-16 ***
AbsBy.Mean[first_pt:last_pt]     -0.0220999  0.0024863  -8.889   <2e-16 ***
```

**Slope, Midlatitude region, Southern Hemisphere**

Model has been replaced from version 3.1

```
                                     Estimate Std. Error t value Pr(>|t|)
```

```
(Intercept)                         4.6280672  0.0218641 211.675  < 2e-16 ***
abs(CoMLATQD[first_pt:last_pt])    -0.0256796  0.0003534 -72.664  < 2e-16 ***
CorrMLTfn[first_pt:last_pt]         0.2907516  0.0059186  49.125  < 2e-16 ***
f10_7eff_swarmc[first_pt:last_pt] -0.0035659  0.0001614 -22.100  < 2e-16 ***
DOY_fn_3[first_pt:last_pt]         -0.1476531  0.0059530 -24.803  < 2e-16 ***
Dst[first_pt:last_pt]               0.0057543  0.0003355  17.152  < 2e-16 ***
MLT_fn_comp[first_pt:last_pt]       0.1222217  0.0060123  20.329  < 2e-16 ***
AbsBy.Mean[first_pt:last_pt]       -0.0091325  0.0023833  -3.832 0.000128 ***
```

**Slope, Equatorial region, Both Hemispheres**

Model has been replaced from version 3.1. EIA_fn was not significant, no new model fitted.

```
                                    Estimate Std. Error t value Pr(>|t|)
(Intercept)                        4.672e-01  3.659e-03 127.673  < 2e-16 ***
f10_7eff_swarmc[first_pt:last_pt]  1.908e-03  3.349e-05  56.967  < 2e-16 ***
DOY_fn_3[first_pt:last_pt]         5.263e-02  1.172e-03  44.890  < 2e-16 ***
DOY_fn_2[first_pt:last_pt]         1.927e-02  1.358e-03  14.195  < 2e-16 ***
AbsBy.Mean[first_pt:last_pt]       1.469e-03  5.069e-04   2.898  0.00375 **
DOY_fn[first_pt:last_pt]           1.638e-02  1.267e-03  12.933  < 2e-16 ***
Clock.stdev[first_pt:last_pt]     -3.003e-04  6.819e-05  -4.404 1.06e-05 ***
abs(CoMLATQD[first_pt:last_pt])   -6.500e-04  9.960e-05  -6.526 6.76e-11 ***
Dst[first_pt:last_pt]             -8.485e-04  6.653e-05 -12.753  < 2e-16 ***
Lat_Diff[first_pt:last_pt]        -2.584e-03  1.243e-04 -20.780  < 2e-16 ***
```

## Explanatory Variables & Parameter Estimates: Version 3.3

This shows the output from R (statistical modelling software). These statistics are also available in the Excel spreadsheet Explanatory-Varibles-Parameter-Estimates.xlsx

**Transformation applied: All dependent variables except the one-dimensional spectral index p**

In our implementation, the dependent variable is transformed and the nth root is modelled. A logarithmic link function is used. Therefore, the form of the equation is:

$$\sqrt[n]{E(y)} = exp(\beta_0 + \beta_1 \cdot x_1 + \cdots + \beta_n \cdot x_n)$$

**Transformation applied: One-dimensional spectral index p**

Different transforms were applied for this dependent variable. In the polar, auroral and midlatitude regions, a normal distribution was used. The form of the equation is:

$$E(y) - \min\big(E(y)\big) + 0.01 = exp(\beta_0 + \beta_1 \cdot x_1 + \cdots + \beta_n \cdot x_n)$$

In the equatorial regions, a gamma distribution with a logarithmic link function was used. Therefore, the form of the equation is:

$$-E(y) + \max\big(E(y)\big) + 0.01 = exp(\beta_0 + \beta_1 \cdot x_1 + \cdots + \beta_n \cdot x_n)$$

**Electron Density, Polar region, Northern Hemisphere**

```
                                 Estimate Std. Error t value Pr(>|t|)    
(Intercept)                     3.944e+00  4.760e-02   82.86   <2e-16 ***
F107O[first_pt:last_pt]         9.439e-03  7.877e-05  119.84   <2e-16 ***
DOY_fn_3[first_pt:last_pt]      4.290e-01  4.244e-03  101.10   <2e-16 ***
LT_fn[first_pt:last_pt]         1.427e-01  3.867e-03   36.90   <2e-16 ***
DOY_fn_2[first_pt:last_pt]      7.269e-02  3.894e-03   18.67   <2e-16 ***
LT_fn_comp[first_pt:last_pt]    9.454e-02  3.773e-03   25.06   <2e-16 ***
abs(CoMLATQD[first_pt:last_pt]) 5.200e-03  5.683e-04    9.15   <2e-16 ***
Hp30[first_pt:last_pt]          5.185e-02  1.963e-03   26.42   <2e-16 ***
swden.Mean[first_pt:last_pt]    9.583e-03  5.860e-04   16.35   <2e-16 ***
By.Mean[first_pt:last_pt]       8.896e-03  7.888e-04   11.28   <2e-16 ***
```

**Electron Density, Polar region, Southern Hemisphere**

```
                                  Estimate Std. Error t value Pr(>|t|)    
(Intercept)                      4.1187027  0.0523794  78.632  < 2e-16 ***
F107O[first_pt:last_pt]          0.0060930  0.0001046  58.232  < 2e-16 ***
DOY_fn_3[first_pt:last_pt]       0.6128316  0.0054906 111.615  < 2e-16 ***
LT_fn[first_pt:last_pt]          0.2578271  0.0049534  52.050  < 2e-16 ***
DOY_fn_2[first_pt:last_pt]       0.1565417  0.0052169  30.007  < 2e-16 ***
LT_fn_comp[first_pt:last_pt]     0.1258049  0.0050144  25.089  < 2e-16 ***
abs(CoMLATQD[first_pt:last_pt])  0.0058358  0.0006437   9.066  < 2e-16 ***
Hp30[first_pt:last_pt]           0.0269408  0.0025961  10.377  < 2e-16 ***
SYM_D[first_pt:last_pt]          0.0062399  0.0010005   6.237 4.60e-10 ***
swden.Mean[first_pt:last_pt]     0.0096110  0.0007506  12.804  < 2e-16 ***
By.Mean[first_pt:last_pt]       -0.0045591  0.0010460  -4.359 1.32e-05 ***
```

**Electron Density, Auroral region, Northern Hemisphere**

```
                                  Estimate Std. Error t value Pr(>|t|)    
(Intercept)                      4.490e+00  2.624e-02 171.117  < 2e-16 ***
F107O[first_pt:last_pt]          8.353e-03  5.831e-05 143.254  < 2e-16 ***
DOY_fn_3[first_pt:last_pt]       3.537e-01  2.719e-03 130.070  < 2e-16 ***
LT_fn[first_pt:last_pt]          1.594e-01  2.627e-03  60.689  < 2e-16 ***
LT_fn_comp[first_pt:last_pt]     7.148e-02  2.456e-03  29.109  < 2e-16 ***
DOY_fn_2[first_pt:last_pt]       3.412e-02  2.596e-03  13.141  < 2e-16 ***
Bt.Mean[first_pt:last_pt]        1.776e-02  7.327e-04  24.244  < 2e-16 ***
abs(CoMLATQD[first_pt:last_pt])  1.391e-03  2.863e-04   4.858 1.19e-06 ***
Clock.Mean[first_pt:last_pt]    -4.523e-04  5.113e-05  -8.845  < 2e-16 ***
```

```
swvel.Mean[first_pt:last_pt]    -8.164e-05  1.906e-05  -4.284 1.85e-05 ***
```

**Electron Density, Auroral region, Southern Hemisphere**

```
                              Estimate Std. Error t value Pr(>|t|)    
(Intercept)                  4.795e+00  1.123e-02 427.117  < 2e-16 ***
F107O[first_pt:last_pt]      5.920e-03  6.479e-05  91.370  < 2e-16 ***
DOY_fn_3[first_pt:last_pt]   5.559e-01  3.290e-03 168.966  < 2e-16 ***
LT_fn[first_pt:last_pt]      2.438e-01  2.938e-03  82.990  < 2e-16 ***
LT_fn_comp[first_pt:last_pt] 6.854e-02  2.844e-03  24.102  < 2e-16 ***
DOY_fn_2[first_pt:last_pt]   1.094e-01  3.032e-03  36.074  < 2e-16 ***
Bt.Mean[first_pt:last_pt]    8.371e-03  8.153e-04  10.268  < 2e-16 ***
Clock.Mean[first_pt:last_pt] -5.760e-04  5.800e-05  -9.931  < 2e-16 ***
swvel.Mean[first_pt:last_pt] -1.629e-04  2.128e-05  -7.656 1.97e-14 ***
SYM_D[first_pt:last_pt]      8.107e-03  6.700e-04  12.099  < 2e-16 ***
Bx.Mean[first_pt:last_pt]    6.043e-03  6.460e-04   9.354  < 2e-16 ***
```

**Electron Density, Midlatitude region, Northern Hemisphere**

```
                                 Estimate Std. Error  t value Pr(>|t|)    
(Intercept)                     5.782e+00  8.469e-03  682.762   <2e-16 ***
F107O[first_pt:last_pt]         5.288e-03  5.325e-05   99.293   <2e-16 ***
DOY_fn_3[first_pt:last_pt]      1.616e-01  2.395e-03   67.500   <2e-16 ***
abs(CoMLATQD[first_pt:last_pt]) -1.490e-02  1.341e-04 -111.120   <2e-16 ***
LT_fn_comp[first_pt:last_pt]    5.468e-02  2.227e-03   24.553   <2e-16 ***
AbsBy.Mean[first_pt:last_pt]    1.557e-02  9.104e-04   17.105   <2e-16 ***
swvel.stdev[first_pt:last_pt]  -9.284e-04  3.058e-04   -3.036   0.0024 ** 
DOY_fn_2[first_pt:last_pt]      2.616e-02  2.358e-03   11.091   <2e-16 ***
Bx.Mean[first_pt:last_pt]       4.771e-03  5.279e-04    9.037   <2e-16 ***
```

**Electron Density, Midlatitude region, Southern Hemisphere**

```
                                 Estimate Std. Error  t value Pr(>|t|)    
(Intercept)                     5.739e+00  8.684e-03  660.792  < 2e-16 ***
F107O[first_pt:last_pt]         6.361e-03  5.459e-05  116.526  < 2e-16 ***
DOY_fn_3[first_pt:last_pt]      3.317e-01  2.468e-03  134.413  < 2e-16 ***
abs(CoMLATQD[first_pt:last_pt]) -1.670e-02  1.367e-04 -122.113  < 2e-16 ***
LT_fn_comp[first_pt:last_pt]    6.649e-02  2.295e-03   28.967  < 2e-16 ***
AbsBy.Mean[first_pt:last_pt]    1.203e-02  9.130e-04   13.179  < 2e-16 ***
DOY_fn_2[first_pt:last_pt]      2.178e-02  2.450e-03    8.887  < 2e-16 ***
Bz.Mean[first_pt:last_pt]      -2.841e-03  7.143e-04   -3.977 6.98e-05 ***
SYM_D[first_pt:last_pt]         1.203e-02  5.971e-04   20.140  < 2e-16 ***
Bx.Mean[first_pt:last_pt]       5.797e-03  5.518e-04   10.505  < 2e-16 ***
```

**Electron Density, Equatorial region, Both Hemispheres(exclude terms which are not available in near real time)**

```
                                 Estimate Std. Error t value Pr(>|t|)    
(Intercept)                     1.131e+01  4.668e-03 2422.54   <2e-16 ***
F107O[first_pt:last_pt]         1.301e-02  4.382e-05  296.82   <2e-16 ***
LT_fn_comp[first_pt:last_pt]    6.031e-01  1.892e-03  318.75   <2e-16 ***
LT_fn[first_pt:last_pt]         3.649e-01  1.850e-03  197.21   <2e-16 ***
abs(CoMLATQD[first_pt:last_pt]) -1.985e-02  2.537e-04  -78.25   <2e-16 ***
DOY_fn_3[first_pt:last_pt]      2.279e-01  1.820e-03  125.21   <2e-16 ***
DOY_fn_2[first_pt:last_pt]      1.367e-01  2.002e-03   68.29   <2e-16 ***
DOY_fn[first_pt:last_pt]       -1.533e-01  2.046e-03  -74.92   <2e-16 ***
Kp[first_pt:last_pt]            2.690e-03  1.114e-04   24.14   <2e-16 ***
Bx.Mean[first_pt:last_pt]       1.031e-02  4.448e-04   23.18   <2e-16 ***
Lat_Diff[first_pt:last_pt]     -3.006e-03  1.893e-04  -15.88   <2e-16 ***
EIA_fn[first_pt:last_pt]       -1.707e-02  3.807e-04  -44.85   <2e-16 ***
```

**|Grad Ne@100km|, Polar region, Northern Hemisphere**

```
                                 Estimate Std. Error t value Pr(>|t|)    
(Intercept)                    -1.536e+00  4.565e-02 -33.657  < 2e-16 ***
F107O[first_pt:last_pt]         4.481e-03  7.534e-05  59.479  < 2e-16 ***
DOY_fn[first_pt:last_pt]        7.077e-02  4.060e-03  17.434  < 2e-16 ***
Kp[first_pt:last_pt]            3.720e-03  1.927e-04  19.307  < 2e-16 ***
```

```
LT_fn[first_pt:last_pt]           9.558e-02  3.710e-03  25.761  < 2e-16 ***
DOY_fn_2[first_pt:last_pt]        2.990e-02  3.737e-03   8.000 1.32e-15 ***
abs(CoMLATQD[first_pt:last_pt])  3.287e-03  5.450e-04   6.031 1.66e-09 ***
LT_fn_comp[first_pt:last_pt]      4.133e-02  3.615e-03  11.432  < 2e-16 ***
swden.Mean[first_pt:last_pt]      4.952e-03  5.622e-04   8.809  < 2e-16 ***
```

**|Grad Ne@100km|, Polar region, Southern Hemisphere**

```
                                  Estimate Std. Error t value Pr(>|t|)
(Intercept)                      -1.808e+00  4.733e-02 -38.195  < 2e-16 ***
F107O[first_pt:last_pt]           2.609e-03  9.428e-05  27.671  < 2e-16 ***
DOY_fn[first_pt:last_pt]         -2.129e-01  4.924e-03 -43.240  < 2e-16 ***
Kp[first_pt:last_pt]              2.732e-03  2.387e-04  11.444  < 2e-16 ***
LT_fn[first_pt:last_pt]           7.474e-02  4.469e-03  16.726  < 2e-16 ***
DOY_fn_2[first_pt:last_pt]        6.304e-02  4.681e-03  13.466  < 2e-16 ***
abs(CoMLATQD[first_pt:last_pt])  9.083e-03  5.817e-04  15.615  < 2e-16 ***
LT_fn_comp[first_pt:last_pt]      6.504e-02  4.505e-03  14.437  < 2e-16 ***
swden.Mean[first_pt:last_pt]      4.894e-03  6.773e-04   7.226 5.28e-13 ***
```

**|Grad Ne@100km|, Auroral region, Northern Hemisphere**

```
                                  Estimate Std. Error t value Pr(>|t|)
(Intercept)                      -1.709e+00  2.268e-02 -75.382  < 2e-16 ***
F107O[first_pt:last_pt]           3.433e-03  6.042e-05  56.829  < 2e-16 ***
DOY_fn_3[first_pt:last_pt]        5.563e-03  2.752e-03   2.022   0.0432 *
Bt.Mean[first_pt:last_pt]         1.500e-02  7.593e-04  19.752  < 2e-16 ***
LT_fn_comp[first_pt:last_pt]      4.047e-02  2.548e-03  15.885  < 2e-16 ***
abs(CoMLATQD[first_pt:last_pt])  7.608e-03  2.828e-04  26.904  < 2e-16 ***
DOY_fn_2[first_pt:last_pt]        3.112e-02  2.675e-03  11.633  < 2e-16 ***
Bz.Mean[first_pt:last_pt]        -5.739e-03  7.579e-04  -7.572 3.77e-14 ***
```

**|Grad Ne@100km|, Auroral region, Southern Hemisphere**

```
                                  Estimate Std. Error t value Pr(>|t|)
(Intercept)                      -1.428e+00  2.582e-02 -55.313  < 2e-16 ***
F107O[first_pt:last_pt]           2.575e-03  6.311e-05  40.802  < 2e-16 ***
DOY_fn[first_pt:last_pt]         -1.676e-01  3.175e-03 -52.790  < 2e-16 ***
Bt.Mean[first_pt:last_pt]         9.164e-03  7.897e-04  11.603  < 2e-16 ***
LT_fn_comp[first_pt:last_pt]      3.882e-02  2.777e-03  13.980  < 2e-16 ***
abs(CoMLATQD[first_pt:last_pt])  5.251e-03  3.314e-04  15.846  < 2e-16 ***
DOY_fn_2[first_pt:last_pt]        3.619e-02  2.936e-03  12.327  < 2e-16 ***
LT_fn[first_pt:last_pt]           4.964e-02  2.870e-03  17.298  < 2e-16 ***
Bz.Mean[first_pt:last_pt]        -4.444e-03  7.890e-04  -5.633 1.79e-08 ***
```

**|Grad Ne@100km|, Midlatitude region, Northern Hemisphere**

```
                                Estimate Std. Error  t value Pr(>|t|)
(Intercept)                    -7.349e-01  4.057e-03 -181.154  < 2e-16 ***
F107O[first_pt:last_pt]         8.731e-04  3.658e-05   23.870  < 2e-16 ***
Newell.Mean[first_pt:last_pt]  4.237e-06  3.670e-07   11.546  < 2e-16 ***
CoMLT[first_pt:last_pt]        -2.075e-03  1.737e-04  -11.943  < 2e-16 ***
DOY_fn_2[first_pt:last_pt]     -9.134e-03  1.691e-03   -5.401 6.67e-08 ***
```

**|Grad Ne@100km|, Midlatitude region, Southern Hemisphere**

```
                                Estimate Std. Error  t value Pr(>|t|)
(Intercept)                    -8.231e-01  3.804e-03 -216.364  < 2e-16 ***
F107O[first_pt:last_pt]         1.799e-03  3.469e-05   51.856  < 2e-16 ***
DOY_fn_3[first_pt:last_pt]      5.851e-02  1.575e-03   37.144  < 2e-16 ***
Newell.Mean[first_pt:last_pt]  2.982e-06  3.412e-07    8.740  < 2e-16 ***
Bt.stdev[first_pt:last_pt]      1.221e-02  2.441e-03    5.001 5.71e-07 ***
CoMLT[first_pt:last_pt]        -4.639e-04  1.516e-04   -3.059  0.00222 **
DOY_fn_2[first_pt:last_pt]     -1.429e-02  1.564e-03   -9.139  < 2e-16 ***
```

**|Grad Ne@100km|, Equatorial region, Both Hemispheres (exclude terms which are not available in near real time)**

```
                               Estimate Std. Error  t value Pr(>|t|)
(Intercept)                  -1.027e+00  2.961e-03 -346.884  < 2e-16 ***
F107O[first_pt:last_pt]       3.577e-03  2.291e-05  156.088  < 2e-16 ***
```

```
CoSZA[first_pt:last_pt]        9.005e-04  7.603e-06  118.437  < 2e-16 ***
DOY_fn[first_pt:last_pt]      -4.136e-02  1.072e-03  -38.583  < 2e-16 ***
DOY_fn_2[first_pt:last_pt]     2.837e-02  1.044e-03   27.162  < 2e-16 ***
DOY_fn_3[first_pt:last_pt]     3.991e-02  9.538e-04   41.842  < 2e-16 ***
Hp30[first_pt:last_pt]         7.089e-03  5.721e-04   12.391  < 2e-16 ***
Bx.Mean[first_pt:last_pt]      2.643e-03  2.335e-04   11.319  < 2e-16 ***
Lat_Diff[first_pt:last_pt]    -6.483e-04  9.903e-05   -6.547 5.88e-11 ***
swden.Mean[first_pt:last_pt] -9.691e-04  1.535e-04   -6.312 2.75e-10 ***
EIA_fn[first_pt:last_pt]      -2.256e-02  1.420e-04 -158.846  < 2e-16 ***
```

**|RODI10s|, Polar region, Northern Hemisphere**

```
                                 Estimate Std. Error t value Pr(>|t|)
(Intercept)                     1.103e+00  4.998e-02  22.062  < 2e-16 ***
F107O[first_pt:last_pt]         5.854e-03  8.249e-05  70.958  < 2e-16 ***
DOY_fn[first_pt:last_pt]        1.015e-01  4.445e-03  22.829  < 2e-16 ***
LT_fn[first_pt:last_pt]         1.166e-01  4.063e-03  28.699  < 2e-16 ***
Kp[first_pt:last_pt]            4.590e-03  2.109e-04  21.763  < 2e-16 ***
abs(CoMLATQD[first_pt:last_pt]) 4.713e-03  5.967e-04   7.898 2.99e-15 ***
DOY_fn_2[first_pt:last_pt]      3.192e-02  4.092e-03   7.800 6.54e-15 ***
LT_fn_comp[first_pt:last_pt]    5.048e-02  3.958e-03  12.754  < 2e-16 ***
swden.Mean[first_pt:last_pt]    4.916e-03  6.155e-04   7.986 1.48e-15 ***
```

**|RODI10s|, Polar region, Southern Hemisphere**

```
                                  Estimate Std. Error t value Pr(>|t|)
(Intercept)                      0.7005283  0.0531501  13.180   <2e-16 ***
F107O[first_pt:last_pt]          0.0036011  0.0001059  34.014   <2e-16 ***
DOY_fn[first_pt:last_pt]        -0.2551780  0.0055283 -46.159   <2e-16 ***
LT_fn[first_pt:last_pt]          0.0902181  0.0050178  17.980   <2e-16 ***
Kp[first_pt:last_pt]             0.0032859  0.0002681  12.257   <2e-16 ***
abs(CoMLATQD[first_pt:last_pt])  0.0125128  0.0006533  19.154   <2e-16 ***
DOY_fn_2[first_pt:last_pt]       0.0784761  0.0052564  14.930   <2e-16 ***
LT_fn_comp[first_pt:last_pt]     0.0856566  0.0050589  16.932   <2e-16 ***
swden.Mean[first_pt:last_pt]     0.0063517  0.0007606   8.351   <2e-16 ***
```

**|RODI10s|, Auroral region, Northern Hemisphere**

```
                                  Estimate Std. Error t value Pr(>|t|)
(Intercept)                      5.906e-01  2.589e-02  22.813  < 2e-16 ***
F107O[first_pt:last_pt]          4.588e-03  6.660e-05  68.895  < 2e-16 ***
DOY_fn[first_pt:last_pt]         2.090e-02  3.060e-03   6.829 8.71e-12 ***
abs(CoMLATQD[first_pt:last_pt])  1.365e-02  3.214e-04  42.480  < 2e-16 ***
Bt.Mean[first_pt:last_pt]        1.927e-02  8.608e-04  22.382  < 2e-16 ***
DOY_fn_2[first_pt:last_pt]       3.701e-02  2.961e-03  12.500  < 2e-16 ***
LT_fn_comp[first_pt:last_pt]     4.590e-02  2.799e-03  16.398  < 2e-16 ***
CorrMLTfn[first_pt:last_pt]     -2.641e-02  2.938e-03  -8.989  < 2e-16 ***
Elya.stdev[first_pt:last_pt]     7.743e-05  1.733e-05   4.468 7.91e-06 ***
SYM_D[first_pt:last_pt]         -2.139e-03  7.004e-04  -3.054  0.00226 **
ief.Mean[first_pt:last_pt]       1.362e-02  1.930e-03   7.054 1.77e-12 ***
```

**|RODI10s|, Auroral region, Southern Hemisphere**

```
                                  Estimate Std. Error t value Pr(>|t|)
(Intercept)                      8.562e-01  2.985e-02  28.682  < 2e-16 ***
F107O[first_pt:last_pt]          3.512e-03  7.103e-05  49.444  < 2e-16 ***
DOY_fn[first_pt:last_pt]        -1.889e-01  3.498e-03 -54.004  < 2e-16 ***
abs(CoMLATQD[first_pt:last_pt])  1.178e-02  3.840e-04  30.663  < 2e-16 ***
Bt.Mean[first_pt:last_pt]        1.347e-02  8.868e-04  15.190  < 2e-16 ***
DOY_fn_2[first_pt:last_pt]       4.378e-02  3.286e-03  13.323  < 2e-16 ***
LT_fn_comp[first_pt:last_pt]     4.708e-02  3.126e-03  15.062  < 2e-16 ***
CorrMLTfn[first_pt:last_pt]     -2.728e-02  3.200e-03  -8.525  < 2e-16 ***
ief.Mean[first_pt:last_pt]       1.494e-02  2.012e-03   7.426 1.15e-13 ***
```

**|RODI10s|, Midlatitude region, Northern Hemisphere**

```
                                  Estimate Std. Error t value Pr(>|t|)
(Intercept)                      1.426e+00  1.019e-02 139.875  < 2e-16 ***
```

```
CorrMLTfn[first_pt:last_pt]     -1.889e-01  2.672e-03 -70.685  < 2e-16 ***
F107O[first_pt:last_pt]          1.360e-03  6.464e-05  21.045  < 2e-16 ***
abs(CoMLATQD[first_pt:last_pt])  2.512e-03  1.641e-04  15.313  < 2e-16 ***
DOY_fn_3[first_pt:last_pt]      -5.809e-03  2.897e-03  -2.005    0.045 *
SYM_H[first_pt:last_pt]         -1.115e-03  1.454e-04  -7.672 1.73e-14 ***
MLT_fn_comp[first_pt:last_pt]   -6.339e-02  2.699e-03 -23.485  < 2e-16 ***
Bz.stdev[first_pt:last_pt]       1.589e-02  1.901e-03   8.360  < 2e-16 ***
DOY_fn_2[first_pt:last_pt]      -3.242e-02  2.847e-03 -11.385  < 2e-16 ***
```

**|RODI10s|, Midlatitude region, Southern Hemisphere**

```
                                  Estimate Std. Error t value Pr(>|t|)
(Intercept)                      1.220e+00  9.576e-03 127.382  < 2e-16 ***
CorrMLTfn[first_pt:last_pt]     -1.730e-01  2.531e-03 -68.349  < 2e-16 ***
F107O[first_pt:last_pt]          2.928e-03  6.155e-05  47.570  < 2e-16 ***
abs(CoMLATQD[first_pt:last_pt])  4.219e-03  1.522e-04  27.723  < 2e-16 ***
DOY_fn_3[first_pt:last_pt]       7.378e-02  2.721e-03  27.119  < 2e-16 ***
SYM_H[first_pt:last_pt]         -1.017e-03  1.376e-04  -7.393 1.46e-13 ***
MLT_fn_comp[first_pt:last_pt]   -3.597e-02  2.543e-03 -14.144  < 2e-16 ***
Bz.stdev[first_pt:last_pt]       1.560e-02  1.795e-03   8.691  < 2e-16 ***
DOY_fn_2[first_pt:last_pt]      -4.201e-02  2.694e-03 -15.597  < 2e-16 ***
```

**|RODI10s|, Equatorial region, Both Hemispheres(exclude terms which are not available in near real time)**

```
                               Estimate Std. Error t value Pr(>|t|)
(Intercept)                   1.616e+00  3.482e-03 463.974   <2e-16 ***
F107O[first_pt:last_pt]       2.637e-03  2.784e-05  94.722   <2e-16 ***
ST_LPEXT[first_pt:last_pt]    8.439e-03  1.243e-04  67.906   <2e-16 ***
DOY_fn_3[first_pt:last_pt]    5.377e-02  1.154e-03  46.571   <2e-16 ***
DOY_fn[first_pt:last_pt]     -3.072e-02  1.312e-03 -23.414   <2e-16 ***
Hp30[first_pt:last_pt]        8.275e-03  6.947e-04  11.911   <2e-16 ***
DOY_fn_2[first_pt:last_pt]   -1.273e-02  1.289e-03  -9.878   <2e-16 ***
Bx.Mean[first_pt:last_pt]     2.673e-03  2.836e-04   9.424   <2e-16 ***
swden.Mean[first_pt:last_pt] -4.130e-04  1.868e-04  -2.211   0.0271 *
Lat_Diff[first_pt:last_pt]   -2.215e-03  1.202e-04 -18.424   <2e-16 ***
EIA_fn[first_pt:last_pt]     -1.659e-02  1.724e-04 -96.196   <2e-16 ***
```

**|RODI1s FP|, Polar region, Northern Hemisphere**

```
                                 Estimate Std. Error t value Pr(>|t|)
(Intercept)                     2.428e+00  1.165e-02 208.394  < 2e-16 ***
F107O[first_pt:last_pt]         6.859e-03  1.252e-04  54.792  < 2e-16 ***
DOY_fn[first_pt:last_pt]        7.903e-02  4.478e-03  17.650  < 2e-16 ***
DOY_fn_2[first_pt:last_pt]      4.889e-02  4.543e-03  10.761  < 2e-16 ***
CorrMLTfn[first_pt:last_pt]     7.998e-02  4.233e-03  18.892  < 2e-16 ***
Newell.stdev[first_pt:last_pt]  2.026e-05  1.943e-06  10.426  < 2e-16 ***
ief.Mean[first_pt:last_pt]     -6.769e-03  2.831e-03  -2.391 0.016808 *
swden.Mean[first_pt:last_pt]    6.539e-03  6.791e-04   9.630  < 2e-16 ***
LT_fn_comp[first_pt:last_pt]    2.734e-02  4.321e-03   6.328  2.6e-10 ***
Bx.Mean[first_pt:last_pt]      -3.824e-03  1.007e-03  -3.799 0.000146 ***
```

**|RODI1s FP|, Polar region, Southern Hemisphere**

```
                                  Estimate Std. Error t value Pr(>|t|)
(Intercept)                      2.009e+00  6.325e-02  31.760  < 2e-16 ***
F107O[first_pt:last_pt]          4.522e-03  1.701e-04  26.584  < 2e-16 ***
DOY_fn[first_pt:last_pt]        -1.755e-01  6.045e-03 -29.025  < 2e-16 ***
DOY_fn_2[first_pt:last_pt]       7.434e-02  6.221e-03  11.951  < 2e-16 ***
CorrMLTfn[first_pt:last_pt]      5.915e-02  6.241e-03   9.479  < 2e-16 ***
Newell.stdev[first_pt:last_pt]   1.863e-05  2.488e-06   7.488 8.00e-14 ***
abs(CoMLATQD[first_pt:last_pt])  8.057e-03  7.660e-04  10.518  < 2e-16 ***
swden.Mean[first_pt:last_pt]     6.205e-03  8.788e-04   7.061 1.84e-12 ***
LT_fn_comp[first_pt:last_pt]     5.975e-02  6.062e-03   9.857  < 2e-16 ***
SYM_D[first_pt:last_pt]          4.009e-03  1.202e-03   3.335 0.000859 ***
```

**|RODI1s FP|, Auroral region, Northern Hemisphere**

```
                                  Estimate Std. Error t value Pr(>|t|)
(Intercept)                      1.4360647  0.0300049  47.861  < 2e-16 ***
F107O[first_pt:last_pt]          0.0058150  0.0001067  54.521  < 2e-16 ***
DOY_fn[first_pt:last_pt]         0.0557752  0.0033358  16.720  < 2e-16 ***
abs(CoMLATQD[first_pt:last_pt])  0.0145439  0.0003715  39.152  < 2e-16 ***
Bt.Mean[first_pt:last_pt]        0.0157475  0.0010767  14.626  < 2e-16 ***
DOY_fn_2[first_pt:last_pt]       0.0495967  0.0036158  13.717  < 2e-16 ***
LT_fn_comp[first_pt:last_pt]     0.0181558  0.0033089   5.487 4.15e-08 ***
Kp[first_pt:last_pt]             0.0026764  0.0002210  12.112  < 2e-16 ***
```

**|RODI1s FP|, Auroral region, Southern Hemisphere**

```
                                  Estimate Std. Error t value Pr(>|t|)
(Intercept)                      1.463e+00  3.705e-02  39.490  < 2e-16 ***
F107O[first_pt:last_pt]          4.123e-03  1.191e-04  34.620  < 2e-16 ***
DOY_fn[first_pt:last_pt]        -1.506e-01  3.992e-03 -37.730  < 2e-16 ***
abs(CoMLATQD[first_pt:last_pt])  1.644e-02  4.683e-04  35.112  < 2e-16 ***
Bt.Mean[first_pt:last_pt]        1.619e-02  1.090e-03  14.853  < 2e-16 ***
DOY_fn_2[first_pt:last_pt]       6.961e-02  4.070e-03  17.103  < 2e-16 ***
LT_fn_comp[first_pt:last_pt]     2.459e-02  3.783e-03   6.501 8.24e-11 ***
CorrMLTfn[first_pt:last_pt]     -2.950e-02  3.853e-03  -7.658 2.01e-14 ***
Elya.stdev[first_pt:last_pt]     8.016e-05  2.232e-05   3.591 0.000331 ***
SYM_D[first_pt:last_pt]          2.987e-03  8.858e-04   3.372 0.000749 ***
ief.Mean[first_pt:last_pt]       2.211e-02  2.478e-03   8.922  < 2e-16 ***
```

**|RODI1s FP|, Midlatitude region, Northern Hemisphere**

```
                                  Estimate Std. Error t value Pr(>|t|)
(Intercept)                      2.443e+00  5.923e-03 412.421  < 2e-16 ***
abs(CoMLATQD[first_pt:last_pt])  4.354e-03  1.095e-04  39.755  < 2e-16 ***
CorrMLTfn[first_pt:last_pt]     -3.969e-02  1.765e-03 -22.489  < 2e-16 ***
DOY_fn_3[first_pt:last_pt]       1.226e-02  1.828e-03   6.705 2.07e-11 ***
Dst[first_pt:last_pt]           -1.940e-03  9.784e-05 -19.828  < 2e-16 ***
Bz.stdev[first_pt:last_pt]       1.357e-02  1.246e-03  10.894  < 2e-16 ***
MLT_fn_comp[first_pt:last_pt]   -4.551e-03  1.830e-03  -2.487   0.0129 *
Bx.Mean[first_pt:last_pt]        2.474e-03  4.385e-04   5.642 1.70e-08 ***
DOY_fn_2[first_pt:last_pt]       2.017e-02  1.940e-03  10.396  < 2e-16 ***
```

**|RODI1s FP|, Midlatitude region, Southern Hemisphere**

```
                                  Estimate Std. Error t value Pr(>|t|)
(Intercept)                      2.3305078  0.0067300 346.287  < 2e-16 ***
abs(CoMLATQD[first_pt:last_pt])  0.0068622  0.0001221  56.217  < 2e-16 ***
CorrMLTfn[first_pt:last_pt]     -0.0629984  0.0020401 -30.880  < 2e-16 ***
DOY_fn_3[first_pt:last_pt]       0.0456546  0.0020679  22.078  < 2e-16 ***
Dst[first_pt:last_pt]           -0.0019549  0.0001117 -17.504  < 2e-16 ***
Bz.stdev[first_pt:last_pt]       0.0161422  0.0014393  11.215  < 2e-16 ***
MLT_fn_comp[first_pt:last_pt]   -0.0162301  0.0020799  -7.803 6.29e-15 ***
DOY_fn_2[first_pt:last_pt]       0.0300880  0.0022341  13.468  < 2e-16 ***
```

**|RODI1s FP|, Equatorial region, Both Hemispheres(exclude terms which are not available in near real time)**

```
                                  Estimate Std. Error t value Pr(>|t|)
(Intercept)                      2.463e+00  3.822e-03 644.347  < 2e-16 ***
F107O[first_pt:last_pt]          3.458e-03  3.319e-05 104.192  < 2e-16 ***
abs(CoMLATQD[first_pt:last_pt]) -5.897e-03  1.005e-04 -58.674  < 2e-16 ***
ST_LPEXT[first_pt:last_pt]       5.670e-03  1.058e-04  53.582  < 2e-16 ***
DOY_fn[first_pt:last_pt]        -1.197e-02  1.084e-03 -11.043  < 2e-16 ***
DOY_fn_3[first_pt:last_pt]       2.399e-02  9.868e-04  24.311  < 2e-16 ***
Kp[first_pt:last_pt]             7.415e-04  6.403e-05  11.581  < 2e-16 ***
DOY_fn_2[first_pt:last_pt]       4.025e-02  1.157e-03  34.790  < 2e-16 ***
Bx.Mean[first_pt:last_pt]        1.760e-03  2.544e-04   6.917 4.64e-12 ***
Lat_Diff[first_pt:last_pt]      -1.102e-03  1.050e-04 -10.497  < 2e-16 ***
swden.Mean[first_pt:last_pt]     6.813e-04  1.641e-04   4.151 3.31e-05 ***
Clock.stdev[first_pt:last_pt]   -3.356e-04  5.749e-05  -5.837 5.32e-09 ***
```

```
EIA_fn[first_pt:last_pt]          -3.764e-04  1.796e-04  -2.095   0.0362 *
```

**Slope, Polar region, Northern Hemisphere**

```
                                   Estimate Std. Error t value Pr(>|t|)
(Intercept)                       3.028e+00  1.067e-01  28.376  < 2e-16 ***
DOY_fn_3[first_pt:last_pt]       -4.641e-01  8.600e-03 -53.963  < 2e-16 ***
F107O[first_pt:last_pt]          -5.450e-03  2.299e-04 -23.706  < 2e-16 ***
abs(CoMLATQD[first_pt:last_pt]) -5.919e-03  1.223e-03  -4.839 1.33e-06 ***
LT_fn[first_pt:last_pt]          -1.658e-01  8.414e-03 -19.703  < 2e-16 ***
swvel.Mean[first_pt:last_pt]     -7.686e-04  5.948e-05 -12.921  < 2e-16 ***
ief.Mean[first_pt:last_pt]       -3.748e-02  5.167e-03  -7.254 4.35e-13 ***
Bx.stdev[first_pt:last_pt]       -5.348e-02  6.685e-03  -8.001 1.38e-15 ***
LT_fn_comp[first_pt:last_pt]     -3.030e-02  7.968e-03  -3.802 0.000144 ***
```

**Slope, Polar region, Southern Hemisphere**

```
                                   Estimate Std. Error t value Pr(>|t|)
(Intercept)                       3.730e+00  1.095e-01  34.065  < 2e-16 ***
DOY_fn_3[first_pt:last_pt]       -5.333e-01  1.043e-02 -51.132  < 2e-16 ***
F107O[first_pt:last_pt]          -2.592e-03  2.852e-04  -9.087  < 2e-16 ***
abs(CoMLATQD[first_pt:last_pt]) -1.464e-02  1.264e-03 -11.587  < 2e-16 ***
LT_fn[first_pt:last_pt]          -2.044e-01  9.841e-03 -20.772  < 2e-16 ***
swvel.Mean[first_pt:last_pt]     -5.615e-04  7.096e-05  -7.913 2.95e-15 ***
ief.Mean[first_pt:last_pt]       -3.980e-02  6.124e-03  -6.499 8.72e-11 ***
DOY_fn_2[first_pt:last_pt]       -9.095e-02  1.060e-02  -8.581  < 2e-16 ***
Bx.stdev[first_pt:last_pt]       -4.336e-02  8.024e-03  -5.404 6.76e-08 ***
LT_fn_comp[first_pt:last_pt]     -1.047e-01  1.007e-02 -10.401  < 2e-16 ***
```

**Slope, Auroral region, Northern Hemisphere**

```
                                   Estimate Std. Error t value Pr(>|t|)
(Intercept)                       3.4931266  0.0595691  58.640   <2e-16 ***
abs(CoMLATQD[first_pt:last_pt]) -0.0180201  0.0007011 -25.701   <2e-16 ***
DOY_fn[first_pt:last_pt]         -0.1478849  0.0061022 -24.235   <2e-16 ***
CorrMLTfn[first_pt:last_pt]       0.0801890  0.0063035  12.721   <2e-16 ***
Bz.stdev[first_pt:last_pt]       -0.0644261  0.0039645 -16.251   <2e-16 ***
swvel.Mean[first_pt:last_pt]     -0.0006698  0.0000485 -13.810   <2e-16 ***
ief.Mean[first_pt:last_pt]       -0.0371718  0.0040631  -9.149   <2e-16 ***
```

**Slope, Auroral region, Southern Hemisphere**

```
                                   Estimate Std. Error t value Pr(>|t|)
(Intercept)                       3.901e+00  6.396e-02  60.986  < 2e-16 ***
DOY_fn_3[first_pt:last_pt]       -3.234e-01  6.530e-03 -49.525  < 2e-16 ***
abs(CoMLATQD[first_pt:last_pt]) -2.224e-02  7.763e-04 -28.642  < 2e-16 ***
CorrMLTfn[first_pt:last_pt]       1.405e-01  6.415e-03  21.894  < 2e-16 ***
Bz.stdev[first_pt:last_pt]       -4.984e-02  3.954e-03 -12.602  < 2e-16 ***
swvel.Mean[first_pt:last_pt]     -4.353e-04  4.919e-05  -8.850  < 2e-16 ***
ief.Mean[first_pt:last_pt]       -3.329e-02  4.096e-03  -8.126  4.8e-16 ***
By.Mean[first_pt:last_pt]        -6.564e-03  1.346e-03  -4.875  1.1e-06 ***
```

**Slope, Midlatitude region, Northern Hemisphere**

```
                                   Estimate Std. Error t value Pr(>|t|)
(Intercept)                       3.8696576  0.0213824 180.974   <2e-16 ***
abs(CoMLATQD[first_pt:last_pt]) -0.0169025  0.0003913 -43.191   <2e-16 ***
CorrMLTfn[first_pt:last_pt]       0.2907435  0.0063058  46.107   <2e-16 ***
DOY_fn_3[first_pt:last_pt]       -0.0996266  0.0063813 -15.612   <2e-16 ***
Dst[first_pt:last_pt]             0.0071497  0.0003450  20.726   <2e-16 ***
MLT_fn_comp[first_pt:last_pt]     0.1191529  0.0065376  18.226   <2e-16 ***
AbsBy.Mean[first_pt:last_pt]     -0.0220999  0.0024863  -8.889   <2e-16 ***
```

**Slope, Midlatitude region, Southern Hemisphere**

```
                                   Estimate Std. Error t value Pr(>|t|)
(Intercept)                       4.7240811  0.0242625 194.707  < 2e-16 ***
abs(CoMLATQD[first_pt:last_pt]) -0.0254578  0.0003514 -72.439  < 2e-16 ***
CorrMLTfn[first_pt:last_pt]       0.2902671  0.0058776  49.385  < 2e-16 ***
F107O[first_pt:last_pt]          -0.0043875  0.0001986 -22.095  < 2e-16 ***
```

```
DOY_fn_3[first_pt:last_pt]      -0.1425521  0.0059129 -24.109  < 2e-16 ***
Dst[first_pt:last_pt]            0.0060884  0.0003305  18.420  < 2e-16 ***
MLT_fn_comp[first_pt:last_pt]    0.1223234  0.0059895  20.423  < 2e-16 ***
AbsBy.Mean[first_pt:last_pt]    -0.0109279  0.0023572  -4.636 3.57e-06 ***
```

**Slope, Equatorial region, Both Hemispheres, EIA fn was not significant, no new model fitted.**

```
                                  Estimate Std. Error t value Pr(>|t|)
(Intercept)                      4.129e-01  4.197e-03  98.381  < 2e-16 ***
F107O[first_pt:last_pt]          2.335e-03  4.061e-05  57.511  < 2e-16 ***
DOY_fn_3[first_pt:last_pt]       5.279e-02  1.165e-03  45.323  < 2e-16 ***
DOY_fn_2[first_pt:last_pt]       2.245e-02  1.344e-03  16.703  < 2e-16 ***
AbsBy.Mean[first_pt:last_pt]     2.422e-03  5.015e-04   4.829 1.37e-06 ***
DOY_fn[first_pt:last_pt]         1.879e-02  1.259e-03  14.924  < 2e-16 ***
Clock.stdev[first_pt:last_pt]   -3.651e-04  6.790e-05  -5.378 7.54e-08 ***
abs(CoMLATQD[first_pt:last_pt]) -6.344e-04  9.913e-05  -6.400 1.56e-10 ***
Dst[first_pt:last_pt]           -1.027e-03  6.554e-05 -15.675  < 2e-16 ***
Lat_Diff[first_pt:last_pt]      -2.574e-03  1.237e-04 -20.803  < 2e-16 ***
```